\documentclass{article}
\usepackage{amsmath, amssymb}
\usepackage{amsfonts}
\usepackage{appendix}
\usepackage{graphicx}
\usepackage{color}
\usepackage{url}
\usepackage{bm}
\usepackage[all]{xy}
\usepackage{arydshln}
\usepackage{enumerate}
\usepackage{hyperref} 
\usepackage{tcolorbox} 
\tcbuselibrary{breakable}

\usepackage{booktabs}
\usepackage{threeparttable}
\usepackage{multirow}
\allowdisplaybreaks

\newcommand{\F}{{\mathbb F}}

\newcommand{\R}{{\mathbb R}}
\newcommand{\C}{{\mathbb C}}

\newcommand{\xbu}{{\bf{x}}}
\newcommand{\ybu}{{\bf{y}}}

\newcommand{\ubu}{{\bf{u}}}
\newcommand{\vbu}{{\bf{v}}}
\newcommand{\Xbu}{{\bf{X}}}
\newcommand{\Ybu}{{\bf{Y}}} 
\newcommand{\Ubu}{{\bf{U}}}
\newcommand{\Vbu}{{\bf{V}}} 
\newcommand{\gmbu}{{\bm{\gamma}}} 
\newcommand{\thbu}{{\bm{\theta}}} 
\newcommand{\albu}{{\bm{\alpha}}} 
\newcommand{\bebu}{{\bm{\beta}}}
\newcommand{\ztbu}{{\bm{\zeta}}} 
\newcommand{\ldbu}{{\bm{\lambda}}}
\newcommand{\xibu}{{\bm{\xi}}} 
\newcommand{\ombu}{{\bm{\omega}}}

\newtheorem{lemma}{Lemma}
\newtheorem{definition}{Definition}
\newtheorem{theorem}{Theorem}
\newtheorem{remark}{Remark} 
\newtheorem{corollary}{Corollary}
\newtheorem{example}{Example}
\newenvironment{proof}{{\noindent\it Proof}\quad}{\hfill $\square$\par}
\newcommand{\tabincell}[2]{\begin{tabular}{@{}#1@{}}#2\end{tabular}}

\DeclareMathOperator{\W}{W} 
\DeclareMathOperator{\N}{N}
\DeclareMathOperator{\wt}{wt} 
\newcommand{\ls}[1]
    {\dimen0=\fontdimen6\the\font\lineskip=#1\dimen0
     \advance\lineskip.5\fontdimen5\the\font
     \advance\lineskip-\dimen0
     \lineskiplimit=0.9\lineskip
     \baselineskip=\lineskip
     \advance\baselineskip\dimen0
     \normallineskip\lineskip\normallineskiplimit\lineskiplimit
     \normalbaselineskip\baselineskip
     \ignorespaces}

\begin{document}

\bibliographystyle{abbrv}

\title{Systematic Constructions of Bent-Negabent Functions, 2-Rotation Symmetric Bent-Negabent Functions and Their Duals\footnote{Fei Guo is currently with the Department of Electrical and Computer Engineering, University of Waterloo, as an exchange Ph.D student. }}  

\author{Fei Guo$^{1}$, 
	Zilong Wang$^1$, 
	Guang Gong$^2$\\
	\small $^1$ State Key Laboratory of Integrated Service Networks, Xidian University \\[-0.8ex]
	\small Xi'an, 710071, China\\
	\small $^2$Department of Electrical and Computer Engineering, University of Waterloo \\[-0.8ex]
	\small Waterloo, Ontario N2L 3G1, Canada  \\
	\small\tt guofei006@gmail.com, zlwang@xidian.edu.cn, ggong@uwaterloo.ca \\ 
}

\maketitle
\thispagestyle{plain} 
\setcounter{page}{1}

\ls{1.5}

\begin{abstract} 
	Bent-negabent functions have many important properties for their application in cryptography since they have the flat absolute spectrum under the both Walsh-Hadamard transform and nega-Hadamard transform. 
	In this paper, 
	we present four new systematic constructions of bent-negabent functions on $4k, 8k, 4k+2$ and $8k+2$ variables, respectively, by modifying the truth tables of two classes of quadratic bent-negabent functions with simple form. 
	The algebraic normal forms and duals of these constructed functions are also determined. 
	We further identify necessary and sufficient conditions for those bent-negabent functions which have the maximum algebraic degree. 
	At last, by modifying the truth tables of a class of quadratic 2-rotation symmetric bent-negabent functions, we present a construction of 2-rotation symmetric bent-negabent functions with any possible algebraic degrees. 
	Considering that there are probably no bent-negabent functions in the rotation symmetric class, it is the first significant attempt to construct bent-negabent functions in the generalized rotation symmetric class. 
\end{abstract} 


\section{Introduction} 
\label{sec:Introduction} 

\emph{Bent functions}, firstly proposed by Rothaus in 1976 \cite{Rothaus1976}, have been extensively investigated during the past few decades due to their important applications in cryptography \cite{Carlet2021}, the design of sequence \cite{Olsen1982, Abdukhalikov2021} and coding theory \cite{Calderbank1986, Ding2016, Pott2011}. 
%Until now, there are a large number of methods for constructing bent functions. 
The most distinct and useful characterization of bent functions is the so-called flat absolute Walsh spectrum, i.e., all spectral values under the Walsh-Hadamard transform have the same absolute value. 
As cryptographic primitives, bent functions have the maximum distance to the set of all affine functions. 
This implies that bent functions can contribute to the best confusion effect in cryptosystems. 
It is known that bent functions only exist on even numbers of variables, and their algebraic degrees are at most $\frac{n}{2}$ ($n$ is the number of variables, similarly hereinafter). 
Up to now, many methods for constructing bent functions have been proposed, of which a non-exhaustive list is \cite{Carlet1994, Carlet2016, Dillon1974, Dobbertin1995, Gao2012, Hodzic2020, McFarland1973, Mesnager2014, Su2017, Su2020, Tang2017, Zhang2020, Zhang2017}. 
The book \cite{Mesnager2016} provides a detailed survey of the results on bent functions.

In \cite{Parker2000, Riera2006}, the bent criterion was generalized by using a transform composed of the tensor product of the identity matrix,  the Walsh-Hadamard matrix and the nega-Hadamard matrix. 
A Boolean function is called \emph{negabent} if it has a flat absolute spectrum under the nega-Hadamard transform. 
Interestingly, negabent functions exist on both even and odd numbers of variables, and all affine functions 
%(i.e., Boolean functions of algebraic degree at most one) 
are negabent \cite{Parker2007}. 
Like bent functions, the maximum possible algebraic degree of any negabent function is $\lceil \frac{n}{2} \rceil$ \cite{Stanica2012}. 
Some constructions and characterizations of negabent functions have been addressed in \cite{Parker2007, Sarkar2012, Schmidt2008, Stanica2010, Stanica2012, Su2013, Zhou2017}. 

A Boolean function is called \emph{bent-negabent} if it is both bent and negabent. 
Some constructions of bent-negabent functions in the Maiorana-McFarland class have been proposed in \cite{Parker2007, Schmidt2008, Stanica2012}. 
The algebraic degrees of these functions are upper bounded by $\lfloor \frac{n}{4} \rfloor +1$. 
In \cite{Su2013}, a construction of $n$-variable bent-negabent functions with any possible algebraic degrees ranging from $2$ to maximum $\frac{n}{2}$ has been presented. 
All bent-negabent functions generated from this construction are in the completed Maiorana-McFarland class. 
In \cite{Zhang2015}, bent-negabent functions outside the completed Maiorana-McFarland class have been constructed under the framework of the indirect sum construction. 

\emph{Symmetric} Boolean functions are a subclass of Boolean functions whose outputs are invariant for all permutations of the inputs. 
It has been proved that a symmetric function is bent if and only if it is quadratic \cite{Savicky1994}, and a symmetric function is negabent if and only if it is affine \cite{Sarkar2009}. 
This directly implies the nonexistence of bent-negabent functions in the symmetric class. 

\emph{Rotation symmetric} Boolean functions are a subclass of Boolean functions whose outputs are invariant under the cyclic shift of the inputs. 
%Obviously, the set of rotation symmetric functions contains all symmetric functions as a special subclass. 
%This class of Boolean functions are of great significance in cryptography. 
%On one hand, search technique for finding functions with good cryptographic properties is allowed in the rotation symmetric class on suitable number of variables. 
%For example, $9$-variable Boolean functions with nonlinearity 241 was firstly found in the rotation symmetric class \cite{Kavut20071743}, which solved a long-standing problem about the bound of nonlinearity of Boolean functions on odd number of variables. 
%On the other hand, researches show that various cryptographically significant functions are covered in the rotation symmetric class \cite{Patterson1983, Mesnager2021, Tang2017}. 
To date, several constructions of rotation symmetric bent functions have been addressed in \cite{Su2017, Gao2012, Tang2017}. 
Nevertheless, whether there exist bent-negabent functions in the rotation symmetric class is still an open problem. 
The nonexistence of rotation symmetric bent-negabent functions has been investigated under several conditions \cite{Mandal2018, Mandal2020, Sarkar2015}. 
Moreover, recently in \cite{Sun2022}, it has been proved that there do not exist any rotation symmetric bent-negabent functions for almost all even numbers of variables. 

In \cite{Kavut2007321}, the rotation symmetric property was generalized to \emph{$k$-rotation symmetric }property. 
A Boolean function is called $k$-rotation symmetric if it is invariant under the $k$-cyclic shift of the inputs, but not the $l$-cyclic shift for all $1 \le l \le k-1$. 
Several constructions of 2-rotation symmetric bent functions have been proposed in \cite{Su2017, Su2020}. 
However, there are no constructions of bent-negabent functions in the generalized rotation symmetric until now. 

Among all methods for constructing bent functions, an effective one is to modify the truth tables of known bent functions. 
That is to say, given an input set and a known bent function, the outputs of the new function are complements of those of the given function for inputs in the set, and same for inputs outside the set. 
%Let $f_0$ be a bent function on $n$ variables, $S$ a nonempty subset of $\mathbb{F}_2^n$, and $\chi_S$ the characteristic function of $S$, i.e.,  
%$$\chi_S(\bm{\mathrm{x}}) = \begin{cases} 
%	1,\ \bm{\mathrm{x}} \in S, \\ 
%	0,\ \text{otherwise} 
%\end{cases}$$ 
%for all $\xbu \in \F_2^n$. 
%The construction of new Boolean functions by using $S$ to modify the truth table of $f_0$ is given by  
%\begin{align} 
%	f(\xbu) = f_0(\xbu) + \chi_S(\xbu) = \begin{cases} 
%		f_0(\xbu) + 1,\ \xbu \in S, \\ 
%		f_0(\xbu),\ \hspace{0.6cm} \text{otherwise}.  
%	\end{cases} 
%	\label{align:f_general} 
%\end{align} 
%For suitable $S$ and known bent functions $f_0$ with simple form, one can obtain bent functions from (\ref{align:f_general}). 
This method was first proposed in \cite{Su2017} to construct rotation symmetric bent functions with any possible algebraic degrees by modifying the truth table of Rothaus' bent function. 
In \cite{Zhang2020}, a different construction was also given by modifying the truth table of Rothaus' bent function. 
%Recently in \cite{Su2020}, three generic constructions of bent functions were presented: the first two are to modify the truth table of Rothaus' bent function, which contains the constructions in \cite{Su2017} and \cite{Zhang2020} as special cases; 
%the third one is to modify the truth table of the Maiorana-McFarland class of bent functions. 
Recently in \cite{Su2020}, three generic constructions of bent functions were presented by using the linear subspace and their orthogonal complement subspace to construct the inputs sets, and modifying the truth tables of Rothaus' bent function and Maiorana-McFarland class of bent functions, among which the first construction contains the constructions in \cite{Su2017} and \cite{Zhang2020} as special cases. 
However, the negabentness has not been considered  in \cite{Su2020}. 
We find that the first construction cannot give rise to negabent functions. 
For the other two constructions, 
we have done some simulations which show that some bent functions from these constructions are not negabent. 

In this paper, we generalize the constructions in \cite{Su2020} in order to get bent-negabent functions and 2-rotation symmetric bent-negabent functions. 
For the constructions in terms of modifying the truth tables of known bent-negabent functions, we analyze some sufficient conditions for the \emph{fragmentary Walsh-Hadamard transform} and the \emph{fragmentary nega-Hadamard transform}, which will be formally defined in Section \ref{section:insight}, such that the produced functions are bent-negabent. 
Using the linear subspace and the coset leader, we construct four vector sets, over which the two fragmentary transforms of given quadratic bent-negabent functions satisfy the required conditions, so we obtain four constructions of bent-negabent functions. 
First, based on a class of quadratic bent-negabent functions on $4t$ variables, we propose two methods for modifying its truth table to obtain new bent-negabent functions on $4k$ ($t = k$) and $8k$ ($t=2k$) variables, respectively. 
Second, based on a class of quadratic bent-negabent functions on $4t+2$ variables, we give two constructions of bent-negabent functions on $4k+2$ ($t=k$) and $8k+2$ ($t=2k$) variables, respectively. 
All constructions of bent-negabent functions mentioned above use quadratic bent-negabent functions instead of Rothaus' bent functions, so they are not special cases of the former two generic constructions in \cite{Su2020}. 
Although those starting functions we used are in the Maiorana-McFarland class, our constructions are still not special cases of the third generic constriction in \cite{Su2020}, because the required conditions in \cite{Su2020} are not satisfied. 
We will address this in details in Section \ref{section:comparison}. 
We also investigate the necessary and sufficient conditions such that those constructed bent-negabent functions have the maximum algebraic degree.  
Finally, we present a construction of 2-rotation symmetric bent-negabent functions with any possible algebraic degrees by modifying the truth tables of a class of quadratic 2-rotation symmetric bent-negabent functions.  
Furthermore, the algebraic normal forms and duals of all these newly constructed bent-negabent functions are determined. 

The reminder of this paper is organized as follows. 
In Section \ref{section:preliminaries}, we review some definitions and notations of Boolean functions, and some basic properties of linear subspaces and cosets. 
In Section \ref{section:insight}, we introduce a new insight into the construction of bent-negabent functions. 
In Section \ref{section:bent_negabent_4k}, we present two constructions of bent-negabent functions on $4k$ and $8k$ variables by modifying the truth tables of a class of quadratic bent-negabent functions. 
In Section \ref{section:bent_negabent_4k+2}, we provide two constructions of bent-negabent functions on $4k+2$ and $8k+2$ variables.  
We also analyze the algebraic normal forms, algebraic degrees and duals of these bent-negabent functions in their respective corresponding sections. 
In Section \ref{section:2_RS_bent_negabent}, we give a construction of 2-rotation symmetric bent-negabent functions with any possible algebraic degrees. 
In Section \ref{section:comparison}, we compare our constructions to some known results. 
Section \ref{section:conclusion} concludes this paper. 

\section{Preliminaries} 
\label{section:preliminaries} 

%In this paper, we shall denote the addition over $\mathbb{F}_2$, the real numbers set $\mathbb{R}$ and the complex numbers set $\mathbb{C}$ by $+$, since there will be no ambiguity. 
Let $\F_2, \R$ and $\C$ be the binary field, the real number field and the complex field, respectively. 
We shall use $+$ to denote the addition in $\F_2$, $\R$ and $\C$, and the actual addition is determined by the context. 
Let $\mathbb{F}_2^n$ be the $n$-dimensional vector space of $\mathbb{F}_2$, where $n$ is a positive integer.  
Given vectors $\bm{\alpha} = (a_0, \cdots, a_{n-1})$ and $\bm{\beta} = (b_0, \cdots, b_{n-1})$ in $\mathbb{F}_2^n$, we say that $\bm{\alpha}$ covers $\bm{\beta}$ if $a_i \ge b_i$ for all $0 \le i \le n-1$, and denote this relation by $\bm{\alpha} \succeq \bm{\beta}$. 
The usual scalar (or dot) product over $\mathbb{F}_2$ and the Hadamard (or term-wise) product of $\bm{\alpha}$ and $\bm{\beta}$, are respectively defined by 
\begin{align*}
& \bm{\alpha} \cdot \bm{\beta} = a_0b_0 + \cdots + a_{n-1}b_{n-1}, \\ 
& \bm{\alpha} * \bm{\beta} = (a_0b_0, \cdots, a_{n-1}b_{n-1}). 
\end{align*}
%The Hamming weight of $\bm{\alpha}$ is defined by $\wt(\alpha) = \sum_{i=0}^{n-1} a_i$.
We shall denote by $\bm{0}_n$ ($\bm{1}_n$, respectively) the all-zero vector (all-one vector, respectively) in $\mathbb{F}_2^n$, and $\bm{\mathrm{e}}_n^{\varepsilon}$ the vector in $\mathbb{F}_2^n$ with $\varepsilon \in \F_2$ in the first position and $0$ elsewhere, i.e., $\bm{\mathrm{e}}_n^{\varepsilon} = (\varepsilon, \bm{0}_{n-1})$. 
%For simplicity, we denote $\underbrace{S\times \cdots \times S}_{k\ \text{times}}$ by $S^k$, for a primitive integer $k$. 
Given a complex number $c = s + r \imath \in \mathbb{C}$, where $a, b \in \mathbb{R}$ and $\imath = \sqrt{-1}$, we denote its absolute value by $|c| = \sqrt{s^2 + r^2}$. 

%\begin{definition} \rm 
%	For a nonempty subset $H$ of $\mathbb{F}_2^n$, 
%	if $\albu + \bebu \in H$ for any vectors $\albu, \bebu \in H$, 
%	then $H$ is called a \emph{linear subspace} of $\mathbb{F}_2^n$. 
%	And $H^{\perp} = \{\xbu \in \mathbb{F}_2^n : \albu \cdot \xbu = 0,\ \forall \ \albu \in H \}$ is called the \emph{orthogonal complement subspace} of $H$. 
%\end{definition}
For a nonempty subset $H$ of $\mathbb{F}_2^n$, 
if $\albu + \bebu \in H$ for any vectors $\albu, \bebu \in H$, 
then $H$ is called a \emph{linear subspace} of $\mathbb{F}_2^n$. 
And $H^{\perp} = \{\xbu \in \mathbb{F}_2^n : \albu \cdot \xbu = 0,\ \forall \ \albu \in H \}$ is called the \emph{orthogonal complement subspace} of $H$.  
Given $\bm{\alpha} \in \mathbb{F}_2^n$ and a linear subspace $H$ of $\mathbb{F}_2^n$, a \emph{coset} of $H$ in $\F_2^n$, denoted by $C_{\albu} (H)$, is defined as the affine subspace  
\begin{align*}
	C_{\albu} (H) = \bm{\alpha} + H = \{\albu + \bebu : \bebu \in H \}. 
\end{align*} 
Then we know $H$ partitions $\mathbb{F}_2^n$ as a union of the cosets of $H$. 
Let $\dim H$ represent the dimension of $H$, i.e., $2^{\dim H} = |H|$, where $|H|$ denotes the cadinality of $H$. 
Then, we have 
\begin{align*}
	\mathbb{F}_2^n = \bigcup_{i=1}^{2^{n-\dim H}} C_{\albu_i} (H), 
\end{align*}
where $C_{\albu_i} (H) \cap C_{\albu_j} (H) = \emptyset$ if $i \ne j$, and $\{\bm{\alpha}_i \}$ is called a \emph{complete set of coset representatives} of $H$ in $\F_2^n$, denoted by $R_H = \{\bm{\alpha}_i : i = 1, \cdots, 2^{n - \dim H} \}$. 

A Boolean function on $n$ variables is a mapping from $\mathbb{F}_2^n$ to $\mathbb{F}_2$. 
By convention, we shall denote the set of all $n$-variable Boolean functions by $\mathcal{B}_n$. 
The most basic representation of $f \in \mathcal{B}_n$ is the truth table, which is a sequence of all outputs of $f$ with inputs in lexicographic order, i.e., 
\[
f = [f(0, \cdots, 0), f(1, 0, \cdots, 0), f(0, 1, \cdots, 0), \cdots, f(1, \cdots, 1)]. 
\]
For any $f \in \mathcal{B}_n$, it can be uniquely expressed by the multivariate polynomial representation, called the algebraic normal form (ANF): 
\[
f(\bm{\mathrm{x}}) = \sum_{\bm{\mathrm{u}} \in \mathbb{F}_2^n} c_{\bm{\mathrm{u}}} \bm{\mathrm{x}}^{\bm{\mathrm{u}}}, 
\]
where $\bm{\mathrm{x}} = (x_0, \cdots, x_{n-1}), \bm{\mathrm{u}} = (u_0, \cdots, u_{n-1}) \in \mathbb{F}_2^n$, $c_{\bm{\mathrm{u}}} \in \mathbb{F}_2$, and $\bm{\mathrm{x}}^{\bm{\mathrm{u}}} = \prod_{i=0}^{n-1} {x_i}^{u_i}$. 
The algebraic degree of $f$ is defined as $\deg(f) = \max_{\bm{\mathrm{u}} \in \mathbb{F}_2^n} \{\wt(\bm{\mathrm{u}}) : c_{\bm{\mathrm{u}}} \ne 0 \}$, where $\wt(\bm{\mathrm{u}}) = \sum_{i=0}^{n-1} u_i$ is the Hamming weight of $\bm{\mathrm{u}}$. 
%The support of $f$, denoted by $\sup(f)$, is defined as the set of all vectors in $\mathbb{F}_2^n$ at which $f$ is equal to one, i.e, $\sup(f) = \{\bm{\mathrm{x}} \in \mathbb{F}_2^n : f(\bm{\mathrm{x}}) = 1 \}$. 
Given a subset $S$ of $\mathbb{F}_2^n$, its characteristic function, denoted by $\chi_S$, is defined as the $n$-variable Boolean function 
$
\chi_S(\bm{\mathrm{x}}) = \begin{cases} 
	1,\ \bm{\mathrm{x}} \in S, \\ 
	0,\ \text{otherwise}. 
\end{cases}
$ 

The Walsh-Hadamard transform and the nega-Hadamard transform of $f \in \mathcal{B}_n$ at $\bm{\mathrm{u}} \in \mathbb{F}_2^n$, denoted by $\W_f(\bm{\mathrm{u}})$ and $\N_f(\bm{\mathrm{u}})$, are respectively defined by 
\begin{align} 
& \W_f(\bm{\mathrm{u}}) = \sum_{\bm{\mathrm{x}} \in \mathbb{F}_2^n} (-1)^{f(\bm{\mathrm{x}}) + \bm{\mathrm{u}} \cdot \bm{\mathrm{x}}}, \label{align:WHT}\\ 
& \N_f(\bm{\mathrm{u}}) = \sum_{\bm{\mathrm{x}} \in \mathbb{F}_2^n} (-1)^{f(\bm{\mathrm{x}}) + \bm{\mathrm{u}} \cdot \bm{\mathrm{x}}} \cdot  \imath^{\wt(\bm{\mathrm{x}})}. \label{align:NHT}
\end{align} 

\begin{definition} \rm 
	Let $n$ be a positive even integer. 
	A Boolean function $f$ on $n$ variables is called \emph{bent} if $|\W_f(\bm{\mathrm{u}})| = 2^{\frac{n}{2}}$ for all $\ubu \in \F_2^n$. 
\end{definition}

\begin{definition} \rm 
	Let $n$ be a positive integer. 
	A Boolean function $f$ on $n$ variables is called \emph{negabent} if $|\N_f(\bm{\mathrm{u}})| = 2^{\frac{n}{2}}$ for all $\ubu \in \F_2^n$. 
\end{definition}

For any bent function $f \in \mathcal{B}_n$, its \emph{dual}, denoted by $\tilde{f}\in \mathcal{B}_n$ and defined by 
\begin{align}
	2^{\frac{n}{2}}(-1)^{\tilde{f}(\bm{\mathrm{x}})} = \W_f (\bm{\mathrm{x}}),\ \text{for all}\ \bm{\mathrm{x}} \in \mathbb{F}_2^n, 
	\label{align:Dual_def}
\end{align} 
is also bent. 
A Boolean function is called \emph{bent-negabent} if it is both bent and negabent. 
From \cite[Theorem 11]{Parker2007} we know that the dual of a bent-negabent function is also bent-negabent. 

The Maiorana-McFarland class of bent functions \cite{McFarland1973, Dillon1974} contains all $2m$-variable functions of the following form: 
\begin{align} 
	f(\bm{\mathrm{x}}, \bm{\mathrm{y}}) = \bm{\mathrm{x}} \cdot \pi(\bm{\mathrm{y}}) + \varphi(\bm{\mathrm{y}}),\ \text{for all}\ \bm{\mathrm{x}}, \bm{\mathrm{y}} \in \mathbb{F}_2^{m}, 
	\label{align:MM_class}
\end{align}  
where $\pi$ is any permutation on $\mathbb{F}_2^{m}$ and $\varphi$ is any Boolean function on $m$ variables. 
It is known that the dual of $f$ in (\ref{align:MM_class}) is given by 
\begin{align}
	\tilde{f}(\bm{\mathrm{x}}, \bm{\mathrm{y}}) = \bm{\mathrm{y}} \cdot \pi^{-1}(\bm{\mathrm{x}}) + \varphi(\pi^{-1}(\bm{\mathrm{x}})), 
	\label{align:dual_MM}
\end{align}
where $\pi^{-1}$ is the inverse of $\pi$ \cite{Dillon1974}. 

Given a vector $\bm{\mathrm{x}} = (x_0, \cdots, x_{n-1}) \in \mathbb{F}_2^n$ and $0 \le l \le n-1$, we shall denote the $l$-cyclic shift of $\bm{\mathrm{x}}$ by 
$\rho_n^{l}(\bm{\mathrm{x}}) = (x_l, \cdots, x_{n-1}, x_0, \cdots, x_{l-1})$. 

\begin{definition} \rm 
	For $f \in \mathcal{B}_n$, given an integer $k|n$, if $f(\rho_n^k(\bm{\mathrm{x}})) = f(\bm{\mathrm{x}})$ for all $\bm{\mathrm{x}} \in \mathbb{F}_2^n$, and at least one $\bm{\mathrm{x}} \in \mathbb{F}_2^n$ such that $f(\rho_n^l(\bm{\mathrm{x}})) \ne f(\bm{\mathrm{x}})$ for each integer $0 < l < k$, then $f$ is called a \emph{$k$-rotation symmetric Boolean function}. 
	Especially, 1-rotation symmetric Boolean functions are the usual rotation symmetric Boolean functions. 
\end{definition} 

\section{New Insight into the Construction of Bent-Negabent Functions}
\label{section:insight}

In this section, we first give the definitions of the fragmentary Walsh-Hadamard transform and the fragmentary nega-Hadamard transform of an $n$-variable Boolean function $f$ over $T$, where $T\subset \F_2^n$. 
Then, based on these notions, we provide a new insight into the construction of bent-negabent functions.  

The notion fragmentary Walsh-Hadamard transform is presented as follows, which was introduced in \cite{Zhang2019} to construct resilient Boolean functions on an odd number of variables with strictly almost optimal nonlinearity. 

\begin{definition} \rm (\cite[Definition 1]{Zhang2019}) 
	\label{definition:fragWalsh}
	Given a function $f \in \mathcal{B}_n$ and a subset $T$ of $\F_2^n$, the \emph{fragmentary Walsh-Hadamard transform} of $f$ over $T$ at $\ubu \in \F_2^n$, denoted by $\W_{f, T}(\ubu)$, is defined by 
	\begin{align} 
		& \W_{f, T}(\bm{\mathrm{u}}) = \sum_{\bm{\mathrm{x}} \in T} (-1)^{f(\bm{\mathrm{x}}) + \bm{\mathrm{u}} \cdot \bm{\mathrm{x}}}. \label{align:PWHT} 
	\end{align} 
\end{definition} 

Similarly to Definition \ref{definition:fragWalsh}, we define the fragmentary nega-Hadamard transform. 

\begin{definition} \rm 
	Given a function $f \in \mathcal{B}_n$ and a subset $T$ of $\F_2^n$, the \emph{fragmentary nega-Hadamard transform} of $f$ over $T$ at $\ubu \in \F_2^n$, denoted by $\N_{f, T}(\ubu)$, is defined by 
	\begin{align} 
		\N_{f, T}(\bm{\mathrm{u}}) = \sum_{\bm{\mathrm{x}} \in T} (-1)^{f(\bm{\mathrm{x}}) + \bm{\mathrm{u}} \cdot \bm{\mathrm{x}}} \cdot \imath^{\wt(\bm{\mathrm{x}})}. \label{align:PNHT}
	\end{align} 
\end{definition} 

Let $n$ be an even integer and $\xbu \in \F_2^n$. 
Given a Boolean function $f_0 \in \mathcal{B}_n$ and a nonempty subset $T$ of $\mathbb{F}_2^n$, we use $T$ to modify the truth table of $f_0$ to present a construction of $n$-variable Boolean functions as 
\begin{align} 
	\label{align:f_frame_construction}
	f(\bm{\mathrm{x}}) = f_0(\bm{\mathrm{x}}) + \chi_T(\bm{\mathrm{x}}) = \begin{cases}
		f_0(\bm{\mathrm{x}}) + 1,\ \bm{\mathrm{x}} \in T, \\ 
		f_0(\bm{\mathrm{x}}),\ \hspace{0.7cm}\text{otherwise}.
	\end{cases} 
\end{align}

\begin{theorem} \rm 
	\label{theorem:frame_construction}
	With the above notations, we have the following results. 
	\begin{enumerate} [(1)]
	\item  Given a bent function $f_0$, $f$ in (\ref{align:f_frame_construction}) is bent if $\W_{f_0, T} (\ombu) = c_{\ombu} \W_{f_0} (\ombu)$, where $c_{\ombu}\in \{0, 1 \}$, for any $\ombu \in \F_2^n$. 
	Moreover, if $f$ is bent, the dual of $f$ is given by $\tilde{f}(\xbu) = \tilde{f}_0(\xbu) + \chi_{\{\ombu \in \F_2^n : c_{\ombu} = 1\}} (\xbu)$. 
	
	\item Given a negabent functions $f_0$, $f$ in (\ref{align:f_frame_construction}) is negabent if $\N_{f_0, T} (\ombu) = c_{\ombu} \N_{f_0} (\ombu)$, where $c_{\ombu} \in \{0, 1, \frac{1\pm \imath}{2} \}$, for any $\ombu \in \F_2^n$. 
	\end{enumerate}
\end{theorem}

\begin{proof} 
	By (\ref{align:WHT}), the Walsh-Hadamard transform of $f$ at $\ombu \in \F_2^n$ is given by 
	\begin{align*}
		\W_f(\ombu) =& \sum_{\xbu \in \F_2^n \setminus T} (-1)^{f_0(\xbu) + \ombu \cdot \xbu} + \sum_{\xbu \in T} (-1)^{f_0(\xbu) + 1 + \ombu\cdot \xbu} \notag \\ 
		=& \W_{f_0} (\ombu) - 2 \W_{f_0, T} (\ombu). 
	\end{align*} 
Then we have 
\begin{align*}
	\W_f(\ombu) = \begin{cases}
		\W_{f_0} (\ombu),\ \hspace{0.3cm} \W_{f_0, T} (\ombu) = 0, \mathrm{i.e.}, c_{\ombu} = 0, \\ 
		- \W_{f_0} (\ombu),\ \W_{f_0, T} (\ombu) = \W_{f_0} (\ombu), \mathrm{i.e.}, c_{\ombu} = 1. 
	\end{cases}
\end{align*} 
Hence, $f$ is bent if $c_{\ombu} \in \{0, 1 \}$ for any $\ombu \in \F_2^n$. 
Together with the definition of dual in (\ref{align:Dual_def}), we know 
\begin{align*} 
	\tilde{f}(\xbu) = \begin{cases}
		\tilde{f}_0(\xbu),\ \hspace{0.6cm} c_{\xbu} = 0, \\ 
		\tilde{f}_0(\xbu) + 1,\ c_{\xbu} = 1. 
	\end{cases}
\end{align*}
Then the assertion (1) is established. 

Similarly, $\N_f$ can be expressed by $\N_{f_0}$ and $\N_{f_0, T}$ as 
\begin{align*} 
	\N_f(\ombu) = \N_{f_0}(\ombu) - 2\N_{f_0, T}(\ombu),\ \text{for all}\ \ombu \in \F_2^n. 
\end{align*} 
Then we have 
\begin{align*}
	\N_f(\ombu) = \begin{cases}
		\N_{f_0} (\ombu),\ \ \ \ \N_{f_0, T} (\ombu) = 0, \\ 
		- \N_{f_0} (\ombu),\ \N_{f_0, T} (\ombu) = \N_{f_0}(\ombu), \\
		\imath \N_{f_0}(\ombu),\ \ \ \N_{f_0, T} (\ombu) = \frac{1-\imath}{2}\N_{f_0}(\ombu), \\ 
		- \imath \N_{f_0}(\ombu),\ \N_{f_0, T} (\ombu) = \frac{1+\imath}{2}\N_{f_0}(\ombu). 
	\end{cases}
\end{align*} 
Hence, the assertion (2) is established. 
\end{proof}

In the following two sections, with quadratic bent-negabent functions serving as $f_0$, we construct suitable sets $T$, over which the fragmentary Walsh-Hadamard transform and the fragmentary nega-Hadamard transform of $f_0$ satisfy the both conditions in Theorem \ref{theorem:frame_construction}, so that those functions produced from (\ref{align:f_frame_construction}) are bent-negabent. 

\section{Constructions of Bent-Negabent Functions on $4t$ Variables}
\label{section:bent_negabent_4k}

In this section, we present two constructions of bent-negabent functions on $4k$ variables and $8k$ variables by modifying the truth tables of a class of quadratic bent-negabent in the Maiorana-McFarland class. 
The ANFs, algebraic degrees and duals of the constructed bent-negabent functions are also analyzed. 

We first review a characterization of bent-negabent functions in the Maiorana-McFarland class \cite{Stanica2012}. 

\begin{theorem} \rm (\cite[Theorem 17]{Stanica2012})  
	\label{theorem:MM_BentNegabent} 
	Let $\pi$ be a weight-sum invariant permutation on $\F_2^m$, i.e., $\wt(\bm{\mathrm{x}}+ \bm{\mathrm{y}}) = \wt(\pi(\bm{\mathrm{x}}) + \pi(\bm{\mathrm{y}}))$ for all $\bm{\mathrm{x}}, \bm{\mathrm{y}} \in \mathbb{F}_2^m$. 
	Then $f$ in (\ref{align:MM_class}) is bent-nagabent if and only if $\varphi$ is bent. 
\end{theorem} 

Since bent functions only exist on even number of variables, in Theorem \ref{theorem:MM_BentNegabent}, $m$ has to be even. 
Hence, bent-negabent functions are produced from Theorem \ref{theorem:MM_BentNegabent} only on $4t$ variables, where $t$ is an integer. 

Let $m = 2t$.   
We shall denote $\bm{\mathrm{x}}' = (x_0, \cdots, x_{t-1}),\ \bm{\mathrm{x}}'' = (x_t, \cdots, x_{m-1}), \bm{\mathrm{y}}' = (y_0, \cdots, y_{t-1}),\ \bm{\mathrm{y}}'' = (y_t, \cdots, y_{m-1}) \in \mathbb{F}_2^t$, and $\bm{\mathrm{x}} = (\bm{\mathrm{x}}', \bm{\mathrm{x}}''),\ \bm{\mathrm{y}} = (\bm{\mathrm{y}}', \bm{\mathrm{y}}'') \in \mathbb{F}_2^m$. 
In Theorem \ref{theorem:MM_BentNegabent}, by setting $\pi$ as the identical mapping, i.e., $\pi(\ybu) = \ybu$ (a weight-sum invariant permutation obviously) and $\varphi$ as Rothaus' bent function, i.e., $\varphi(\bm{\mathrm{y}}) = \bm{\mathrm{y}}'\cdot \bm{\mathrm{y}}''$, we immediately obtain a class of quadratic $4t$-variable bent-negabent functions of the following form: 
\begin{align}
	g_0(\bm{\mathrm{x}}, \bm{\mathrm{y}}) = \bm{\mathrm{x}}\cdot \bm{\mathrm{y}} + \bm{\mathrm{y}}'\cdot \bm{\mathrm{y}}'' = \sum_{i=0}^{m-1}x_i y_i + \sum_{i=0}^{t-1}y_i y_{t+i}.   
	\label{align:g_0}
\end{align} 

Similarly, we shall denote $\bm{\mathrm{u}}' = (u_0, \cdots, u_{t-1}), \bm{\mathrm{u}}'' = (u_t, \cdots, u_{m-1}), \bm{\mathrm{v}}' = (v_0, \cdots, v_{t-1}), \bm{\mathrm{v}}'' = (v_t, \cdots, v_{m-1}) \in \mathbb{F}_2^t$, and $\bm{\mathrm{u}} = (\bm{\mathrm{u}}', \bm{\mathrm{u}}''), \bm{\mathrm{v}} = (\bm{\mathrm{v}}', \bm{\mathrm{v}}'') \in \mathbb{F}_2^{m}$. 
From (\ref{align:dual_MM}) and the proof of \cite[Theorem 17]{Stanica2012}, we know that the Walsh-Hadamard transform and the nega-Hadamard transform of $g_0$ at $(\bm{\mathrm{u}}, \bm{\mathrm{v}}) \in \mathbb{F}_2^{4t}$ are respectively given by 
\begin{align} 
	& \W_{g_0}(\bm{\mathrm{u}}, \bm{\mathrm{v}}) = 2^{2t} (-1)^{\bm{\mathrm{u}}'\cdot \bm{\mathrm{u}}''+ \bm{\mathrm{u}}\cdot \bm{\mathrm{v}}}, 
	\label{align:Walsh_{g_0}} \\ 
	& \N_{g_0}(\bm{\mathrm{u}}, \bm{\mathrm{v}}) = 2^{2t}(-1)^{(\bm{\mathrm{u}}'+ \bm{\mathrm{v}}')\cdot (\bm{\mathrm{u}}''+ \bm{\mathrm{v}}'')} \imath^{t - \wt(\bm{\mathrm{u}})}.  
	\label{align:Nega_{g_0}}
\end{align}

Given a nonempty subset $S$ of $\mathbb{F}_2^{4t}$, using it to modify the truth table of $g_0$, in the sequel, 
we obtain a systematic construction of $4t$-variable Boolean functions as 
\begin{align} 
	\label{align:g_BentNegabent}
	g(\bm{\mathrm{x}}, \bm{\mathrm{y}}) = g_0(\bm{\mathrm{x}}, \bm{\mathrm{y}}) + \chi_S(\bm{\mathrm{x}}, \bm{\mathrm{y}}) = \begin{cases}
		g_0(\bm{\mathrm{x}}, \bm{\mathrm{y}}) + 1,\ (\bm{\mathrm{x}}, \bm{\mathrm{y}}) \in S, \\ 
		g_0(\bm{\mathrm{x}}, \bm{\mathrm{y}}),\ \hspace{0.7cm}\text{otherwise}.
	\end{cases} 
\end{align}

In the following subsections, we will present two methods to define $S$  such that $g$ in (\ref{align:g_BentNegabent}) is bent-negabent. 
To avoid confusion, we will use $S_1$ and $S_2$ instead of $S$. 

To investigate the fragmentary Walsh-Hadamard transforms and the fragmentary nega-Hadamard transform, we will frequently use the exponential sum of linear functions, as shown in the following lemma. 

\begin{lemma} \rm 
	\label{lemma:ex_sum} 
	For any $\bm{\mathrm{\alpha}} \in \mathbb{F}_2^k$, we have  
	\[
	\sum_{\bm{\mathrm{x}} \in \mathbb{F}_2^k} (-1)^{\bm{\mathrm{\alpha}} \cdot \bm{\mathrm{x}}} = \begin{cases}
		2^k,\ \bm{\mathrm{\alpha}} = \bm{\mathrm{0}}_k, \\ 
		0,\ \hspace{0.2cm} \text{otherwise}. 
	\end{cases}
	\]
\end{lemma} 

\subsection{Bent-Negabent Functions on $4k$ Variables} 

In this subsection, let $k$ be an integer, and $t = k$. 
For $\gmbu = (\gmbu_1, \gmbu_2) \in \F_2^{2k}$, where $\gmbu_i \in \mathbb{F}_2^{k}$ for $i = 1, 2$,  
we shall define 
\begin{align*} 
	L_{\gmbu} = \{(\xbu, \ybu) \in \mathbb{F}_2^{4k}: (\xbu', \ybu') \in \mathbb{F}_2^{2k}, \xbu'' = \xbu' + \gmbu_1, \ybu'' = \ybu' + \gmbu_2 \}. 
\end{align*}
Let $\Gamma$ be a nonempty subset of $\mathbb{F}_2^{2k}$, 
and $S_1$ be a subset of $\mathbb{F}_2^{4k}$ defined by 
\begin{align} 
S_1 = \bigcup_{\gmbu \in \Gamma} L_{\gmbu}. 
\label{align:S1_set} 
\end{align} 

We have the following result. 

\begin{theorem} \rm 
	\label{theorem:4k} 
	Given the subset $S_1$ of $\mathbb{F}_2^{4k}$ defined in (\ref{align:S1_set}) and $g_0 \in \mathcal{B}_{4k}$ defined in (\ref{align:g_0}), 
	the $4k$-variable function $g$ in (\ref{align:g_BentNegabent})
	is bent-negabent. 
\end{theorem}

In order to prove this theorem, we need the following lemma, 
which gives the fragmentary Walsh-Hadamard transform and the fragmentary nega-Hadamard transform of $g_0$ over $S_1$. 

\begin{lemma} \rm 
	\label{lemma:4k_WHT&NHT} 
	Given the subset $S_1$ of $\mathbb{F}_2^{4k}$ defined in (\ref{align:S1_set}) and $g_0 \in \mathcal{B}_{4k}$ defined in (\ref{align:g_0}), 
	the fragmentary Walsh-Hadamard transform and the fragmentary nega-Hadamard transform of $g_0$ over $S_1$ at $(\bm{\mathrm{u}}, \bm{\mathrm{v}}) \in \mathbb{F}_2^{4k}$ are respectively given by 
	\begin{align}
		\W_{g_0, S_1}(\bm{\mathrm{u}}, \bm{\mathrm{v}}) 
		=& \begin{cases}
			\W_{g_0}(\bm{\mathrm{u}}, \bm{\mathrm{v}}),\ \bm{\gamma}_1 = \bm{\mathrm{u}}' + \bm{\mathrm{u}}'' + \bm{\mathrm{v}}'+ \bm{\mathrm{v}}'' + \bm{1}_k, \bm{\gamma}_2 = \bm{\mathrm{u}}'+ \bm{\mathrm{u}}''\ \text{for}\ \gmbu \in \Gamma, \\ 
			0,\ \hspace{1.4cm}\text{otherwise},  
			\label{align:WHT_4k} 
		\end{cases} \\ 
		\N_{g_0, S_1}(\bm{\mathrm{u}}, \bm{\mathrm{v}}) =& \begin{cases} 
			\N_{g_0}(\bm{\mathrm{u}}, \bm{\mathrm{v}}),\ \bm{\gamma}_1 = \bm{\mathrm{v}}' + \bm{\mathrm{v}}'', \bm{\gamma}_2 = \bm{\mathrm{u}}'+ \bm{\mathrm{u}}'' + \bm{\mathrm{v}}' + \bm{\mathrm{v}}'' + \bm{1}_k\ \text{for}\ \gmbu \in \Gamma, \\ 
		0,\ \hspace{1.3cm}\text{otherwise}. 
			\label{align:NHT_4k}
		\end{cases} 
	\end{align}
\end{lemma}

\begin{proof}  
	By (\ref{align:PWHT}), the fragmentary Walsh-Hadamard transform of $g_0$ over $S_1$ at $(\bm{\mathrm{u}}, \bm{\mathrm{v}}) \in \mathbb{F}_2^{4k}$ is given by 
	\begin{align*}
	\W_{g_0, S_1}(\bm{\mathrm{u}}, \bm{\mathrm{v}}) 
	%=& \sum_{(\bm{\mathrm{x}}, \bm{\mathrm{y}})\in S_1}(-1)^{\bm{\mathrm{x}}\cdot \bm{\mathrm{y}}+ \bm{\mathrm{y}}'\cdot \bm{\mathrm{y}}'' + \bm{\mathrm{u}}\cdot \bm{\mathrm{x}} + \bm{\mathrm{v}} \cdot \bm{\mathrm{y}}}  \\ 
	=& \sum_{\gmbu \in \Gamma} \sum_{(\bm{\mathrm{x}}, \bm{\mathrm{y}})\in L_{\gmbu}}(-1)^{\bm{\mathrm{x}}\cdot \bm{\mathrm{y}}+ \bm{\mathrm{y}}'\cdot \bm{\mathrm{y}}'' + \bm{\mathrm{u}}\cdot \bm{\mathrm{x}} + \bm{\mathrm{v}} \cdot \bm{\mathrm{y}}}  \\ 
	=& \sum_{\bm{\gamma} \in \Gamma } \sum_{(\xbu', \ybu')\in \mathbb{F}_2^{2k}} (-1)^{((\bm{\mathrm{y}}', \bm{\mathrm{y}}'+ \bm{\gamma}_2)+ \bm{\mathrm{u}})\cdot (\bm{\mathrm{x}}', \bm{\mathrm{x}}' + \bm{\gamma}_1) + \bm{\mathrm{y}}'\cdot (\bm{\mathrm{y}}'+ \bm{\gamma}_2) + \bm{\mathrm{v}}\cdot (\bm{\mathrm{y}}', \bm{\mathrm{y}}'+ \bm{\gamma}_2) } \\ 
	=& \sum_{\bm{\gamma} \in \Gamma }(-1)^{\bm{\mathrm{u}}'' \cdot \bm{\gamma}_1 + \bm{\mathrm{v}}'' \cdot \bm{\gamma}_2 + \bm{\gamma}_1 \cdot \bm{\gamma}_2} \sum_{\bm{\mathrm{x}}'\in \mathbb{F}_2^k} (-1)^{(\bm{\mathrm{u}}' + \bm{\mathrm{u}}'' + \bm{\gamma}_2)\cdot \bm{\mathrm{x}}'} \sum_{\bm{\mathrm{y}}'\in \mathbb{F}_2^k}(-1)^{(\bm{\mathrm{v}}'+ \bm{\mathrm{v}}''+ \bm{\gamma}_1 + \bm{\gamma}_2  + \bm{1}_k)\cdot \bm{\mathrm{y}}' }. 
	\end{align*} 

	We consider the following two cases. 
	
	(1) If there does not exist a $\gmbu$ in $\Gamma$ such that $\bm{\mathrm{u}}' + \bm{\mathrm{u}}'' + \bm{\gamma}_2 = \bm{0}_k$ and $\bm{\mathrm{v}}'+ \bm{\mathrm{v}}''+ \bm{\gamma}_1 + \bm{\gamma}_2  + \bm{1}_k = \bm{0}_k$, then we have $\W_{g_0} (\bm{\mathrm{u}}, \bm{\mathrm{v}}) = 0$ by Lemma \ref{lemma:ex_sum}. 
	
	(2) If there exists a $\gmbu$ in $\Gamma$ such that $\bm{\mathrm{u}}' + \bm{\mathrm{u}}'' + \bm{\gamma}_2 = \bm{0}_k$ and $\bm{\mathrm{v}}'+ \bm{\mathrm{v}}''+ \bm{\gamma}_1 + \bm{\gamma}_2  + \bm{1}_k = \bm{0}_k$, 
	i.e., $\bm{\gamma}_1 = \bm{\mathrm{u}}' + \bm{\mathrm{u}}'' + \bm{\mathrm{v}}'+ \bm{\mathrm{v}}'' + \bm{1}_k$, $\bm{\gamma}_2 = \bm{\mathrm{u}}'+ \bm{\mathrm{u}}''$, 
	%For $(\bm{\gamma}_1, \bm{\gamma}_2) \in \Gamma$ satisfying $\bm{\gamma}_1 = \bm{\mathrm{u}}' + \bm{\mathrm{u}}'' + \bm{\mathrm{v}}'+ \bm{\mathrm{v}}'' + \bm{1}_k$ and $\bm{\gamma}_2 = \bm{\mathrm{u}}'+ \bm{\mathrm{u}}''$, 
	it holds that  
	\begin{align*}
	\bm{\mathrm{u}}'' \cdot \bm{\gamma}_1 + \bm{\mathrm{v}}''\cdot \bm{\gamma}_2 + \bm{\gamma}_1 \cdot \bm{\gamma}_2 =& \bm{\mathrm{u}}' \cdot \bm{\gamma}_1 + \bm{\mathrm{v}}'' \cdot \bm{\gamma}_2 \\ 
	=& \bm{\mathrm{u}}'\cdot \bm{\mathrm{u}}'' + \bm{\mathrm{u}}' \cdot (\bm{\mathrm{v}}' + \bm{\mathrm{v}}'') + \bm{\mathrm{v}}'' \cdot (\bm{\mathrm{u}}'+ \bm{\mathrm{u}}'')\\ 
	=& \bm{\mathrm{u}}'\cdot \bm{\mathrm{u}}''+ \bm{\mathrm{u}}\cdot \bm{\mathrm{v}}. 
	\end{align*} 
	Together with (\ref{align:Walsh_{g_0}}), we have 
	$\W_{g_0, S_1} (\bm{\mathrm{u}}, \bm{\mathrm{v}}) =  2^{2k}(-1)^{\bm{\mathrm{u}}'\cdot \bm{\mathrm{u}}''+ \bm{\mathrm{u}}\cdot \bm{\mathrm{v}}} =  \W_{g_0}(\bm{\mathrm{u}}, \bm{\mathrm{v}})$. 

	From the two cases discussed above, (\ref{align:WHT_4k}) follows immediately. 
	%From the cases discussed above, the Walsh-Hadamard transform of $g$ at $(\bm{\mathrm{u}}, \bm{\mathrm{v}}) \in \mathbb{F}_2^{4k}$ is given by  
%	\begin{align}
%		\W_g(\bm{\mathrm{u}}, \bm{\mathrm{v}}) 
%		=& \begin{cases}
%			-\W_{g_0}(\bm{\mathrm{u}}, \bm{\mathrm{v}}),\ \bm{\gamma}_1 = \bm{\mathrm{u}}' + \bm{\mathrm{u}}'' + \bm{\mathrm{v}}'+ \bm{\mathrm{v}}'' + \bm{1}_k, \bm{\gamma}_2 = \bm{\mathrm{u}}'+ \bm{\mathrm{u}}''\ \text{for}\ (\bm{\gamma}_1, \bm{\gamma}_2) \in \Gamma, \\ 
%			\W_{g_0}(\bm{\mathrm{u}}, \bm{\mathrm{v}}),\ \hspace{0.4cm}\text{otherwise}, 
%			\label{align:WHT_4k}
%		\end{cases} 
%	\end{align}
%	Hence, $|\W_g(\bm{\mathrm{u}}, \bm{\mathrm{v}})| = |\W_{g_0}(\bm{\mathrm{u}}, \bm{\mathrm{v}})| = 2^{2k}$ for all $(\bm{\mathrm{u}}, \bm{\mathrm{v}}) \in \mathbb{F}_2^{4k}$. 
	%That is to say, $g$ is a bent function. 
	
	%Next, we prove that $g$ is negabent. 
%	From the proof of \cite[Theorem 17]{Stanica2012}, we know the nega-Hadamard transform of $g_0$ at $(\bm{\mathrm{u}}, \bm{\mathrm{v}}) \in \mathbb{F}_2^n$ is 
%	\begin{align}
%	\N_{g_0}(\bm{\mathrm{u}}, \bm{\mathrm{v}}) = 2^m (-1)^{(\bm{\mathrm{u}}'+ \bm{\mathrm{v}}')\cdot (\bm{\mathrm{u}}''+ \bm{\mathrm{v}}'')}\imath^{\frac{m}{2}-\wt(\bm{\mathrm{u}})}. 
%	\label{align:Nega_{g_0}}
%	\end{align}
	By (\ref{align:PNHT}), the fragmentary nega-Hadamard transform of $g_0$ over $S_1$ at $(\bm{\mathrm{u}}, \bm{\mathrm{v}}) \in \mathbb{F}_2^{4k}$ is given by 
	\begin{align*}
	\N_{g_0, S_1}(\bm{\mathrm{u}}, \bm{\mathrm{v}}) 
	%=& \sum_{(\bm{\mathrm{x}}, \bm{\mathrm{y}})\in S_1}(-1)^{\bm{\mathrm{x}}\cdot \bm{\mathrm{y}}+ \bm{\mathrm{y}}'\cdot \bm{\mathrm{y}}'' + \bm{\mathrm{u}}\cdot \bm{\mathrm{x}} + \bm{\mathrm{v}} \cdot \bm{\mathrm{y}}}\imath^{\wt(\bm{\mathrm{x}}, \bm{\mathrm{y}})} \\ 
	=& \sum_{\gmbu \in \Gamma} \sum_{(\bm{\mathrm{x}}, \bm{\mathrm{y}})\in L_{\gmbu}}(-1)^{\bm{\mathrm{x}}\cdot \bm{\mathrm{y}}+ \bm{\mathrm{y}}'\cdot \bm{\mathrm{y}}'' + \bm{\mathrm{u}}\cdot \bm{\mathrm{x}} + \bm{\mathrm{v}} \cdot \bm{\mathrm{y}}}\imath^{\wt(\bm{\mathrm{x}}, \bm{\mathrm{y}})} \\ 
	=& \sum_{\bm{\gamma} \in \Gamma} \sum_{(\xbu', \ybu') \in \mathbb{F}_2^{2k}} (-1)^{((\bm{\mathrm{y}}', \bm{\mathrm{y}}'+ \bm{\gamma}_2)+ \bm{\mathrm{u}})\cdot (\bm{\mathrm{x}}', \bm{\mathrm{x}}' + \bm{\gamma}_1) + \bm{\mathrm{y}}'\cdot (\bm{\mathrm{y}}'+ \bm{\gamma}_2) + \bm{\mathrm{v}}\cdot (\bm{\mathrm{y}}', \bm{\mathrm{y}}'+ \bm{\gamma}_2) } \imath^{\wt(\bm{\mathrm{x}}', \bm{\mathrm{x}}' + \bm{\gamma}_1) + \wt(\bm{\mathrm{y}}', \bm{\mathrm{y}}'+ \bm{\gamma}_2)} \\ 
	=& \sum_{\bm{\gamma} \in \Gamma} (-1)^{\bm{\mathrm{u}}'' \cdot \bm{\gamma}_1 + \bm{\mathrm{v}}'' \cdot \bm{\gamma}_2 + \bm{\gamma}_1\cdot \bm{\gamma}_2} \sum_{\bm{\mathrm{x}}'\in \mathbb{F}_2^k} (-1)^{(\bm{\mathrm{u}}' + \bm{\mathrm{u}}'' + \bm{\gamma}_2)\cdot \bm{\mathrm{x}}' } \imath^{\wt(\bm{\mathrm{x}}', \bm{\mathrm{x}}'+ \bm{\gamma}_1)} \\ 
	& \hspace{2.0cm}  \sum_{\bm{\mathrm{y}}'\in \mathbb{F}_2^k}(-1)^{(\bm{\mathrm{v}}'+ \bm{\mathrm{v}}''+ \bm{\gamma}_1 + \bm{\gamma}_2  + \bm{1}_k)\cdot \bm{\mathrm{y}}' } \imath^{\wt(\bm{\mathrm{y}}', \bm{\mathrm{y}}'+ \bm{\gamma}_2)} \\ 
	=& \sum_{\bm{\gamma} \in \Gamma} (-1)^{\bm{\mathrm{u}}'' \cdot \bm{\gamma}_1 + \bm{\mathrm{v}}'' \cdot \bm{\gamma}_2 + \bm{\gamma}_1\cdot \bm{\gamma}_2} \imath^{\wt(\bm{\gamma}_1) + \wt(\gamma_2)} \sum_{\bm{\mathrm{x}}'\in \mathbb{F}_2^k} (-1)^{(\bm{\mathrm{u}}' + \bm{\mathrm{u}}'' + \bm{\gamma}_1 + \bm{\gamma}_2 + \bm{1}_k)\cdot \bm{\mathrm{x}}'} \\ 
	& \hspace{2.0cm} \sum_{\bm{\mathrm{y}}'\in \mathbb{F}_2^k}(-1)^{(\bm{\mathrm{v}}'+ \bm{\mathrm{v}}'' + \bm{\gamma}_1 ) \cdot \bm{\mathrm{y}}'}, 
	\end{align*} 
	where the last identity holds by $\imath^{\wt(\bm{\mathrm{y}}', \bm{\mathrm{y}}'+ \bm{\gamma}_2)} = \imath^ {2\wt(\bm{\mathrm{y}}') + \wt(\bm{\gamma}_2) - 2\wt(\bm{\gamma}_2 * \bm{\mathrm{y}}')} = (-1)^{(\bm{\gamma}_2 + \bm{1}_k) \cdot \bm{\mathrm{y}}'} \imath^{\wt(\bm{\gamma}_2)}$. 
	%and the last identity holds by Lemma \ref{lemma:ex_sum}. 
	We consider the following two cases. 
	
	(1) If there dose not exist a $\bm{\gamma} \in \Gamma$ such that $\bm{\mathrm{u}}'+ \bm{\mathrm{u}}'' + \gmbu_1 + \gmbu_2 + \bm{1}_k = \bm{0}_k$ and $\bm{\mathrm{v}}' + \bm{\mathrm{v}}'' + \gmbu_1 = \bm{0}_k$, then we have $\N_{g_0, S_1}(\bm{\mathrm{u}}, \bm{\mathrm{v}}) = 0$, by Lemma \ref{lemma:ex_sum}. 
	
	(2) If there exists a $\bm{\gamma} \in \Gamma$ such that $\bm{\mathrm{u}}'+ \bm{\mathrm{u}}'' + \gmbu_1 + \gmbu_2 + \bm{1}_k = \bm{0}_k$ and $\bm{\mathrm{v}}' + \bm{\mathrm{v}}'' + \gmbu_1 = \bm{0}_k$, i.e., $\bm{\gamma}_1 = \bm{\mathrm{v}}' + \bm{\mathrm{v}}''$ and $\bm{\gamma}_2 = \bm{\mathrm{u}}'+ \bm{\mathrm{u}}'' + \bm{\mathrm{v}}' + \bm{\mathrm{v}}'' + \bm{1}_k$, 
	then we have the following derivation: 
	\begin{align}
	(-1)^{\bm{\mathrm{u}}'' \cdot \bm{\gamma}_1 + \bm{\mathrm{v}}'' \cdot \bm{\gamma}_2 + \bm{\gamma}_1\cdot \bm{\gamma}_2} \imath^{\wt(\bm{\gamma}_1) + \wt(\bm{\gamma}_2)} =& (-1)^{\bm{\mathrm{u}}'' \cdot \bm{\gamma}_1 + \bm{\mathrm{v}}'\cdot \bm{\gamma}_2} \imath^{\wt(\bm{\gamma}_1 + \bm{\gamma}_2) + 2\wt(\bm{\gamma}_1 * \bm{\gamma}_2)} \notag \\ 
	=& (-1)^{\bm{\mathrm{u}}'' \cdot(\bm{\mathrm{v}}' + \bm{\mathrm{v}}'') + \bm{\mathrm{v}}'\cdot (\bm{\mathrm{u}}'+ \bm{\mathrm{u}}''+ \bm{\mathrm{v}}'+ \bm{\mathrm{v}}''+ \bm{1}_k)} \imath^{\wt(\bm{\mathrm{u}}' + \bm{\mathrm{u}}'' + \bm{1}_k) + 2\wt((\bm{\mathrm{v}}'+ \bm{\mathrm{v}}'') * (\bm{\mathrm{u}}' + \bm{\mathrm{u}}''))} \notag \\ 
	=& (-1)^{\bm{\mathrm{u}}'\cdot \bm{\mathrm{v}}'' + \bm{\mathrm{u}}''\cdot \bm{\mathrm{v}}' +  \bm{\mathrm{v}}' \cdot \bm{\mathrm{v}}''} \imath^{k - \wt(\bm{\mathrm{u}}'+ \bm{\mathrm{u}}'')} \notag \\ 
	=& (-1)^{\bm{\mathrm{u}}'\cdot \bm{\mathrm{v}}'' + \bm{\mathrm{u}}''\cdot \bm{\mathrm{v}}' + \bm{\mathrm{v}}' \cdot \bm{\mathrm{v}}''} \imath^{k - \wt(\bm{\mathrm{u}}) + 2\wt(\bm{\mathrm{u}}' * \bm{\mathrm{u}}'')} \notag \\ 
	=& (-1)^{(\bm{\mathrm{u}}'+ \bm{\mathrm{v}}')\cdot (\bm{\mathrm{u}}''+ \bm{\mathrm{v}}'')} \imath^{k - \wt(\bm{\mathrm{u}})}.  
	\label{align:inter_Res1}
	\end{align} 
	Together with (\ref{align:Nega_{g_0}}), we have $\N_{g_0, S_1}(\bm{\mathrm{u}}, \bm{\mathrm{v}}) = 2^{2k} (-1)^{(\bm{\mathrm{u}}'+ \bm{\mathrm{v}}')\cdot (\bm{\mathrm{u}}''+ \bm{\mathrm{v}}'')} \imath^{k - \wt(\bm{\mathrm{u}})} = \N_{g_0} (\bm{\mathrm{u}}, \bm{\mathrm{v}})$. 
	
	Then (\ref{align:WHT_4k}) follows from the cases discussed above. 
\end{proof}

\textbf{Proof of Theorem \ref{theorem:4k}}: 
It is an immediate consequence of Lemma \ref{lemma:4k_WHT&NHT} and Theorem \ref{theorem:frame_construction}. 
{\hfill $\square$\par} 

Next, we analyze the ANF and the algebraic degree of $g$ in (\ref{align:g_BentNegabent}). 
We need the following lemma, which comes from the proof of \cite[Lemma 4]{Su2017}. 

\begin{lemma} \rm 
	\label{lemma:chi_S_beta}
	Given $\bm{\beta} \in \mathbb{F}_2^k$, we denote by $S_{\bm{\beta}}$ the set $\{\bm{\mathrm{x}} \in \mathbb{F}_2^{2k} : \bm{\mathrm{x}}' \in \mathbb{F}_2^k, \bm{\mathrm{x}}'' = \bm{\mathrm{x}}' + \bm{\beta} \}$. 
	Then the ANF of  the characteristic function  of $S_{\bm{\beta}}$ is given by 
	\[
	\chi_{S_{\bm{\beta}}} (\bm{\mathrm{x}})= 
	\sum_{\mbox{\tiny $\begin{array} {c} \bm{\mathrm{u}}' * \bm{\mathrm{u}}'' = \bm{0}_k \\ \bm{\mathrm{u}}' + \bm{\mathrm{u}}'' \succeq \bm{\beta} \end{array}$}} \bm{\mathrm{x}}^{\bm{\mathrm{u}}}. 
	\] 
\end{lemma}

By Lemma \ref{lemma:chi_S_beta}, we give the ANF of $g$ in (\ref{align:g_BentNegabent}) in the following theorem. 

\begin{theorem} \rm 
	\label{theorem:ANF_4k}
	Given the set $S_1$ defined in (\ref{align:S1_set}), the ANF of $g \in \mathcal{B}_{4k}$ in (\ref{align:g_BentNegabent}) is given by 
	\[
	g(\bm{\mathrm{x}}, \bm{\mathrm{y}}) = g_0(\bm{\mathrm{x}}, \bm{\mathrm{y}}) + \sum_{\gmbu \in \Gamma} \left(\sum_{\mbox{\tiny $\begin{array} {c} \bm{\mathrm{u}}' * \bm{\mathrm{u}}'' = \bm{0}_k \\ \bm{\mathrm{u}}' + \bm{\mathrm{u}}'' \succeq \bm{\gamma}_1 \end{array}$}} \bm{\mathrm{x}}^{\bm{\mathrm{u}}}\right) 
	\left(\sum_{\mbox{\tiny $\begin{array} {c} \bm{\mathrm{v}}' * \bm{\mathrm{v}}'' = \bm{0}_k \\ \bm{\mathrm{v}}' + \bm{\mathrm{v}}'' \succeq \bm{\gamma}_2 \end{array}$}} \bm{\mathrm{y}}^{\bm{\mathrm{v}}} \right). 
 	\]
\end{theorem}

In the following corollary, we show the necessary and sufficient condition under which the algebraic degree of $g$ is the maximum. 

\begin{corollary} \rm 
	\label{corollary:Deg_4k}
	Given the set $S_1$ defined in (\ref{align:S1_set}), the algebraic degree of $g\in \mathcal{B}_{4k}$ in (\ref{align:g_BentNegabent}) is $2k$ if and only if $|\Gamma|$ is odd. 
\end{corollary}

\begin{proof} 
	For $\bm{\mathrm{u}} \in \mathbb{F}_2^{2k}$ satisfying $\bm{\mathrm{u}}' * \bm{\mathrm{u}}'' = \bm{0}_k$, it is clear that $\wt(\bm{\mathrm{u}}) \le k$. 
	Furthermore, $\bm{\mathrm{u}}' * \bm{\mathrm{u}}'' = \bm{0}_k$ and $\wt(\bm{\mathrm{u}}) = k$ if and only if $\bm{\mathrm{u}}'+ \bm{\mathrm{u}}'' = \bm{1}_k$. 
	In this case, $\bm{\mathrm{u}}'+ \bm{\mathrm{u}}'' \succeq \bm{\gamma}_1$ holds for arbitrary $\bm{\gamma}_1 \in \mathbb{F}_2^k$. 
	For this reason, for any two vectors $(\bm{\gamma}_1, \bm{\gamma}_2)$ and $(\bm{\theta}_1, \bm{\theta}_2)$ in $\Gamma$, where $\gmbu_i, \thbu_i \in \F_2^k$ for $i = 1, 2$, we know that both the functions $\chi_{S_{\bm{\gamma}_1}}(x)\chi_{S_{\bm{\gamma}_2}}(y)$ and $\chi_{S_{\bm{\theta}_1}}(x)\chi_{S_{\bm{\theta}_2}}(y)$ have the degree $2k$, and their monomial terms with degree $2k$ are the same. 
	So, if $|\Gamma|$ is even, all the monomial terms with algebraic degree $2k$ are canceled. 
	Thus, the algebraic degree of $g$ is $2k$ if and only if $|\Gamma|$ is odd. 
\end{proof}

\begin{lemma} \rm (\cite[Theorem 11]{Parker2007}) 
	\label{lemma:dual_bentnegabent}
	Let $n$ be an even integer, and $\varphi \in \mathcal{B}_n$ be a bent-negabent function. 
	Then $\tilde{\varphi}$ (the dual of $\varphi$) is also bent-negabent. 
\end{lemma} 

The dual of $g$ is given in the following theorem. 

\begin{theorem} \rm 
	\label{theorem:dual_4k}
	Given the set $S_1$ defined in (\ref{align:S1_set}), the dual of $g\in \mathcal{B}_{4k}$ in (\ref{align:g_BentNegabent}) is still bent-negabent and given by 
	\begin{align} 
		\label{align:dual_g}
		\tilde{g}(\bm{\mathrm{x}}, \bm{\mathrm{y}}) = \bm{\mathrm{x}}'\cdot \bm{\mathrm{x}}'' + \bm{\mathrm{x}} \cdot \bm{\mathrm{y}} + \chi_{\widetilde{S}_1}(\bm{\mathrm{x}}, \bm{\mathrm{y}}), 
	\end{align}
	where $\widetilde{S}_1$ is a subset of $\mathbb{F}_2^{4k}$ defined by 
	\[
	\widetilde{S}_1 = \bigcup_{\gmbu \in \Gamma} \{(\bm{\mathrm{x}}, \bm{\mathrm{y}}) \in \mathbb{F}_2^{4k} : (\xbu', \ybu') \in \mathbb{F}_2^{2k},  \bm{\mathrm{x}}'' = \bm{\mathrm{x}}' + \bm{\gamma}_2, \bm{\mathrm{y}}'' = \bm{\mathrm{y}}' + \bm{\gamma}_1 + \bm{\gamma}_2 + \bm{1}_k\}. 
	\] 
\end{theorem}

\begin{proof} 
	From Lemma \ref{lemma:dual_bentnegabent} we know that $\tilde{g}$ is also a bent-negabent function. 
	Form (\ref{align:dual_MM}) we know that the dual of $g_0$ is given by 
	$\tilde{g}_0 (\bm{\mathrm{x}}, \bm{\mathrm{y}}) = \bm{\mathrm{x}}' \cdot \bm{\mathrm{x}}'' + \bm{\mathrm{x}} \cdot \bm{\mathrm{y}}$.  
	Then the dual of $g$ is obtained from Theorem \ref{theorem:frame_construction}-(1) and (\ref{align:WHT_4k}). 
%	By (\ref{align:WHT_f_frame}) and (\ref{align:WHT_4k}),  the Walsh-Hadamard transform of $g$ at $(\ubu, \vbu) \in \F_2^{4k}$ is given by 
%	\begin{align} 
%		\label{align:WHT_4k_New}
%		\W_g(\ubu, \vbu) = \begin{cases} 
%			-\W_{g_0} (\ubu, \vbu),\ \bm{\gamma}_1 = \bm{\mathrm{u}}' + \bm{\mathrm{u}}'' + \bm{\mathrm{v}}'+ \bm{\mathrm{v}}'' + \bm{1}_k, \bm{\gamma}_2 = \bm{\mathrm{u}}'+ \bm{\mathrm{u}}''\ \text{for}\ \gmbu \in \Gamma, \\ 
%			\W_{g_0} (\ubu, \vbu),\ \ \ \ \text{otherwise}. 
%		\end{cases}
%	\end{align}
%	Together with (\ref{align:Dual_def}), $\tilde{g}$ and $\tilde{g}_0$ have the following relation: 
%	\begin{align*}
%	\tilde{g}(\bm{\mathrm{x}}, \bm{\mathrm{y}}) = \begin{cases} 
%		\tilde{g}_0(\bm{\mathrm{x}}, \bm{\mathrm{y}}) + 1,\ \bm{\mathrm{x}}'' = \bm{\mathrm{x}}' + \bm{\gamma}_2, \bm{\mathrm{y}}'' = \bm{\mathrm{y}}' + \bm{\gamma}_1 + \bm{\gamma}_2 + \bm{1}_k\ \text{for}\ \gmbu \in \Gamma, \\ 
%		\tilde{g}_0(\bm{\mathrm{x}}, \bm{\mathrm{y}}),\ \hspace{0.6cm} \text{otherwise}, 
%	\end{cases}
%	\end{align*}
%	which implies that $\tilde{g}(\bm{\mathrm{x}}, \bm{\mathrm{y}}) = \tilde{g}_0 (\bm{\mathrm{x}}, \bm{\mathrm{y}}) + \chi_{\tilde{S}_1}(\bm{\mathrm{x}}, \bm{\mathrm{y}})$. 
	%Hence, the assertion is established. 
\end{proof}

We now show an example of an $8$-variable bent-negabent function with the maximum algebraic degree to illustrate this construction. 

\begin{example} \rm 
	Let $k=2$ and $\Gamma = \{(0, 0, 0, 1) \}$. 
	By (\ref{align:S1_set}), $S_1$ is given by 
	\begin{align*} 
		S_1 =& \{(\bm{\mathrm{x}}, \bm{\mathrm{y}}) \in \mathbb{F}_2^8 : \bm{\mathrm{x}}'' = \bm{\mathrm{x}}' \in \mathbb{F}_2^2, \bm{\mathrm{y}}' \in \mathbb{F}_2^2, \bm{\mathrm{y}}'' = \bm{\mathrm{y}}' + (0, 1)\} \\ 
		=&\{(0,0, 0,0), (1, 0, 1, 0), (0, 1, 0, 1), (1, 1, 1, 1) \} \times \{(0, 0, 0, 1), (1, 0, 1, 1), (0, 1, 0, 0), (1, 1, 1, 0) \}, 
	\end{align*} 
	where $\times$ expresses that Cartesian product of two sets, i.e., $S\times T = \{(\bm{\alpha}, \bm{\beta}) \in \F_2^{4k} : \bm{\alpha} \in S, \bm{\beta} \in T\}$ for two subsets $S$ and $T$ of $\mathbb{F}_2^{2k}$. 
	Using a SageMath program, we verified that the $8$-variable function $g$ generated by (\ref{align:g_BentNegabent}) is bent-negabent with algebraic degree $4$, and its ANF is given by 
	\begin{center} 
		\fcolorbox{white}{white}{
			\parbox{.95\linewidth}
			{$g(x_0,\cdots, x_3, y_0, \cdots, y_3) = x_0x_1y_0y_1 + x_0x_1y_0y_3 + x_0x_1y_1y_2 + x_0x_1y_1 + x_0x_1y_2y_3 + x_0x_1y_3 + x_0x_3y_0y_1 + x_0x_3y_0y_3 + x_0x_3y_1y_2 + x_0x_3y_1 + x_0x_3y_2y_3 + x_0x_3y_3 + x_0y_0y_1 + x_0y_0y_3 + x_0y_0 + x_0y_1y_2 + x_0y_1 + x_0y_2y_3 + x_0y_3 + x_1x_2y_0y_1 + x_1x_2y_0y_3 + x_1x_2y_1y_2 + x_1x_2y_1 + x_1x_2y_2y_3 + x_1x_2y_3 + x_1y_0y_1 + x_1y_0y_3 + x_1y_1y_2 + x_1y_2y_3 + x_1y_3 + x_2x_3y_0y_1 + x_2x_3y_0y_3 + x_2x_3y_1y_2 + x_2x_3y_1 + x_2x_3y_2y_3 + x_2x_3y_3 + x_2y_0y_1 + x_2y_0y_3 + x_2y_1y_2 + x_2y_1 + x_2y_2y_3 + x_2y_2 + x_2y_3 + x_3y_0y_1 + x_3y_0y_3 + x_3y_1y_2 + x_3y_1 + x_3y_2y_3 + y_0y_1 + y_0y_2 + y_0y_3 + y_1y_2 + y_1y_3 + y_1 + y_2y_3 + y_3$.} 
		}
	\end{center}
\end{example}

\subsection{Bent-Negabent Functions on $8k$ Variables} 

In this subsection, let $k$ be a positive integer, $t = 2k$  and $m=2t$. 
%We shall denote the subsets $\{(0,0), (1, 1)\}$ and $\{(0, 1), (1, 0)\}$ of $\mathbb{F}_2^2$ by $A$ and $B$, respectively. 
 Let us define a repetition code of length $2d$: 
\begin{align*}
	&A_{2d} = \{ \underbrace{0 \cdots 0}_{2d}, \underbrace{1 \cdots 1}_{2d} \}, \\ 
	&B_{2d} = \{ \underbrace{0 \cdots 0}_{d} \underbrace{1 \cdots 1}_{d}, \underbrace{1 \cdots 1}_{d} \underbrace{0 \cdots 0}_{d} \}. 
\end{align*} 
For example, 
\begin{align*}
	& d = 1,\ A_2 = \{00, 11 \},\ B_2 = \{01, 10 \}, \\ 
	& d = 2,\ A_4 = \{0000, 1111 \},\ B_4 = \{0011, 1100 \}. 
\end{align*} 
We shall define 
\begin{align*} 
	& A_{2d}^r = \{(\xbu_1, \cdots, \xbu_r) : \xbu_i \in A_{2d}\ \text{for}\ 1 \le i \le r \}, \\ 
	& B_{2d}^r = \{(\xbu_1, \cdots, \xbu_r) : \xbu_i \in B_{2d}\ \text{for}\ 1 \le i \le r \}. 
\end{align*}
In the following, we give the result for $d = 1$ in detail. 
For $d = 1$, $A_2^r$ is a subspace of $\mathbb{F}_2^{2r}$ and $B_2^r$ is a coset of $A_2^r$ in $\F_2^{2r}$. 
For $\bm{\gamma} = (\bm{\gamma}_1, \bm{\gamma}_2) \in \mathbb{F}_2^{4k}$, where $\bm{\gamma}_i \in \mathbb{F}_2^{2k}$ for $i=1, 2$, let us define 
\begin{align}
	C_{\gmbu, A_2^{2k}} = \{(\xbu, \ybu) \in \mathbb{F}_2^{8k} : \xbu \in A_2^{2k}, \ybu \in C_{\gmbu} (A_2^{2k}) \}. 
\end{align} 

Let $\Gamma$ be a nonempty subset of $R_{A_2^{2k}}$, i.e., a complete set of coset representatives of $A_2^{2k}$ in $\F_2^{4k}$. 
We shall define a subset $S_2$ of $\mathbb{F}_2^{8k}$ by 
\begin{align} 
	S_2 = \bigcup_{\gmbu \in \Gamma} C_{\gmbu, A_2^{2k}}, 
\label{align:S2_set} 
\end{align} 
for which we have the following result. 

\begin{theorem} \rm 
	\label{theorem:8k_variable}
	Given the subset $S_2$ of $\mathbb{F}_2^{8k}$ defined in (\ref{align:S2_set}) and $g_0 \in \mathcal{B}_{8k}$ defined in (\ref{align:g_0}), the $8k$-variable function $g$ in (\ref{align:g_BentNegabent}) is bent-negabent. 
\end{theorem}

In order to prove Theorem \ref{theorem:8k_variable}, we need the following two lemmas, which are the fragmentary Walsh-Hadamard transform and the fragmentary nega-Hadamard transform of linear functions over $A_2^{k}$ and $g_0$ over $S_2$, respectively. 

\begin{lemma} \rm 
	\label{lemma:ex_sum_A^k}
	For any $\bm{\mathrm{u}} \in \mathbb{F}_2^{2k}$, we have the following results on the fragmentary Walsh-Hadamard transform and the fragmentary nega-Hadamard transform of a linear function: 
	\begin{align} 
		& \sum_{\bm{\mathrm{x}} \in A_2^k}(-1)^{\bm{\mathrm{u}}\cdot \bm{\mathrm{x}}} = \begin{cases}
			2^k,\ \bm{\mathrm{u}} \in A_2^k, \\ 
			0,\ \hspace{0.2cm} \text{otherwise},  
		\end{cases} 
		\label{align:ex_sum_A^k} \\ 
		& \sum_{\bm{\mathrm{x}} \in A_2^k}(-1)^{\bm{\mathrm{u}}\cdot \bm{\mathrm{x}}}\imath^{\wt(\bm{\mathrm{x}})} = \begin{cases}
			2^k,\ \bm{\mathrm{u}} \in B_2^k, \\ 
			0,\ \hspace{0.2cm}\text{otherwise}. 
		\end{cases} 
		\label{align:nega_ex_sum_A^k}
	\end{align} 
\end{lemma} 

\begin{proof} 
	First, for any $\ubu \in A_2^k$, we have $\sum_{\bm{\mathrm{x}} \in A_2^k}(-1)^{\bm{\mathrm{u}}\cdot \bm{\mathrm{x}}} = \sum_{\bm{\mathrm{x}} \in A_2^k}(-1)^0 = |A_2^k| = 2^k$.  
	On the other hand, for any $\ubu \notin A_2^k$, we know that $\sum_{\bm{\mathrm{x}} \in A_2^k}(-1)^{\bm{\mathrm{u}}\cdot \bm{\mathrm{x}}} = 0$ since $|\{\xbu \in A_2^k : \ubu \cdot \xbu = 0 \}| = \frac{|A_2^k|}{2}$. 
	Hence, 
	 (\ref{align:ex_sum_A^k}) holds. 
	
	Let $\bm{\mathrm{u}} = (\bm{\mathrm{u}}_0, \cdots, \bm{\mathrm{u}}_{k-1})$ and $\bm{\mathrm{x}} = (\bm{\mathrm{x}}_0, \cdots, \bm{\mathrm{x}}_{k-1})$, where $\bm{\mathrm{u}}_i, \bm{\mathrm{x}}_i \in \mathbb{F}_2^2$ for $i = 0, \cdots, k-1$. 
	Then we have 
	\[
	\sum_{\bm{\mathrm{x}} \in A_2^k}(-1)^{\bm{\mathrm{u}}\cdot \bm{\mathrm{x}}}\imath^{\wt(\bm{\mathrm{x}})} =\prod_{i=0}^{k-1}\left(\sum_{\bm{\mathrm{x}}_i\in A_2}(-1)^{\bm{\mathrm{u}}_i\cdot \bm{\mathrm{x}}_i}\imath^{\wt(\bm{\mathrm{x}}_i)}\right). 
	\]
	Clearly, it holds that  
	$\sum_{\bm{\mathrm{x}}_i\in A_2}(-1)^{\bm{\mathrm{u}}_i\cdot \bm{\mathrm{x}}_i}\imath^{\wt(\bm{\mathrm{x}}_i)} = 1 - (-1)^{\bm{1}_2\cdot \bm{\mathrm{u}}_i}$, 
	which equals $2$ for $\bm{\mathrm{u}}_i\in B_2$, and $0$, otherwise. 
	Hence, (\ref{align:nega_ex_sum_A^k}) holds. 
\end{proof}

The following lemma gives the fragmentary Walsh-Hadamard transform and the fragmentary nega-Hadamard transform of $g_0$ over $S_2$. 

\begin{lemma} \rm 
	\label{lemma:8k_WHT&NHT}
	Given the subset $S_2$ of $\mathbb{F}_2^{8k}$ defined in (\ref{align:S2_set}) and $g_0 \in \mathcal{B}_{8k}$ defined in (\ref{align:g_0}), the fragmentary Walsh-Hadamard transform and the fragmentary nega-Hadamard transform of $g_0$ over $S_2$ at $(\bm{\mathrm{u}}, \bm{\mathrm{v}}) \in \mathbb{F}_2^{8k}$ are respectively given by 
	 	\begin{align}
	 	& \W_{g_0, S_2}(\bm{\mathrm{u}}, \bm{\mathrm{v}}) = \begin{cases}
	 		\W_{g_0}(\bm{\mathrm{u}}, \bm{\mathrm{v}}),\ \bm{\mathrm{u}} \in C_{\bm{\gamma}} (A_2^{2k}), \bm{\mathrm{v}} \in C_{(\gmbu_2, \gmbu_1)} (A_2^{2k})\ \text{for}\ \gmbu \in \Gamma, \\ 
	 		0,\ \hspace{1.4cm}\text{otherwise}, 
	 	\end{cases} 
	 	\label{align:WHT_8k} \\ 
	 	& \N_{g_0, S_2}(\bm{\mathrm{u}}, \bm{\mathrm{v}}) = \begin{cases}
	 		\N_{g_0}(\bm{\mathrm{u}}, \bm{\mathrm{v}}),\ \ubu + \gmbu \in B_2^{2k}, \vbu + \gmbu + (\gmbu_2, \gmbu_1) \in B_2^{2k}\ \text{for}\ \gmbu \in \Gamma, \\ 
	 		0,\ \hspace{1.3cm}\text{otherwise}. 
	 	\end{cases} 
 	\label{align:NHT_8k}
	 \end{align}
\end{lemma} 

\begin{proof} 
	%We first prove that $g$ is bent. 
	By (\ref{align:PWHT}), the fragmentary Walsh-Hadamard transform of $g_0$ over $S_2$ at $(\bm{\mathrm{u}}, \bm{\mathrm{v}}) \in \mathbb{F}_2^{8k}$ is given by 
	\begin{align*}
	\W_{g_0, S_2}(\bm{\mathrm{u}}, \bm{\mathrm{v}}) 
	%=& \sum_{(\bm{\mathrm{x}}, \bm{\mathrm{y}})\in S_2}(-1)^{\bm{\mathrm{x}}\cdot \bm{\mathrm{y}}+ \bm{\mathrm{y}}'\cdot \bm{\mathrm{y}}'' + \bm{\mathrm{u}}\cdot \bm{\mathrm{x}} + \bm{\mathrm{v}} \cdot \bm{\mathrm{y}}} \\ 
	=& \sum_{\gmbu \in \Gamma} \sum_{(\bm{\mathrm{x}}, \bm{\mathrm{y}})\in C_{\gmbu, A_2^{2k}}}(-1)^{\bm{\mathrm{x}}\cdot \bm{\mathrm{y}}+ \bm{\mathrm{y}}'\cdot \bm{\mathrm{y}}'' + \bm{\mathrm{u}}\cdot \bm{\mathrm{x}} + \bm{\mathrm{v}} \cdot \bm{\mathrm{y}}} \\ 
	=& \sum_{\gmbu \in \Gamma} \sum_{\xbu \in A_2^{2k} } \sum_{\ztbu \in A_2^{2k} } (-1)^{\xbu \cdot (\gmbu + \ztbu) + (\gmbu_1 + \ztbu_1)\cdot (\gmbu_2 + \ztbu_2) + \bm{\mathrm{u}}\cdot \xbu + \bm{\mathrm{v}} \cdot (\gmbu + \ztbu) } \\ 
	=& \sum_{\gmbu \in \Gamma} (-1)^{\gmbu_1\cdot \gmbu_2 + \vbu \cdot \gmbu} \sum_{\ztbu \in A_2^{2k}}(-1)^{[\vbu + (\gmbu_2, \gmbu_1)] \cdot \ztbu} \sum_{\xbu \in A_2^{2k}}(-1)^{(\ubu + \gmbu) \cdot \xbu} \\ 
	=& 2^{4k} \sum_{\gmbu \in \Gamma_1(\bm{\mathrm{u}}, \bm{\mathrm{v}})}(-1)^{\bm{\gamma}_1 \cdot \bm{\gamma}_2 + \vbu \cdot \gmbu}, 
	\end{align*}
	where $\ztbu_i \in \F_2^{2k}$ for $i = 1, 2$ and $\ztbu = (\ztbu_1, \ztbu_2)$, and  $\Gamma_1(\bm{\mathrm{u}}, \bm{\mathrm{v}})$ is a subset of $\Gamma$ defined by 
	\[
	\Gamma_1(\bm{\mathrm{u}}, \bm{\mathrm{v}}) = \{\gmbu \in \Gamma : \bm{\mathrm{u}} \in C_{\bm{\gamma}} (A_2^{2k}), \bm{\mathrm{v}} \in C_{(\gmbu_2, \gmbu_1)} (A_2^{2k}) \}, 
	\]
	and the second identity holds since $\bm{\mathrm{y}} \in C_{\bm{\gamma}} (A_2^{2k})$ if and only if $\bm{\mathrm{y}} = \bm{\gamma} + \bm{\zeta}$ for $\bm{\zeta} \in A_2^{2k}$, 
	and the third identity holds by the fact that $\bm{\mathrm{x}} \cdot \bm{\zeta} = 0$ and $\ztbu_1 \cdot \ztbu_2 = 0$ for $\bm{\mathrm{x}}, \bm{\zeta} \in A_2^{2k}$, 
	and the last identity holds by (\ref{align:ex_sum_A^k}). 
	
	For $\gmbu \in \Gamma_1(\bm{\mathrm{u}}, \bm{\mathrm{v}})$, we know $(\ubu' + \vbu'') \cdot (\ubu'' + \vbu') = 0$ since both $\ubu' + \vbu''$ and $\ubu'' + \vbu'$ are in $A_2^k$. 
	Then, we have 
	\begin{align}
		\bm{\gamma}_1 \cdot \bm{\gamma}_2 + \vbu \cdot \gmbu =& (\vbu' + \bm{\gamma}_2) \cdot (\vbu'' + \bm{\gamma}_1) + \vbu' \cdot \vbu'' \notag \\ 
		=& \vbu' \cdot \vbu''  \notag \\ 
		=& \vbu' \cdot \vbu'' + (\ubu' + \vbu'') \cdot (\ubu'' + \vbu') \notag \\ 
		=& \bm{\mathrm{u}}'\cdot \bm{\mathrm{u}}'' + \bm{\mathrm{u}} \cdot \bm{\mathrm{v}}. \label{align:inter_Res3}
	\end{align}
	Recalling the Walsh-Hadamard transform of $g_0$ in (\ref{align:Walsh_{g_0}}), we have 
	\[
	\W_{g_0, S_2}(\bm{\mathrm{u}}, \bm{\mathrm{v}}) = |\Gamma_1(\bm{\mathrm{u}}, \bm{\mathrm{v}})| \W_{g_0}(\bm{\mathrm{u}}, \bm{\mathrm{v}}). 
	\]
	
	To prove (\ref{align:WHT_8k}), it is sufficient to prove that the cardinality of $\Gamma_1(\bm{\mathrm{u}}, \bm{\mathrm{v}})$ is less than or equal to $1$. 
	Assume that there are two elements $\gmbu$ and $\thbu$ in $\Gamma_1(\bm{\mathrm{u}}, \bm{\mathrm{v}})$. 
	Then, there exist two vectors $\bm{\alpha}, \bm{\beta} \in A_2^k$ such that 
	$\begin{cases} 
		\bm{\gamma} = \bm{\mathrm{u}}+ \bm{\alpha}, \\ 
		\bm{\theta} = \bm{\mathrm{u}}+ \bm{\beta}, 
	\end{cases}
	$
	 which implies 
	$
		\gmbu + A_2^{2k} = \ubu + A_2^{2k} = \thbu + A_2^{2k}. 
	$ 
	Then, we have $\gmbu = \thbu$ since they are coset representatives of $A_2^{2k}$ in $\F_2^{4k}$. 
	Hence, we have $|\Gamma_1(\bm{\mathrm{u}}, \bm{\mathrm{v}})| \le 1$,  
	and (\ref{align:WHT_8k}) follows immediately. 
	
	%Next, we prove that $g$ is negabent. 
	By (\ref{align:PNHT}), the fragmentary nega-Hadamard transform of $g_0$ over $S_2$ at $(\bm{\mathrm{u}}, \bm{\mathrm{v}}) \in \mathbb{F}_2^{8k}$ is given by 
	\begin{align*}
	\N_{g_0, S_2}(\bm{\mathrm{u}}, \bm{\mathrm{v}}) 
	=& \sum_{\gmbu \in \Gamma} \sum_{(\bm{\mathrm{x}}, \bm{\mathrm{y}})\in C_{\gmbu, A_2^{2k}}}(-1)^{\bm{\mathrm{x}}\cdot \bm{\mathrm{y}}+ \bm{\mathrm{y}}'\cdot \bm{\mathrm{y}}'' + \bm{\mathrm{u}}\cdot \bm{\mathrm{x}} + \bm{\mathrm{v}} \cdot \bm{\mathrm{y}}}\imath^{\wt(\bm{\mathrm{x}}, \bm{\mathrm{y}})} \\ 
	=& \sum_{\gmbu \in \Gamma} \sum_{\xbu \in A_2^{2k} } \sum_{ \ztbu \in A_2^{2k} } (-1)^{\xbu \cdot (\bm{\zeta} + \bm{\gamma}) +(\ztbu_1 + \gmbu_1) \cdot (\ztbu_2 + \gmbu_2) + \bm{\mathrm{u}} \cdot \xbu + \bm{\mathrm{v}}\cdot (\bm{\zeta} + \bm{\gamma}) } \imath^{\wt(\xbu) + \wt(\bm{\zeta} + \bm{\gamma})} \\ 
	=& \sum_{\gmbu \in \Gamma}(-1)^{\vbu \cdot \gmbu + \bm{\gamma}_1\cdot \bm{\gamma}_2} \imath^{\wt(\gmbu)}  \sum_{\ztbu \in A_2^{2k}} (-1)^{(\vbu + \gmbu + (\gmbu_2, \gmbu_1)) \cdot \ztbu} \imath^{\wt(\ztbu)}  \sum_{\xbu \in A_2^{2k}}(-1)^{(\ubu + \gmbu) \cdot \xbu} \imath^{\wt(\xbu)} \\ 
	=& 2^{4k} \sum_{\gmbu \in \Gamma_2(\bm{\mathrm{u}}, \bm{\mathrm{v}})}(-1)^{\vbu \cdot \gmbu + \bm{\gamma}_1\cdot \bm{\gamma}_2} \imath^{\wt(\gmbu)}, 
	\end{align*} 
	where $\ztbu_i \in \F_2^{2k}$ for $i = 1, 2$ and $\ztbu = (\ztbu_1, \ztbu_2)$, and $\Gamma_2(\bm{\mathrm{u}}, \bm{\mathrm{v}})$ is a subset of $\Gamma$ defined by 
	\[
	\Gamma_2(\bm{\mathrm{u}}, \bm{\mathrm{v}}) = \{\gmbu \in \Gamma : \ubu + \gmbu \in B_2^{2k}, \vbu + \gmbu + (\gmbu_2, \gmbu_1) \in B_2^{2k} \}, 
	\]
	and the second identity holds since $\bm{\mathrm{y}}\in  C_{\gmbu} (A_2^{2k})$ if and only if $\ybu = \bm{\gamma} + \bm{\zeta}$ for $\bm{\zeta} \in A_2^{2k}$, 
	and the third identity holds by the fact that $\xbu \cdot \ztbu = \bm{\zeta}_1 \cdot \bm{\zeta}_2 = 0$ for $\xbu, \ztbu \in A_2^{2k}$, 
	and the last identity holds by (\ref{align:nega_ex_sum_A^k}). 
	
	Note that $B_2^k$ is an affine subspace of $\mathbb{F}_2^m$, and can be expressed as $B_2^k = \bm{\xi} + A_2^k$, where $\bm{\xi}$ is an arbitrary vector in $B_2^k$. 
	For $\gmbu \in \Gamma_2(\bm{\mathrm{u}}, \bm{\mathrm{v}})$, there exist $\bm{\lambda}_1, \bm{\lambda}_2, \bm{\lambda}_3, \bm{\lambda}_4 \in A_2^k$ such that 
	$\bm{\mathrm{u}}' = \gmbu_1 + \bm{\xi} + \bm{\lambda}_1, 
	\bm{\mathrm{u}}'' = \gmbu_2 + \bm{\xi} + \bm{\lambda}_2, 
	\bm{\mathrm{v}}' = \gmbu_1 + \bm{\gamma}_2 + \bm{\xi} + \bm{\lambda}_3$ 
	and $\bm{\mathrm{v}}'' =  \bm{\gamma}_1 + \bm{\gamma}_2 + \bm{\xi} + \bm{\lambda}_4$. 
	Then we have 
	\begin{align}
	(-1)^{\vbu \cdot \gmbu + \gmbu_1 \cdot \gmbu_2} \imath^{\wt(\gmbu)}  
	=& (-1)^{(\gmbu_1 + \bm{\gamma}_2 + \bm{\xi} + \bm{\lambda}_3)\cdot \bm{\gamma}_1 + (\bm{\gamma}_1 + \bm{\gamma}_2 + \bm{\xi} + \bm{\lambda}_4)\cdot \bm{\gamma}_2 + \bm{\gamma}_1 \cdot \bm{\gamma}_2} \imath^{\wt(\gmbu)} \notag \\ 
	=& (-1)^{(\ldbu_1 + \ldbu_3 + \ubu')\cdot \bm{\gamma}_1 + (\ldbu_2 + \ldbu_4 + \ubu'')\cdot \bm{\gamma}_2 + \bm{\gamma}_1 \cdot \bm{\gamma}_2} \imath^{\wt(\gmbu)} \notag \\ 
	=& (-1)^{(\ldbu_1 + \ldbu_3)\cdot \bm{\gamma}_1 + (\ldbu_2 + \ldbu_4)\cdot \bm{\gamma}_2 + \bm{\gamma}_1 \cdot \bm{\gamma}_2} \imath^{\wt(\ubu' + \gmbu_1) - \wt(\ubu') + \wt(\ubu'' + \gmbu_2) - \wt(\ubu'')} \notag \\ 
	=& (-1)^{(\ldbu_1 + \ldbu_3)\cdot \bm{\gamma}_1 + (\ldbu_2 + \ldbu_4)\cdot \bm{\gamma}_2 + \bm{\gamma}_1 \cdot \bm{\gamma}_2} \imath^{2k - \wt(\ubu)} \notag \\ 
	=& (-1)^{(\ldbu_1 + \ldbu_3 + \gmbu_2)\cdot (\ldbu_2 + \ldbu_4 + \gmbu_1)} \imath^{2k - \wt(\ubu)} \notag \\ 
	=& (-1)^{(\ubu' + \vbu')\cdot (\ubu'' + \vbu'')} \imath^{2k - \wt(\ubu)}, 
	\label{align:inter_Res4}
	\end{align} 
	where the forth identity holds since both $\ubu' + \gmbu_1$ and $\ubu'' + \gmbu_2$ are in $B_2^k$, and the fifth identity holds since both $\ldbu_1 + \ldbu_3$ and $\ldbu_2 + \ldbu_4$ are in $A_2^k$. 
	Together with (\ref{align:Nega_{g_0}}), we have 
	\[
	\N_{g_0, S_2}(\bm{\mathrm{u}}, \bm{\mathrm{v}}) = |\Gamma_2(\bm{\mathrm{u}}, \bm{\mathrm{v}})| \N_{g_0}(\bm{\mathrm{u}}, \bm{\mathrm{v}}). 
	\]
	
	To prove (\ref{align:NHT_8k}), we should prove that the cardinality of $\Gamma_2(\bm{\mathrm{u}}, \bm{\mathrm{v}})$ is less than or equal to $1$. 
	Assume that there are two elements $\gmbu$ and $\thbu$ in $\Gamma_2(\bm{\mathrm{u}}, \bm{\mathrm{v}})$. 
	Then, there exist two vectors $\bm{\alpha}, \bm{\beta}$ in $A_2^k$ such that 
	$\begin{cases} 
		\bm{\gamma} = \ubu + \bm{\xi} + \bm{\alpha}, \\ 
		\bm{\theta} = \ubu + \bm{\xi} + \bm{\beta}.
	\end{cases}$ 
	Then we have 
	$\gmbu + A_2^{2k} = \ubu + \bm{\xi} + A_2^{2k} = \thbu + A_2^{2k}$, 
	which implies	$\gmbu = \thbu$ by the definition of $\Gamma$. 
	Hence, we have $|\Gamma_2(\bm{\mathrm{u}}, \bm{\mathrm{v}})| \le 1$, and (\ref{align:NHT_8k}) follows. 
\end{proof}

\textbf{Proof of Theorem \ref{theorem:8k_variable}}: 
It follows from Lemma \ref{lemma:8k_WHT&NHT} and Theorem \ref{theorem:frame_construction} directly. 
{\hfill $\square$\par} 

To analyze the ANF of $g$, we need the following lemma. 

\begin{lemma} \rm 
	\label{lemma:chi_S_alpha}
	For $\gmbu = (\gamma_0, \cdots, \gamma_{4k-1}) \in \F_2^{4k}$, the characteristic function of $C_{\gmbu, A_2^{2k}}$ is given by
	\[
	\chi_{C_{\gmbu, A_2^{2k}}} (\xbu, \ybu) = \left(\prod_{i=0}^{2k-1}(x_{2i}+ x_{2i+1} + 1) \right) \left(\prod_{i=0}^{2k-1} (y_{2i} + y_{2i+1} + \gamma_{2i} + \gamma_{2i+1} + 1) \right). 
	\]
\end{lemma}

In the following theorem, we give the ANF of $g$. 

\begin{theorem} \rm 
	\label{theorem:ANF_8k}
	Given the set $S_2$ defined in (\ref{align:S2_set}), the ANF of $g\in \mathcal{B}_{8k}$ in (\ref{align:g_BentNegabent}) is given by 
	\[
	g(\bm{\mathrm{x}}, \bm{\mathrm{y}}) = g_0(\bm{\mathrm{x}}, \bm{\mathrm{y}}) + \sum_{\gmbu \in \Gamma}  \chi_{C_{\gmbu, A_2^{2k}}} (\xbu, \ybu), 
	\]
	where $\chi_{C_{\gmbu, A_2^{2k}}}$ is given by Lemma \ref{lemma:chi_S_alpha}. 
\end{theorem} 

\begin{remark} \rm 
	\label{remark:4k&8k}
	It is obvious that the subset $S_2$ of $\mathbb{F}_2^{8k}$ in (\ref{align:S2_set}) is different from the subset $S_1$ of $\mathbb{F}_2^{8k}$ in (\ref{align:S1_set}). 
	Moreover, from Theorem \ref{theorem:ANF_8k} we know the ANF of $\chi_{S_2}$ does not contain any term with $x_{2i}x_{2i+1}$ or $y_{2i}y_{2i+1}$ for $0 \le i \le 2k-1$, but the ANF of $\chi_{S_1}$ contains some of these terms, by Theorem \ref{theorem:ANF_4k}.  
	Hence, bent-negabent functions constructed by Theorem \ref{theorem:8k_variable} are different from those on $8k$ variables constructed by Theorem \ref{theorem:4k}. 
\end{remark}

From Theorem \ref{theorem:ANF_8k}, we immediately give the necessary and sufficient condition such that $g$ has the maximum algebraic degree. 

\begin{corollary} \rm 
	Given the set $S_2$ defined in (\ref{align:S2_set}), the algebraic degree of $g \in \mathcal{B}_{8k}$ in (\ref{align:g_BentNegabent}) is $4k$ if and only if $|\Gamma|$ is odd. 
\end{corollary}

The dual of $g$ is given in the following theorem, of which the proof can be finished similarly to Theorem \ref{theorem:dual_4k} and we omit it. 

\begin{theorem} \rm 
	Given the set $S_2$ defined in (\ref{align:S2_set}), the dual of $g\in \mathcal{B}_{8k}$ in (\ref{align:g_BentNegabent}) is also bent-negabent, and given by 
	\[
	\tilde{g} (\bm{\mathrm{x}}, \bm{\mathrm{y}}) = \bm{\mathrm{x}}'\cdot \bm{\mathrm{x}}'' + \bm{\mathrm{x}} \cdot \bm{\mathrm{y}} + \chi_{\tilde{S}_2}(\bm{\mathrm{x}}, \bm{\mathrm{y}}), 
	\]
	where $\widetilde{S}_2$ is a subset of $\mathbb{F}_2^{8k}$ defined by 
	\[
	\tilde{S}_2 =  \bigcup_{\gmbu \in \Gamma} \{(\bm{\mathrm{x}}, \bm{\mathrm{y}}) \in \mathbb{F}_2^{8k} :  \xbu \in C_{\bm{\gamma}} (A_2^{2k}), \ybu \in C_{(\gmbu_2, \gmbu_1)} (A_2^{2k}) \}. 
	\]
\end{theorem}

\section{Constructions of Bent-Negabent Functions on $4t+2$ Variables} 
\label{section:bent_negabent_4k+2}

%In this section, we first characterize bent-negabent functions on $4t+2$ variables in a subclass of the Maiorana-McFarland set. 
In this section, we present two constructions of bent-negabent functions on $4k+2$ and $8k+2$ variables by modifying the truth tables of a class of quadratic bent-negabent functions with simple form. 

%We first analyze the characterization of bent-negabent functions on $4t+2$ variables in the Maiorana-McFarland class. 
Let $m=2t$, $\bm{\mathrm{x}}$ and $\bm{\mathrm{y}}$ be of the same meaning as those in Section \ref{section:bent_negabent_4k}. 
We shall denote $x_m, y_m \in \mathbb{F}_2$, and $\bm{\mathrm{X}} = (\bm{\mathrm{x}}, x_m), \bm{\mathrm{Y}} = (\bm{\mathrm{y}}, y_m) \in \mathbb{F}_2^{2t+1}$. 
We present a class of $(4t+2)$-variable Boolean functions of the  following form: 
\begin{align} 
	h_0(\bm{\mathrm{X}}, \bm{\mathrm{Y}}) = \bm{\mathrm{X}} \cdot \bm{\mathrm{Y}} + x_0y_m + \bm{\mathrm{y}}'\cdot \bm{\mathrm{y}}'' = \sum_{i=0}^{m} x_iy_i + x_0y_{m} + \sum_{i = 0}^{t-1} y_iy_{t+i}.  
	\label{align:h_0}
\end{align} 

Similarly, let $\bm{\mathrm{u}}$ and $\bm{\mathrm{v}}$ be of the same meaning as those in Section \ref{section:bent_negabent_4k}, $u_m, v_m \in \mathbb{F}_2$, and $\bm{\mathrm{U}} = (\bm{\mathrm{u}}, u_m), \bm{\mathrm{V}} = (\bm{\mathrm{v}}, v_m) \in \mathbb{F}_2^{2t+1}$. 
%Then we have the following lemma. 
In the following lemma, we show that $h_0$ is a bent-negabent function and provide its Walsh-Hadamard transform and nega-Hadamard transform in terms of $g_0$. 

\begin{lemma} \rm 
	The function $h_0$ in (\ref{align:h_0}) is bent-negabent, and the Walsh-Hadamard transform and the nega-Hadamard transform of $h_0$ at $(\bm{\mathrm{U}}, \bm{\mathrm{V}}) \in \mathbb{F}_2^{4t+2}$ are respectively given by 
	\begin{align}
		\W_{h_0}(\bm{\mathrm{U}}, \bm{\mathrm{V}}) 
		=& 2^{2t+1} (-1)^{\bm{\mathrm{u}}'\cdot \bm{\mathrm{u}}'' + \bm{\mathrm{U}}\cdot \bm{\mathrm{V}} + u_m(v_0 + u_t)}, 
		\label{align:WHT_h0} \\ 
		\N_{h_0}(\bm{\mathrm{U}}, \bm{\mathrm{V}}) =& \N_{g_0}(\bm{\mathrm{u}}, \bm{\mathrm{v}}) \left[1 + \imath (-1)^{u_m} + (-1)^{u_0 + u_t + v_t + v_m} - \imath(-1)^{u_m + u_0 + u_t + v_t + v_m} \right], 
		\label{align:h0_NegaHadaTrans}
	\end{align} 
	where $g_0 \in \mathcal{B}_{4t}$ is defined in (\ref{align:g_0}) and $\N_{g_0}$ is given in (\ref{align:Nega_{g_0}}). 
\end{lemma} 

\begin{proof} 
	The function $h_0$ can be rewritten as $h_0(\bm{\mathrm{X}}, \bm{\mathrm{Y}}) = \bm{\mathrm{X}} \cdot (y_0 + y_m, y_1, \cdots, y_m) + \bm{\mathrm{y}}' \cdot \bm{\mathrm{y}}''$. 
	Hence, $h_0$ is a bent function in the Maiorana-McFarland class. 
	It is easy to verify that the inverse of the permutation $(y_0 + y_m, y_1, \cdots, y_m)$ is itself. 
	By (\ref{align:dual_MM}) and (\ref{align:Dual_def}), the Walsh-Hadamard transform of $h_0$ at $(\bm{\mathrm{U}}, \bm{\mathrm{V}}) \in \mathbb{F}_2^{4t+2}$ is given by 
	\begin{align*}
		\W_{h_0}(\bm{\mathrm{U}}, \bm{\mathrm{V}}) =& 2^{2t+1} (-1)^{\bm{\mathrm{V}} \cdot (u_0 + u_m, u_1, \cdots, u_m) + (u_0+ u_m, u_1, \cdots, u_{t-1}) \cdot \bm{\mathrm{u}}''} \notag \\ 
		=& 2^{2t+1} (-1)^{\bm{\mathrm{u}}'\cdot \bm{\mathrm{u}}'' + \bm{\mathrm{U}}\cdot \bm{\mathrm{V}} + u_m\cdot (v_0 + u_t)}. 
	\end{align*} 

	By (\ref{align:NHT}), the nega-Hadamard transform of $h_0$ at $(\bm{\mathrm{U}}, \bm{\mathrm{V}}) \in \mathbb{F}_2^{4t+2}$ is given by 
	\begin{align*}
		\N_{h_0}(\bm{\mathrm{U}}, \bm{\mathrm{V}}) =& \sum_{(\bm{\mathrm{X}}, \bm{\mathrm{Y}}) \in \mathbb{F}_2^{4t+2}} (-1)^{\bm{\mathrm{X}} \cdot \bm{\mathrm{Y}}+   x_0y_m + \bm{\mathrm{y}}'\cdot \bm{\mathrm{y}}'' + \bm{\mathrm{U}}\cdot \bm{\mathrm{X}}+ \bm{\mathrm{V}} \cdot \bm{\mathrm{Y}}} \imath^{\wt(\bm{\mathrm{X}}, \bm{\mathrm{Y}})} \\ 
		=& \sum_{(\bm{\mathrm{x}}, \bm{\mathrm{y}}) \in \mathbb{F}_2^{4t}} (-1)^{\bm{\mathrm{x}}\cdot \bm{\mathrm{y}} + \bm{\mathrm{y}}'\cdot \bm{\mathrm{y}}'' + \bm{\mathrm{u}}\cdot \bm{\mathrm{x}} + \bm{\mathrm{v}} \cdot \bm{\mathrm{y}}} \imath^{\wt(\bm{\mathrm{x}}, \bm{\mathrm{y}})} \sum_{x_m \in \mathbb{F}_2}(-1)^{u_m\cdot x_m} \imath^{\wt(x_m)} \sum_{y_m \in \mathbb{F}_2}(-1)^{(x_0 + v_m + x_m)\cdot y_m} \imath^{\wt(y_m)} \\ 
		=& \sum_{(\bm{\mathrm{x}}, \bm{\mathrm{y}}) \in \mathbb{F}_2^{4t}} (-1)^{\bm{\mathrm{x}}\cdot \bm{\mathrm{y}} + \bm{\mathrm{y}}'\cdot \bm{\mathrm{y}}'' + \bm{\mathrm{u}}\cdot \bm{\mathrm{x}} + \bm{\mathrm{v}} \cdot \bm{\mathrm{y}}} \imath^{\wt(\bm{\mathrm{x}}, \bm{\mathrm{y}})} \left[1 + (-1)^{x_0 + u_m + v_m} + \imath (-1)^{u_m} + \imath (-1)^{x_0 + v_m}\right] \\ 
		=& \N_{g_0}(\bm{\mathrm{u}}, \bm{\mathrm{v}}) + (-1)^{u_m + v_m}\N_{g_0}(\bm{\mathrm{u}}+ \bm{\mathrm{e}}_m^1, \bm{\mathrm{v}}) + \imath(-1)^{u_m}\N_{g_0}(\bm{\mathrm{u}}, \bm{\mathrm{v}}) + \imath(-1)^{v_m}\N_{g_0}(\bm{\mathrm{u}}+ \bm{\mathrm{e}}_m^1, \bm{\mathrm{v}}). 
	\end{align*} 
	From (\ref{align:Nega_{g_0}}) we know    
	\begin{align*}
		\N_{g_0}(\bm{\mathrm{u}}+ \bm{\mathrm{e}}_m^1, \bm{\mathrm{v}}) =& 2^{2t}(-1)^{(\bm{\mathrm{u}}'+ \bm{\mathrm{v}}'+ \bm{\mathrm{e}}_t^1)\cdot (\bm{\mathrm{u}}''+ \bm{\mathrm{v}}'')} \imath^{t - \wt(\bm{\mathrm{u}}+ \bm{\mathrm{e}}_m^1)} \\ 
		=& 2^{2t}(-1)^{(\bm{\mathrm{u}}'+ \bm{\mathrm{v}}')\cdot (\bm{\mathrm{u}}''+ \bm{\mathrm{v}}'') + u_t + v_t} \imath^{t - \wt(\bm{\mathrm{u}}) - 1 + 2\wt(u_0)} \\ 
		=& (-1)^{u_t + v_t+ u_0 + 1} \imath \N_{g_0}(\bm{\mathrm{u}}, \bm{\mathrm{v}}). 
	\end{align*} 
	Then, 
	(\ref{align:h0_NegaHadaTrans}) follows immediately. 
%	we have 
%	\begin{align}
%		\N_{h_0}(\bm{\mathrm{U}}, \bm{\mathrm{V}}) =& \N_{g_0}(\bm{\mathrm{u}}, \bm{\mathrm{v}}) \left[1 + \imath (-1)^{u_m} + (-1)^{u_0 + u_t + v_t + v_m} - \imath(-1)^{u_m + u_0 + u_t + v_t + v_m} \right]. 
%		\label{align:inter_Nega_h0}
%	\end{align} 
	%which implies (\ref{align:h0_NegaHadaTrans}) immediately. 
	It is easy to verify that $|1 + \imath (-1)^{u_m} + (-1)^{u_0 + u_t + v_t + v_m} - \imath(-1)^{u_m + u_0 + u_t + v_t + v_m}| = 2$ for all $(u_m, u_0 + u_t + v_t + v_m)$ in $\F_2^2$. 
	Hence, $h_0$ is negabent. 
\end{proof} 

Given a nonempty subset $S$ of $\mathbb{F}_2^{4t+2}$, 
our systematic construction of $(4t+2)$-variable Boolean functions by modifying the truth table of $h_0 \in \mathcal{B}_{4t+2}$ in (\ref{align:h_0}) is given by 
\begin{align} 
	\label{align:h_BentNegabent}
	h(\bm{\mathrm{X}}, \bm{\mathrm{Y}}) = h_0(\bm{\mathrm{X}}, \bm{\mathrm{Y}}) + \chi_S(\bm{\mathrm{X}}, \bm{\mathrm{Y}}) = \begin{cases}
		h_0(\bm{\mathrm{X}}, \bm{\mathrm{Y}}) + 1,\ (\bm{\mathrm{X}}, \bm{\mathrm{Y}}) \in S, \\ 
		h_0(\bm{\mathrm{X}}, \bm{\mathrm{Y}}),\ \hspace{0.7cm}\text{otherwise}. 
	\end{cases} 
\end{align} 

In the following subsections, we will provide two methods for defining $S$  such that $h$ in (\ref{align:h_BentNegabent}) is a bent-negabent function. 
To avoid confusion, we will use $S_3$ and $S_4$ instead of $S$.

\subsection{Bent-Negabent Functions on $4k+2$ Variables}

In this subsection, let $k$ be an integer, $t = k$ and $m = 2t$. 
%Let $\bm{\gamma}_1 \in \mathbb{F}_2^{k}$, $\bm{\gamma}_2 \in \mathbb{F}_2^{k}$, and $\bm{\gamma} = (\bm{\gamma}_1, \bm{\gamma}_2)$. 
Given $\bm{\gamma} = (\bm{\gamma}_1, \bm{\gamma}_2) \in \F_2^{2k}$, where $\gmbu_i \in \F_2^k$ for $i=1, 2$, let $E_{\bm{\gamma}}$ be an arbitrary nonempty subset of $\mathbb{F}_2$, i.e., $E_{\bm{\gamma}} = \{0 \}, \{1 \}$ or $\mathbb{F}_2$. 
We shall define 
\begin{align}
	L_{\gmbu, E_{\gmbu}} = \{(\Xbu, \Ybu) \in \mathbb{F}_2^{4k+2} : (\xbu, \ybu) \in L_{\gmbu}, x_m \in \mathbb{F}_2, y_m \in E_{\gmbu} \}. 
\end{align}
Let $\Gamma$ be a nonempty subset of $\mathbb{F}_2^{2k}$, and $S_3$ be a subset of $\mathbb{F}_2^{4k+2}$ defined by 
%\begin{align} 
%R = \bigcup_{\bm{\gamma} \in \Gamma}\{(\bm{\mathrm{X}}, \bm{\mathrm{Y}}) \in \mathbb{F}_2^{4k+2}: \bm{\mathrm{x}}' \in \mathbb{F}_2^k, \bm{\mathrm{x}}'' = \bm{\mathrm{x}}' + \bm{\gamma}_1, x_m \in \mathbb{F}_2, \bm{\mathrm{y}}'\in \mathbb{F}_2^k, \bm{\mathrm{y}}'' = \bm{\mathrm{y}}' + \bm{\gamma}_2, y_m \in E_{\bm{\gamma}} \}. 
%\label{align:R_set}
%\end{align}
\begin{align} 
	S_3 = \bigcup_{\gmbu \in \Gamma} L_{\gmbu, E_{\gmbu}}. 
	\label{align:S3_set} 
\end{align}

We have the following result, of which the proof will be given later. 

\begin{theorem} \rm 
	\label{theorem:4k+2}
	Given the subset $S_3$ of $\mathbb{F}_2^{4k+2}$ defined in (\ref{align:S3_set}) and $h_0 \in \mathcal{B}_{4k+2}$ defined in (\ref{align:h_0}), the $(4k+2)$-variable function $h$ in (\ref{align:h_BentNegabent}) is bent-negabent. 
\end{theorem}

In the following lemma, we give the fragmentary Walsh-Hadamard transform and the fragmentary nega-Hadamard transform of $h_0$ over $S_3$, whose proof is presented in Appendix. 

\begin{lemma} \rm 
	\label{lemma:4k+2_WHT&NHT}
	Given the subset $S_3$ of $\mathbb{F}_2^{4k+2}$ defined in (\ref{align:S3_set}) and $h_0 \in \mathcal{B}_{4k+2}$ defined in (\ref{align:h_0}), the fragmentary Walsh-Hadamard transform and fragmentary nega-Hadamard transform of $h_0$ over $S_3$ at $(\bm{\mathrm{U}}, \bm{\mathrm{V}}) \in \mathbb{F}_2^{4k+2}$ are respectively given by 
	\begin{align}
		& \W_{h_0, S_3}(\bm{\mathrm{U}}, \bm{\mathrm{V}}) = \begin{cases}
			\W_{h_0}(\bm{\mathrm{U}}, \bm{\mathrm{V}}),\ \bm{\gamma}_2 = \bm{\mathrm{u}}'+ \bm{\mathrm{u}}'' + \bm{\mathrm{e}}_{k}^{u_m}, \bm{\gamma}_1 + \bm{\gamma}_2 = \bm{\mathrm{v}}' + \bm{\mathrm{v}}'' + \bm{1}_k, \\ 
			\hspace{2.1cm} \text{and}\ u_m \in E_{\bm{\gamma}}\ \text{for}\ \gmbu \in \Gamma, \\ 
			0,\ \hspace{1.6cm} \text{otherwise}, 
		\end{cases} 
		\label{align:WHT_4k+2} \\ 
		& \N_{h_0, S_3}(\bm{\mathrm{U}}, \bm{\mathrm{V}}) = \begin{cases}
			\frac{1}{2}(1+\imath (-1)^{u_0+u_k+v_k+v_m+u_m+\varepsilon}) \N_{h_0} (\Ubu, \Vbu),\ \text{if there is only}\\  
			\hspace{2.0cm} \text{one vector}\ \bm{\gamma} = (\bm{\gamma}_1, \bm{\gamma}_2) \in \Gamma\ \text{such that}\ \bm{\gamma}_1 = \bm{\mathrm{v}}' + \bm{\mathrm{v}}'',\ \text{and} \\ 
			\hspace{2.0cm} \bm{\gamma}_1 + \bm{\gamma}_2 = \bm{\mathrm{u}}'+ \bm{\mathrm{u}}''+ \bm{1}_k  + \bm{\mathrm{e}}_{k}^{\varepsilon},\ \text{where}\ \varepsilon \in E_{\bm{\gamma}}, \\ 
			\N_{h_0} (\Ubu, \Vbu),\ \text{if there are two vectors}\ \bm{\gamma} = (\bm{\gamma}_1, \bm{\gamma}_2), \hat{\bm{\gamma}} = (\bm{\gamma}_1, \hat{\bm{\gamma}}_2) \in \Gamma\ \\  \hspace{2.0cm} \text{such that}\ \bm{\gamma}_1 = \bm{\mathrm{v}}' + \bm{\mathrm{v}}'',\  \bm{\gamma}_1 + \bm{\gamma}_2 = \bm{\mathrm{u}}'+ \bm{\mathrm{u}}''+ \bm{1}_k + \bm{\mathrm{e}}_{k}^{\varepsilon},\ \text{and} \\ 
			\hspace{2.0cm} \bm{\gamma}_1 + \hat{\bm{\gamma}}_2 = \bm{\mathrm{u}}'+ \bm{\mathrm{u}}''+ \bm{1}_k + \bm{\mathrm{e}}_{k}^{\hat{\varepsilon}},\  \text{where}\ \varepsilon \in E_{\bm{\gamma}}$ and $\hat{\varepsilon} \in E_{\hat{\bm{\gamma}}}, \\
			0,\ \hspace{1.6cm}\text{otherwise}. 
		\end{cases} 
		\label{align:NHT_4k+2} 
	\end{align}
\end{lemma} 

\textbf{Proof of Theorem \ref{theorem:4k+2}}: 
	It is an immediate consequence of Lemma \ref{lemma:4k+2_WHT&NHT} and Theorem \ref{theorem:frame_construction}. 
{\hfill $\square$\par} 

The ANF of $h$ is given in the following theorem. 

\begin{theorem} \rm 
	Given the set $S_3$ defined in (\ref{align:S3_set}), the ANF of $h\in \mathcal{B}_{4k+2}$ in (\ref{align:h_BentNegabent}) is given by 
	\[
	h(\bm{\mathrm{X}}, \bm{\mathrm{Y}}) = h_0(\bm{\mathrm{X}}, \bm{\mathrm{Y}}) + \sum_{\gmbu \in \Gamma} \left[\left(\sum_{\mbox{\tiny $\begin{array} {c} \bm{\mathrm{u}}' * \bm{\mathrm{u}}'' = \bm{0}_k \\ \bm{\mathrm{u}}' + \bm{\mathrm{u}}'' \succeq \bm{\gamma}_1 \end{array}$}} \bm{\mathrm{x}}^{\bm{\mathrm{u}}}\right)  \left(\sum_{\mbox{\tiny $\begin{array} {c} \bm{\mathrm{v}}' * \bm{\mathrm{v}}'' = \bm{0}_k \\ \bm{\mathrm{v}}' + \bm{\mathrm{v}}'' \succeq \bm{\gamma}_2 \end{array}$}} \bm{\mathrm{y}}^{\bm{\mathrm{v}}} \right)\chi_{E_{\bm{\gamma}}}(y_m)\right],  
	\]
	where $\chi_{E_{\bm{\gamma}}}$ is the characteristic function of $E_{\bm{\gamma}}$, i.e., $\chi_{E_{\bm{\gamma}}}(y_m) = \begin{cases}
		y_m,\ \hspace{0.6cm} E_{\bm{\gamma}} = \{1\}, \\ 
		y_m + 1,\ E_{\bm{\gamma}} = \{0\}, \\ 
		1,\ \hspace{0.9cm} E_{\bm{\gamma}} = \mathbb{F}_2. 
	\end{cases}$
\end{theorem} 

\begin{proof} 
	The characteristic function of $S_3$ in (\ref{align:S3_set}) is given by 
	\[
	\chi_{S_3}(\bm{\mathrm{X}}, \bm{\mathrm{Y}}) = \sum_{\gmbu \in \Gamma} \chi_{S_{\bm{\gamma}_1}}(\bm{\mathrm{x}}) \chi_{S_{\bm{\gamma}_2}}(\bm{\mathrm{y}})\chi_{E_{\bm{\gamma}}}(y_m), 
	\] 
	where $\chi_{S_{\bm{\gamma}_i}}$ for $i=1, 2$ are given by Lemma \ref{lemma:chi_S_beta}.
	Then the desired result is reached. 
\end{proof}

In the following corollary, we give the necessary and sufficient condition under which the algebraic degree of $h$ reaches the maximum. 

\begin{corollary} \rm 
	\label{corollary:Deg_4k+2}
	Given the set $S_3$ defined in (\ref{align:S3_set}), the algebraic degree of $h\in \mathcal{B}_{4k+2}$ in (\ref{align:h_BentNegabent}) is 
	$2k+1$ if and only if $\sum_{\bm{\gamma} \in \Gamma} |E_{\bm{\gamma}}|$ is odd. 
\end{corollary}

\begin{proof}
	For the same reason as Corollary \ref{corollary:Deg_4k}, %we know that the algebraic degree of $\chi_{S_{\bm{\gamma}_1}}(\bm{\mathrm{x}}) \chi_{S_{\bm{\gamma}_2}}(\bm{\mathrm{y}})\chi_{S_{E_{\bm{\gamma}}}}(y_m)$ is $2k+1$ if $|E_{\bm{\gamma}}|=1$, and $2k$ if $|E_{\bm{\gamma}}|=2$. 
	for any two vectors $\bm{\gamma} = (\bm{\gamma}_1, \bm{\gamma}_2)$ and $\bm{\theta} = (\bm{\theta}_1, \bm{\theta}_2)$ in $\Gamma$ satisfying $|E_{\bm{\gamma}}| = |E_{\bm{\theta}}| = 1$, we know that the functions $\chi_{S_{\bm{\gamma}_1}}(\bm{\mathrm{x}}) \chi_{S_{\bm{\gamma}_2}}(\bm{\mathrm{y}})\chi_{S_{E_{\bm{\gamma}}}}(y_m)$ and $\chi_{S_{{\thbu}_1}}(\bm{\mathrm{x}}) \chi_{S_{{\thbu}_2}}(\bm{\mathrm{y}})\chi_{S_{E_{\thbu}}}(y_m)$ have the degree $2k+1$, and their monomial terms with degree $2k+1$ are the same. 
	%Hence, the algebraic degree of $\sum_{(\bm{\gamma}_1, \bm{\gamma}_2) \in \Gamma} \chi_{S_{\bm{\gamma}_1}}(\bm{\mathrm{x}}) \chi_{S_{\bm{\gamma}_2}}(\bm{\mathrm{y}})\chi_{S_{E_{\bm{\gamma}}}}(y_m)$ is equal to $2k+1$ if $\sum_{\bm{\gamma} \in \Gamma} |E_{\bm{\gamma}}|$ is odd, and less than or equal to $2k$, otherwise. 
	%Then the desired result is reached. 
	So, if $\sum_{\bm{\gamma} \in \Gamma} |E_{\bm{\gamma}}|$ is even , all the monomial terms with algebraic degree $2k+1$ are canceled. 
	Hence, the algebraic degree of $h$ is $2k+1$ if and only if $\sum_{\bm{\gamma} \in \Gamma} |E_{\bm{\gamma}}|$ is odd. 
\end{proof}

From (\ref{align:Dual_def}) and (\ref{align:WHT_h0}) we know $\tilde{h}_0 (\bm{\mathrm{X}}, \bm{\mathrm{Y}}) = \bm{\mathrm{X}} \cdot \bm{\mathrm{Y}} + \bm{\mathrm{x}}'\cdot \bm{\mathrm{x}}'' + x_m\cdot (x_k + y_0)$. 
The dual of $h$ is given in the following theorem. 

\begin{theorem} \rm 
	Given the set $S_3$ defined in (\ref{align:S3_set}), 
	 the dual of $h\in \mathcal{B}_{4k+2}$ in (\ref{align:h_BentNegabent}) is also bent-negabent and given by 
	 \[
	 \tilde{h} (\bm{\mathrm{X}}, \bm{\mathrm{Y}}) = \bm{\mathrm{X}} \cdot \bm{\mathrm{Y}} + \bm{\mathrm{x}}'\cdot \bm{\mathrm{x}}'' + x_m\cdot (x_k + y_0) + \chi_{\tilde{S}_3}(\bm{\mathrm{X}}, \bm{\mathrm{Y}}), 
	 \]
	 where $\tilde{S}_3$ is a subset of $\mathbb{F}_2^{4k+2}$ defined by 
	 \begin{align*}
	 \tilde{S}_3 = \bigcup_{\gmbu \in \Gamma} \{(\bm{\mathrm{X}}, \bm{\mathrm{Y}}) \in \mathbb{F}_2^{4k+2} :  (\xbu', \ybu') \in \mathbb{F}_2^{2k}, \bm{\mathrm{x}}'' = \bm{\mathrm{x}}' + \bm{\mathrm{e}}_{k}^{x_m} + \bm{\gamma}_2,  
	  x_m \in E_{\bm{\gamma}}, \bm{\mathrm{y}}'' = \bm{\mathrm{y}}' + \bm{1}_k + \bm{\gamma}_1 + \bm{\gamma}_2, y_m \in \mathbb{F}_2 \}. 
	 \end{align*}
\end{theorem} 

Based on Theorem \ref{theorem:4k+2} , we show an example of a $10$-variable bent-negabent function with the maximum algebraic degree. 

\begin{example} \rm 
	Let $k=2$, $\Gamma = \{(1, 0, 0, 0), (0, 1, 0, 1) \}$, $E_{(1, 0, 0, 0)} = \{1 \}$, and $E_{(0, 1, 0, 1)} = \mathbb{F}_2$. 
	By (\ref{align:S3_set}), $S_3$ is given by  
	\begin{align*}
		S_3 =& \{ (0, 0, 1, 0), (1, 0, 0, 0), (0, 1, 1, 1), (1, 1, 0, 1) \} \times \mathbb{F}_2 \times \{(0, 0, 0, 0), (1, 0, 1, 0), (0, 1, 0, 1), (1, 1, 1, 1) \} \times \{1 \} \\
		& \cup \{(0, 0, 0, 1), (1, 0, 1, 1), (0, 1, 0, 0), (1, 1, 1, 0) \} \times \mathbb{F}_2 \times \{(0, 0, 0, 1), (1, 0, 1, 1), (0, 1, 0, 0), (1, 1, 1, 0) \} \times \mathbb{F}_2. 
	\end{align*} 
	Using a SageMath program, we verified that the $10$-variable function generated by (\ref{align:h_BentNegabent}) is bent-negabent with algebraic degree $5$, and its ANF is given by 
	\begin{center} 
	\fcolorbox{white}{white}{
		\parbox{.95\linewidth}
		{$h(x_0, \cdots, x_4, y_0, \cdots, y_4) = x_0x_1y_0y_1y_4 + x_0x_1y_0y_1 + x_0x_1y_0y_3y_4 + x_0x_1y_0y_3 + x_0x_1y_0y_4 + x_0x_1y_1y_2y_4 + x_0x_1y_1y_2 + x_0x_1y_1y_4 + x_0x_1y_1 + x_0x_1y_2y_3y_4 + x_0x_1y_2y_3 + x_0x_1y_2y_4 + x_0x_1y_3y_4 + x_0x_1y_3 + x_0x_1y_4 + x_0x_3y_0y_1y_4 + x_0x_3y_0y_1 + x_0x_3y_0y_3y_4 + x_0x_3y_0y_3 + x_0x_3y_0y_4 + x_0x_3y_1y_2y_4 + x_0x_3y_1y_2 + x_0x_3y_1y_4 + x_0x_3y_1 + x_0x_3y_2y_3y_4 + x_0x_3y_2y_3 + x_0x_3y_2y_4 + x_0x_3y_3y_4 + x_0x_3y_3 + x_0x_3y_4 + x_0y_0y_1y_4 + x_0y_0y_3y_4 + x_0y_0y_4 + x_0y_0 + x_0y_1y_2y_4 + x_0y_1y_4 + x_0y_2y_3y_4 + x_0y_2y_4 + x_0y_3y_4 + x_1x_2y_0y_1y_4 + x_1x_2y_0y_1 + x_1x_2y_0y_3y_4 + x_1x_2y_0y_3 + x_1x_2y_0y_4 + x_1x_2y_1y_2y_4 + x_1x_2y_1y_2 + x_1x_2y_1y_4 + x_1x_2y_1 + x_1x_2y_2y_3y_4 + x_1x_2y_2y_3 + x_1x_2y_2y_4 + x_1x_2y_3y_4 + x_1x_2y_3 + x_1x_2y_4 + x_1y_0y_1 + x_1y_0y_3 + x_1y_1y_2 + x_1y_2y_3 + x_1y_3 + x_2x_3y_0y_1y_4 + x_2x_3y_0y_1 + x_2x_3y_0y_3y_4 + x_2x_3y_0y_3 + x_2x_3y_0y_4 + x_2x_3y_1y_2y_4 + x_2x_3y_1y_2 + x_2x_3y_1y_4 + x_2x_3y_1 + x_2x_3y_2y_3y_4 + x_2x_3y_2y_3 + x_2x_3y_2y_4 + x_2x_3y_3y_4 + x_2x_3y_3 + x_2x_3y_4 + x_2y_0y_1y_4 + x_2y_0y_3y_4 + x_2y_0y_4 + x_2y_1y_2y_4 + x_2y_1y_4 + x_2y_2y_3y_4 + x_2y_2y_4 + x_2y_2 + x_2y_3y_4 + x_2y_4 + x_3y_0y_1 + x_3y_0y_3 + x_3y_1y_2 + x_3y_1 + x_3y_2y_3 + x_4y_4 + y_0y_2 + y_1y_3$. }
	}
	\end{center}
\end{example}

\subsection{Bent-Negabent Functions on $8k+2$ Variables }
\label{subsection:8k+2}

In this subsection, let $k$ be an integer, $t = 2k$ and $m = 2t$. 
%Let $\bm{\gamma}_1 \in \mathbb{F}_2^{2k}$, $\bm{\gamma}_2 \in \mathbb{F}_2^{2k}$, and $\bm{\gamma} = (\bm{\gamma}_1, \bm{\gamma}_2)$. 
Given $\bm{\gamma} = (\bm{\gamma}_1, \bm{\gamma}_2) \in \mathbb{F}_2^{4k}$, where $\gmbu_i \in \F_2^{2k}$ for $i=1, 2$, let $E_{\bm{\gamma}}$ be an arbitrary nonempty subset of $\mathbb{F}_2$, i.e., $E_{\bm{\gamma}} = \{0 \}, \{1 \}$ or $\mathbb{F}_2$. 
We shall define 
\begin{align}
	C_{\gmbu, A_2^{2k}, E_{\gmbu}} = \{(\Xbu, \Ybu)\in \mathbb{F}_2^{8k+2} : (\xbu, \ybu) \in C_{\gmbu, A_2^{2k}}, x_m \in \mathbb{F}_2, y_m \in E_{\gmbu} \}. 
\end{align}
Let $\Gamma$ be a nonempty subset of $R_{A_2^{2k}}$. 
%i.e., the coset leaders representatives relative to $A_2^{2k}$. 
We shall define a subset $S_4$ of $\mathbb{F}_2^{8k+2}$ by 
\begin{align}
	S_4 = \bigcup_{\gmbu \in \Gamma} C_{\gmbu, A_2^{2k}, E_{\gmbu}}. 
	\label{align:S4_set}
\end{align}

We have the following result, of which the proof will be given later. 

\begin{theorem} \rm 
	\label{theorem:8k+2}
	Given the subset $S_4$ of $\mathbb{F}_2^{8k+2}$ defined in (\ref{align:S4_set}) and $h_0 \in \mathcal{B}_{8k+2}$ defined in (\ref{align:h_0}), the $(8k+2)$-variable function $h$ in (\ref{align:h_BentNegabent}) is bent-negabent. 
\end{theorem}

The following lemma gives the fragmentary Walsh-Hadamard transform and the fragmentary nega-Hadamard transform of $h_0$ over $S_4$, whose lengthy proof is given in Appendix.  

\begin{lemma} \rm 
	\label{lemma:8k+2_WHT&NHT}
	Given the subset $S_4$ of $\mathbb{F}_2^{8k+2}$ defined in (\ref{align:S4_set}) and $h_0 \in \mathcal{B}_{8k+2}$ defined in (\ref{align:h_0}), the fragmentary Walsh-Hadamard transform and the fragmentary nega-Hadamard transform of $h_0$ over $S_4$ at $(\bm{\mathrm{U}}, \bm{\mathrm{V}}) \in \mathbb{F}_2^{8k+2}$ are respectively given by 
	\begin{align}
		& \W_{h_0, S_4}(\bm{\mathrm{U}}, \bm{\mathrm{V}}) = \begin{cases} 
			\W_{h_0}(\bm{\mathrm{U}}, \bm{\mathrm{V}}),\ u_m \in E_{\bm{\gamma}}, \bm{\mathrm{u}} + \bm{\mathrm{e}}_{4k}^{u_m} \in C_{\gmbu}(A_2^{2k}), \bm{\mathrm{v}} \in C_{(\gmbu_2, \gmbu_1)} (A_2^{2k})\ \text{for}\ \gmbu \in \Gamma, \\ 
			0,\ \hspace{1.6cm} \text{otherwise}. 
		\end{cases}
		\label{align:WHT_8k+2} \\ 
		& \N_{h_0, S_4}(\bm{\mathrm{U}}, \bm{\mathrm{V}}) = \begin{cases}
			\frac{1}{2}(1+\imath (-1)^{u_0+u_{2k}+v_{2k}+v_m+u_m+\varepsilon}) \N_{h_0} (\Ubu, \Vbu),\\ 
			\hspace{2.0cm} \varepsilon \in E_{\bm{\gamma}}, \bm{\mathrm{u}} + \gmbu+  \bm{\mathrm{e}}_{4k}^{\varepsilon} \in B_2^{2k}, \bm{\mathrm{v}} + \gmbu + (\bm{\gamma}_2, \bm{\gamma}_1) \in B_2^{2k}\ \text{for}\ \gmbu \in \Gamma, \\
			0,\ \hspace{1.6cm}\text{otherwise}. 
		\end{cases} 
		\label{align:NHT_8k+2} 
	\end{align}
\end{lemma} 

\textbf{Proof of Theorem \ref{theorem:8k+2}}: 
	It follows from Lemma \ref{lemma:8k+2_WHT&NHT} and Theorem \ref{theorem:frame_construction} immediately. 
{\hfill $\square$\par} 

In the following theorem, we give the ANF of $h$.  

\begin{theorem} \rm  
	Given the set $S_4$ defined in (\ref{align:S4_set}), the ANF of $h \in \mathcal{B}_{8k+2}$ in (\ref{align:h_BentNegabent}) is given by 
	\[
	h(\bm{\mathrm{X}}, \bm{\mathrm{Y}}) = h_0(\bm{\mathrm{X}}, \bm{\mathrm{Y}}) + \sum_{\gmbu \in \Gamma} \chi_{C_{\gmbu, A_2^{2k}}}(\xbu, \ybu)\chi_{E_{\gmbu}} (y_m), 
	\]
	where $\chi_{C_{\gmbu, A_2^{2k}}}$ is given by Lemma \ref{lemma:chi_S_alpha}. 
\end{theorem} 

\begin{remark} \rm 
	For the same reason as Remark \ref{remark:4k&8k}, 
	we know that bent-negabent functions constructed from Theorem \ref{theorem:8k+2} are different from those on $8k+2$ variables constructed from Theorem \ref{theorem:4k+2}. 
\end{remark}

In the following corollary, we give the necessary and sufficient condition such that the algebraic degree of $h$ reaches the maximum. 
The proof can be completed similarly to Corollary \ref{corollary:Deg_4k+2} and we omit it. 

\begin{corollary} \rm 
	Given the set $S_4$ defined in (\ref{align:S4_set}), the algebraic degree of $h\in \mathcal{B}_{8k+2}$ in (\ref{align:h_BentNegabent}) is $4k+1$ if and only if $\sum_{\bm{\gamma} \in \Gamma} |E_{\bm{\gamma}}|$ is odd. 
\end{corollary}

We have the following result on the dual of $h$. 

\begin{theorem} \rm 
	Given the set $S_4$ defined in (\ref{align:S4_set}), the dual of $h \in \mathcal{B}_{8k+2}$ in (\ref{align:h_BentNegabent}) is also bent-negabent, and given by 
	\[
	\tilde{h} (\bm{\mathrm{X}}, \bm{\mathrm{Y}}) = \bm{\mathrm{X}}\cdot \bm{\mathrm{Y}} + \bm{\mathrm{x}}'\cdot \bm{\mathrm{x}}'' + x_m\cdot (x_{2k} + y_0) + \chi_{\tilde{S}_4}(\bm{\mathrm{X}}, \bm{\mathrm{Y}}),  
	\]
	where $\tilde{S}_4$ is a subset of $\mathbb{F}_2^{8k+2}$ defined by 
	\begin{align*}
	\tilde{S}_4 = \bigcup_{\gmbu \in \Gamma } \{(\bm{\mathrm{X}}, \bm{\mathrm{Y}}) \in \mathbb{F}_2^{8k+2} :\ & x_m \in E_{\bm{\gamma}}, \bm{\mathrm{x}} + \bm{\mathrm{e}}_{4k}^{x_m} \in C_{\gmbu}(A_2^{2k}), \bm{\mathrm{y}} \in C_{(\gmbu_2, \gmbu_1)} (A_2^{2k}), y_m \in \mathbb{F}_2 \}. 
	\end{align*}
\end{theorem}

\section{Construction of 2-Rotation Symmetric Bent-Negabent Functions with Any Possible Algebraic Degrees}
\label{section:2_RS_bent_negabent} 

In this section, we present a construction of 2-rotation symmetric bent-negabent functions with any possible algebraic degrees by modifying the truth tables of a class of quadratic 2-rotation symmetric bent-negabent functions. 

Let $\bm{\mathrm{x}} = (x_0, \cdots, x_{2k-1}), \bm{\mathrm{y}} = (y_0, \cdots, y_{2k-1}) \in \mathbb{F}_2^{2k}$. 
For simplicity, we shall  denote 
\[
\begin{cases} 
	\bm{\mathrm{x}}_{ev} = (x_0, x_2, \cdots, x_{2k-2}), \\ 
	\bm{\mathrm{x}}_{od} = (x_1, x_3, \cdots, x_{2k-1}), \\
	\bm{\mathrm{y}}_{ev} = (y_0, y_2, \cdots, y_{2k-2}), \\
	\bm{\mathrm{y}}_{od} = (y_1, y_3, \cdots, y_{2k-1}). 
\end{cases}
\]
Let $f_0 \in \mathcal{B}_{4k}$ be a 2-rotation symmetric Boolean function, which is affine equivalent to $g_0 \in \mathcal{B}_{4k}$ in (\ref{align:g_0}), with the following ANF:  
\begin{align}
	f_0(\bm{\mathrm{x}}, \bm{\mathrm{y}}) = g_0(\bm{\mathrm{x}}_{ev}, \bm{\mathrm{y}}_{ev}, \bm{\mathrm{x}}_{od}, \bm{\mathrm{y}}_{od}) =& \bm{\mathrm{x}}_{ev} \cdot \bm{\mathrm{x}}_{od} + \bm{\mathrm{y}}_{ev} \cdot \bm{\mathrm{y}}_{od} + \bm{\mathrm{x}}_{od} \cdot \bm{\mathrm{y}}_{od} \notag \\ 
	=& \sum_{i = 0}^{k-1} (x_{2i}x_{2i+1} + y_{2i}y_{2i+1} + x_{2i+1}y_{2i+1}). 
	\label{align:f_0}
\end{align}
From \cite[Theorem 2]{Schmidt2008} we know that $f_0$ is also bent-negabent. 

Next, we propose a method for constructing bent-negabent functions by modifying the truth table of $f_0$. 
Let $\Gamma$ be a nonempty subset of $\mathbb{F}_2^{2k}$, and $T$ a subset of $\mathbb{F}_2^{4k}$ defined by 
\begin{align} 
	T = \bigcup_{\bm{\gamma} \in \Gamma} \{(\bm{\mathrm{x}}, \bm{\mathrm{y}}) \in \mathbb{F}_2^{4k}: \bm{\mathrm{x}}\in \mathbb{F}_2^{2k},\  \bm{\mathrm{y}} = \bm{\mathrm{x}} + \bm{\gamma} \}. 
	\label{align:T_set}
\end{align} 

The following corollary follows from Theorem \ref{theorem:4k} immediately. 

\begin{corollary} \rm 
	\label{corollary:f}
	Given the subset $T$ of $\mathbb{F}_2^{4k}$ defined in (\ref{align:T_set}), the function $f \in \mathcal{B}_{4k}$ defined by 
	\begin{align} 
		\label{align:f_BentNegabent}
		f(\bm{\mathrm{x}}, \bm{\mathrm{y}}) = f_0(\bm{\mathrm{x}}, \bm{\mathrm{y}}) + \chi_T(\bm{\mathrm{x}}, \bm{\mathrm{y}})  = \begin{cases}
			f_0(\bm{\mathrm{x}}, \bm{\mathrm{y}}) + 1,\ (\bm{\mathrm{x}}, \bm{\mathrm{y}}) \in T, \\ 
			f_0(\bm{\mathrm{x}}, \bm{\mathrm{y}}),\ \hspace{0.7cm}\text{otherwise},
		\end{cases} 
	\end{align} 
	is bent-negabent. 
\end{corollary}

We shall denote the orbit generated by $\bm{\mathrm{x}} \in \mathbb{F}_2^{2k}$ by $O_{2k}(\bm{\mathrm{x}}) = \{\rho_n^i(\bm{\mathrm{x}}) : 0 \le i < 2k \}$. 
We choose a representative element from every orbit, and denote the set of all representative elements by $R_{2k}$. 
For example, for $k=2$, if we select the lexicographically first element as the representative element of every obit, then $R_4 = \{(0,0,0,0), (1,0,0,0), (1,1,0,0), (1,0,1,0), (1, 1, 1, 0), (1,1,1,1) \}$. 

Based on Corollary \ref{corollary:f}, we present a construction of 2-rotation symmetric bent-negabent functions in the following theorem. 

\begin{theorem} \rm 
	\label{theorem:RS_BentNegabent}
	Let $P$ be an arbitrary nonempty subset of $R_{2k}$, and 
	$
		\Gamma = \bigcup_{\bm{\beta} \in P} O_{2k}(\bm{\beta}). 
	$
	Then the $4k$-variable function $f$ defined by (\ref{align:f_BentNegabent}) is a 2-rotation symmetric bent-negabent function. 
\end{theorem} 

\begin{proof}
	From Corollary \ref{corollary:f} we know that $f$ is bent-negabent. 
	To prove the 2-rotation symmetric property of $f$, considering that $f_0$ is a 2-rotation symmetric function, it is sufficient to prove that $\chi_T$ is a rotation symmetric function. 
	That is to say, the set $T$ is the union of some orbits. 
	Suppose that $(\bm{\mathrm{x}}, \bm{\mathrm{y}}) \in \mathbb{F}_2^{4k}$ is an arbitrary element of $T$, and $\bm{\mathrm{y}} = \bm{\mathrm{x}} + \bm{\gamma}$, where $\bm{\gamma} = (\gamma_0, \cdots, \gamma_{2k-1}) \in \Gamma$. 
	We have 
	\begin{align*}
	\rho_{4k}^1(\bm{\mathrm{x}}, \bm{\mathrm{y}}) =& \rho_{4k}^1(x_0, \cdots, x_{2k-1}, y_0, \cdots, y_{2k-1}) \\
	=& (x_1, \cdots, x_{2k-1}, y_0, y_1, \cdots, y_{2k-1}, x_0) \\ 
	=& (x_1, \cdots, x_{2k-1}, y_0, x_1 + \gamma_1, \cdots, y_{2k-1} + \gamma_{2k-1}, y_0 + \gamma_0) \\ 
	=& (x_1, \cdots, x_{2k-1}, y_0, (x_1, \cdots, x_{2k-1}, y_0) + \rho_{2k}^1(\bm{\gamma}) ). 
	\end{align*}
	From the definition of $\Gamma$ we know $\rho_{2k}^1(\bm{\gamma})\in \Gamma$. 
	Then, $\rho_{4k}^1(\bm{\mathrm{x}}, \bm{\mathrm{y}})$ is also an element of $T$. 
	Hence, $T$ is the union of some orbits. 
	This completes the proof. 
\end{proof}

We give the ANF of $f$ in Theorem \ref{theorem:RS_BentNegabent} in the following theorem. 

\begin{theorem} \rm 
	\label{theorem:ANF_RS_BentNegabent}
	The ANF of $f \in \mathcal{B}_{4k}$ in Theorem \ref{theorem:RS_BentNegabent} is given by 
	\begin{align*}
	f(\bm{\mathrm{x}}, \bm{\mathrm{y}}) %=& f_0(\bm{\mathrm{x}}, \bm{\mathrm{y}}) + \sum_{\bm{\beta} \in P} \sum_{\bm{\gamma} \in O_{2k}(\bm{\beta})}\chi_{S_{\bm{\gamma}}}(\bm{\mathrm{x}}, \bm{\mathrm{y}}) \\
	=& f_0(\bm{\mathrm{x}}, \bm{\mathrm{y}}) + \sum_{\bm{\beta} \in P} \sum_{\bm{\gamma} \in O_{2k}(\bm{\beta})} \sum_{\mbox{\tiny $\begin{array} {c} \bm{\mathrm{u}} * \bm{\mathrm{v}} = \bm{0}_{2k} \\ \bm{\mathrm{u}} + \bm{\mathrm{v}} \succeq \bm{\gamma} \end{array}$}} (\bm{\mathrm{x}}, \bm{\mathrm{y}})^{(\bm{\mathrm{u}}, \bm{\mathrm{v}})}. 
	\end{align*}
\end{theorem}

In the following corollary, we give the necessary and sufficient condition under which the algebraic degree of $f$ in Theorem \ref{theorem:RS_BentNegabent} reaches the maximum. 

\begin{corollary} \rm 
	The algebraic degree of $f \in \mathcal{B}_{4k}$ in Theorem \ref{theorem:RS_BentNegabent} is $2k$ if and only if $\sum_{\bm{\beta} \in P} |O_{2k}(\bm{\beta})|$ is odd. 
\end{corollary}

By Theorem \ref{theorem:dual_4k}, we directly give the dual of $f$ in Theorem \ref{theorem:RS_BentNegabent}. 

\begin{theorem} \rm 
	The dual of $f \in \mathcal{B}_{4k}$ in Theorem \ref{theorem:RS_BentNegabent} is also a 2-rotation symmetric bent-negabent function, and given by 
	\[
	\tilde{f}(\bm{\mathrm{x}}, \bm{\mathrm{y}}) = \bm{\mathrm{x}}_{ev} \cdot \bm{\mathrm{x}}_{od} + \bm{\mathrm{y}}_{ev}\cdot \bm{\mathrm{y}}_{od} + \bm{\mathrm{x}}_{ev}\cdot \bm{\mathrm{y}}_{ev} + \chi_{\tilde{T}}(\bm{\mathrm{x}}, \bm{\mathrm{y}}), 
	\]
	where $\tilde{T}$ is a subset of $\mathbb{F}_2^{4k}$ defined by 
	\[
	\tilde{T} = \{(\bm{\mathrm{x}}, \bm{\mathrm{y}}) \in \mathbb{F}_2^{4k} : (\bm{\mathrm{x}}_{ev} + \bm{\mathrm{x}}_{od}+ \bm{\mathrm{y}}_{ev} + \bm{\mathrm{y}}_{od} + \bm{1}_{k}, \bm{\mathrm{x}}_{ev} + \bm{\mathrm{y}}_{ev}) \in \bigcup_{\bm{\beta} \in P} O_{2k}(\bm{\beta})\}. 
	\]
\end{theorem}

In what follows, we give two simplified forms of $f$ in Theorem \ref{theorem:RS_BentNegabent}. 
We first present a result on the linear combination of $\sum_{\bm{\mathrm{u}} * \bm{\mathrm{v}} = \bm{0}_{2k}, \bm{\mathrm{u}} + \bm{\mathrm{v}} \succeq \bm{\gamma}} (\bm{\mathrm{x}}, \bm{\mathrm{y}})^{(\bm{\mathrm{u}}, \bm{\mathrm{v}})}$. 

\begin{lemma} \rm (\cite[Lemma 5]{Su2017}) 
	\label{lemma:Su_RS}
	For each $\bm{\alpha} \in R_{2k}$, there exists a nonempty subset $A_{\bm{\alpha}} \subseteq R_{2k}$ such that 
	\begin{align*}
	\sum_{\mbox{\tiny $\begin{array} {c} \bm{\mathrm{u}} * \bm{\mathrm{v}} = \bm{0}_{2k} \\ \bm{\mathrm{u}} + \bm{\mathrm{v}} \in O_{2k}(\bm{\alpha}) \end{array}$}} (\bm{\mathrm{x}}, \bm{\mathrm{y}})^{(\bm{\mathrm{u}}, \bm{\mathrm{v}})} = \sum_{\bm{\beta} \in A_{\bm{\alpha}}} \sum_{\bm{\gamma} \in O_{2k}(\bm{\beta})} \sum_{\mbox{\tiny $\begin{array} {c} \bm{\mathrm{u}} * \bm{\mathrm{v}} = \bm{0}_{2k} \\ \bm{\mathrm{u}} + \bm{\mathrm{v}} \succeq \bm{\gamma} \end{array}$}} (\bm{\mathrm{x}}, \bm{\mathrm{y}})^{(\bm{\mathrm{u}}, \bm{\mathrm{v}})}. 
	\end{align*}
\end{lemma}

From Theorem \ref{theorem:ANF_RS_BentNegabent} and Lemma \ref{lemma:Su_RS}, the first simplified form of $f$ in Theorem \ref{theorem:RS_BentNegabent} is given as follows. 

\begin{corollary}\rm 
	\label{corollary:2RS_BN}
	Let $A$ be a nonempty subset of $R_{2k}$. 
	The function $f \in \mathcal{B}_{4k}$ defined by 
	\begin{align*}
	f(\bm{\mathrm{x}}, \bm{\mathrm{y}}) = f_0(\bm{\mathrm{x}}, \bm{\mathrm{y}}) + \sum_{\bm{\gamma} \in A} \left( \sum_{\mbox{\tiny $\begin{array} {c} \bm{\mathrm{u}} * \bm{\mathrm{v}} = \bm{0}_{2k} \\ \bm{\mathrm{u}} + \bm{\mathrm{v}} \in O_{2k}(\bm{\gamma}) \end{array}$}} (\bm{\mathrm{x}}, \bm{\mathrm{y}})^{(\bm{\mathrm{u}}, \bm{\mathrm{v}})} \right) 
	\end{align*}
	is a 2-rotation symmetric bent-negabent function. 
\end{corollary}

In Corollary \ref{corollary:2RS_BN}, if $A$ only contains one element, then we obtain the following corollary.

\begin{corollary}\rm 
	\label{corollary:2RS_BN_anydegree}
	Let $\bm{\gamma}$ be an arbitrary vector in $ \mathbb{F}_2^{2k}$ and $\wt(\bm{\gamma}) \ge 2$. 
	The function $f \in \mathcal{B}_{4k}$ defined by 
	\begin{align*}
	f(\bm{\mathrm{x}}, \bm{\mathrm{y}}) = f_0(\bm{\mathrm{x}}, \bm{\mathrm{y}}) +  \sum_{\mbox{\tiny $\begin{array} {c} \bm{\mathrm{u}} * \bm{\mathrm{v}} = \bm{0}_{2k} \\ \bm{\mathrm{u}} + \bm{\mathrm{v}} \in O_{2k}(\bm{\gamma}) \end{array}$}} (\bm{\mathrm{x}}, \bm{\mathrm{y}})^{(\bm{\mathrm{u}}, \bm{\mathrm{v}})} 
	\end{align*}
	is a 2-rotation symmetric bent-negabent function, and $\deg(f) = \wt(\bm{\gamma})$. 
\end{corollary}

\begin{remark} \rm 
	Using Corollary \ref{corollary:2RS_BN_anydegree}, we can easily obtain 2-rotation symmetric bent-negabent functions with any possible algebraic degrees ranging from $2$ to $2k$. 
\end{remark}

Next, according to Corollary \ref{corollary:2RS_BN_anydegree}, we give an example of an $8$-variable 2-rotation symmetric bent-negabent function with the maximum algebraic degree. 

\begin{example} \rm 
	In Corollary \ref{corollary:2RS_BN_anydegree}, let $k=2$ and $\bm{\gamma} = \bm{1}_4$. 
	Then the ANF of $f$ is given by 
	\begin{center} 
		\fcolorbox{white}{white}{
			\parbox{.9\linewidth}
			{$f(x_0, \cdots, x_3, y_0, \cdots, y_3) = f_0(x_0, \cdots, x_3, y_0, \cdots, y_3) + x_0x_1x_2x_3 + x_0x_1x_2y_3 + x_0x_1x_3y_2 + x_0x_2x_3y_1 + x_1x_2x_3y_0 + x_0x_1y_2y_3 + x_0x_2y_1y_3 + x_0x_3y_1y_2 + x_1x_2y_0y_3 + x_1x_3y_0y_2 + x_2x_3y_0y_1 + x_0y_1y_2y_3 + x_1y_0y_2y_3 + x_2y_0y_1y_3 + x_3y_0y_1y_2 + y_0y_1y_2y_3$,} 
		}
	\end{center}
	which was verified to be a 2-rotation symmetric bent-negabent function by using a SageMath program. 
\end{example}

\section{Comparisons with Known Results} 
\label{section:comparison}

In this section, we compare some relations on bent and negabent functions, and show that those constructions of bent-negabent functions in Sections \ref{section:bent_negabent_4k} and \ref{section:bent_negabent_4k+2} are not special cases of the third generic construction in \cite{Su2020}. 

Interestingly, all the characteristic functions $\chi_{S_i}$ for $i = 1, 2, 3, 4$ are negabent but not bent, and $\chi_T$ is a rotation symmetric negabent functions but not bent. 
(The proof is not difficult, and we do not include it in this paper.) 
Hence, in our constructions, the sum of a quadratic bent-negabent function and a negabent function not bent function, gives rise to a bent-negabent function. 
Recall the important characterization of negabent functions that a function $f$ on $2k$ variables is negabent if and only if $f + \sigma_2$ is bent, where $\sigma_2$ is the $2k$-variable quadratic homogeneous symmetric function. 
We summarize these results in Table \ref{table:someresults} for showing the differences in the negabent case. 

Next, we explain that our constructions of bent-negabent functions are not special cases of the third generic construction in \cite{Su2020}. 
We present the third generic construction in \cite{Su2020} in the following theorem. 

\begin{table}[t] 
	\caption{Some relations on bent and negabent functions}
	\begin{center}
		\label{table:someresults}
		\small
		\begin{tabular}{lll}
			\toprule
			$f$ & \ \  quadratic function $\delta$  & \ \ $f+\delta$ \cr 
			%heading
			%\noalign{\smallskip}
			\midrule
			bent & \ \ $\sigma_2$: bent not negabent  & \ \ negabent  \cr
			negabent & \ \ $\sigma_2$: bent not negabent  & \ \ bent \cr 
			$\chi_{S_1}$/$\chi_{S_2}$: \emph{negabent not bent}  & \ \ $g_0$: \emph{bent-negabent} & \ \ \emph{bent-negabent} \cr 
			$\chi_{S_3}$/$\chi_{S_4}$: \emph{negabent not bent}  & \ \ $h_0$: \emph{bent-negabent} & \ \ \emph{bent-negabent} \cr 
			\tabincell{l}{$\chi_T$: \emph{rotation symmetric} \\ \hspace{0.6cm} \emph{negabent not bent}}  & \ \ \tabincell{l}{$f_0$: \emph{2-rotation symmetric} \\ \hspace{0.5cm} \emph{bent-negabent}} & \ \ \tabincell{l}{\emph{2-rotation symmetric} \\ \emph{bent-negabent}} \cr 
			\bottomrule 
		\end{tabular}
	\end{center}
\end{table}

\begin{theorem} \rm (\cite[Theorem 8]{Su2020}) 
	\label{theorem:Su3Construction}
	Let $f_0 \in \mathcal{B}_{2m}$ be a bent function of the form (\ref{align:MM_class}), and $L$ be a linear subspace of $\F_2^{m}$, and $\Theta$ be a nonempty subset of $R_L$, i.e., a complete set of coset representatives of $L$ in $\F_2^{m}$. 
	For $\xbu, \ybu \in \F_2^m$, define a subset $S$ of $\F_2^{2m}$ by 
	\begin{align} 
		\label{align:Su_S_set}
		S = \bigcup_{\thbu \in \Theta} \{(\xbu, \ybu) \in \F_2^{2m} : \xbu \in L, \ybu \in C_{\thbu}(L^{\perp}) \}. 
	\end{align} 
	Suppose that $\pi, \varphi$ and $L$ satisfy the following two conditions:  
	\begin{enumerate} [C-1] 
		\item $\pi$ is linear and satisfies $\pi(L^{\perp}) = L^{\perp}$, 
		\item $\varphi(\albu + L^{\perp}) = \varphi(\albu)$ for all $\albu \in \F_2^m$. 
	\end{enumerate} 
	Then the function defined by $f(\xbu, \ybu) = f_0(\xbu, \ybu) + \chi_S(\xbu, \ybu)$ is bent. 
\end{theorem}  

Comparing with Theorem \ref{theorem:Su3Construction}, we have the following discussions about our four constructions, i.e., Theorems \ref{theorem:4k}, \ref{theorem:8k_variable}, \ref{theorem:4k+2} and \ref{theorem:8k+2}, where each of them is not a special case of Theorem \ref{theorem:Su3Construction}. 

(i) {\bf Theorem \ref{theorem:4k}:} 
For $m = 2k$, 
we know that $g_0$ is a function in the Maiorana-McFarland class with $\pi(\ybu) = \ybu$, $\varphi(\ybu) = \ybu'\cdot \ybu''$. 
For $S_1$ in (\ref{align:S1_set}), $\{\xbu \in \F_2^{2k}: \xbu' \in \F_2^k, \xbu'' = \xbu' + \gmbu_1\}$ is not a linear subspace except for $\gmbu_1 = \bm{0}_k$. 
%Hence, $S_1$ cannot be special cases of $S$ in (\ref{align:Su_S_set}) for $\gmbu_1 \ne \bm{0}_k$. 
If $\gmbu_1 = \bm{0}_k$, i.e., $L = \{\xbu \in \F_2^{2k}: \xbu' = \xbu'' \in \F_2^k \}$, then $L^{\perp} = L$ and $\gmbu_2 = \bm{0}_k$. 
That is to say, $\Gamma = \Theta = \{\bm{0}_m \}$, and $S_1 = S = \{(\xbu, \ybu)\in \F_2^{4k} : \xbu' = \xbu'' \in \F_2^k, \ybu' = \ybu'' \in \F_2^k \}$. 
For this case, the condition C-1 is satisfied, while C-2 is not satisfied. 
For instance, let $k = 2$ and $\albu = (1, 0, 0, 0)$. 
We have $\albu + L^{\perp} = \{(1, 0, 0, 0), (0, 0, 1, 0), (1, 1, 0, 1), (0, 1, 1, 1) \}$, and $\varphi(\albu + L^{\perp}) = \F_2 \ne \varphi(\albu) = 0$. 
%Therefore, Theorem \ref{theorem:4k} is not a special case of Theorem \ref{theorem:Su3Construction}. 

(ii) {\bf Theorem \ref{theorem:8k_variable}:} 
It is clear that the set $S_2$ in (\ref{align:S2_set}) is a special cases of $S$ in (\ref{align:Su_S_set}) with $L = L^{\perp} = A_2^{2k}$. 
Then we know that the condition C-1 is satisfied while C-2 is not satisfied. %for the construction of Theorem \ref{theorem:8k_variable}. 
For example, 
let $k = 1$ (i.e., $m=4$) and $\albu = (1, 0, 0, 0)$. 
Then, $\albu + A_2^{2k} = \{(1, 0, 0, 0), (0, 1, 0, 0), (1, 0, 1, 1), (0, 1, 1, 1) \}$, and $\varphi(\albu + A_2^{2k}) = \F_2 \ne \varphi(\albu) = 0$. 
%Hence, Theorem \ref{theorem:8k_variable} is not a special case of Theorem \ref{theorem:Su3Construction}. 

(iii) {\bf Theorem \ref{theorem:4k+2}:}
In accordance with those notations in Section \ref{section:bent_negabent_4k+2}, we rewrite $S$ in (\ref{align:Su_S_set}) as 
\begin{align}
	\label{align:Su_S_set2}
	S = \bigcup_{\thbu \in \Theta} \{(\Xbu, \Ybu) \in \F_2^{2m+2} : \Xbu \in L, \Ybu \in C_{\thbu}(L^{\perp}) \}. 
\end{align}
We know $h_0$ is a bent function in the Maiorana-McFarland class with $\pi(\Ybu) = (y_0 + y_m, y_1, \cdots, y_m)$, $\varphi(\Ybu) = \ybu'\cdot \ybu''$. 
For the similar reason to (i), $S_3$ in (\ref{align:S3_set}) is a special cases of $S$ in (\ref{align:Su_S_set2}) if and only if 
%$\gmbu = \bm{0}_{2t}$, i.e., 
$\Gamma = \Theta = \{\bm{0}_m \}$, and 
$L = \{\Xbu \in \F_2^{2t+1} : \xbu' = \xbu'' \in \F_2^t, x_m \in \F_2 \}$ and $L^{\perp} = \{\Xbu \in \F_2^{2t+1} : \xbu' = \xbu'' \in \F_2^t, x_m=0 \}$. 
For this case, the condition C-1 is satisfied while C-2 is not satisfied. 
For example, let $k=2$ and $\albu = (1, 0, 0, 0, 0)$. 
Then we have $\albu + L^{\perp} = \{(1, 0, 0, 0, 0), (1, 1, 0, 1, 0), (0, 0, 1, 0, 0), (0, 1, 1, 1, 0) \}$ and $\varphi(\albu + L^{\perp}) = \F_2 \ne \varphi(\albu) = 0$. 
%Hence, Theorem \ref{theorem:4k+2} is not a special case of Theorem \ref{theorem:Su3Construction}. 

(iv) {\bf Theorem \ref{theorem:8k+2}:}
It is obvious that $S_4$ in (\ref{align:S4_set}) is a special cases of $S$ in (\ref{align:Su_S_set2}) with $L = A_2^{2k} \times \F_2$ and $L^{\perp} = A_2^{2k} \times \{0 \}$. 
Then the condition C-1 is satisfied while C-2 is not satisfied. 
For example, 
let $k = 1$, i.e., $m=4$, and $\albu = (1, 0, 0, 0, 0)$. 
Then, $\albu + A_2^{2k} \times \{0 \} = \{(1, 0, 0, 0, 0), (0, 1, 0, 0, 0), (1, 0, 1, 1, 0), (0, 1, 1, 1, 0) \}$, and $\varphi(\albu + A_2^{2k} \times \{0 \}) = \F_2 \ne \varphi(\albu) = 0$. 
%Hence, Theorem \ref{theorem:8k+2} is not a special case of Theorem \ref{theorem:Su3Construction}. 

\section{Concluding Remarks} 
\label{section:conclusion}

In this paper, we have focused on systematic methods for constructing bent-negabent functions. 
We have discussed three sets of the new constructions: 
(1) bent-negabent functions on $4k$ and $8k$ variables by using the sets $S_1$ in (\ref{align:S1_set}) and $S_2$ in (\ref{align:S2_set}) to modify the truth table of $g_0$ in (\ref{align:g_0});
(2) bent-negabent functions on $4k+2$ and $8k+2$ variables by using the sets $S_3$ in (\ref{align:S3_set}) and $S_4$ in (\ref{align:S4_set}) to modify the truth table of $h_0$ in (\ref{align:h_0});  
(3) 2-rotation symmetric bent-negabent functions on $4k$ variables with any possible algebraic degrees by modifying the truth table of $f_0$ in (\ref{align:f_0}). 
We also identified the necessary and sufficient conditions under which the algebraic degrees of bent-negabent functions from (1) and (2) reach the maximum. 
The ANFs and duals of all these constructed functions were also determined. 
Moreover, all our constructions of bent-negabent functions mentioned above are not special cases of the generic constructions of bent functions in \cite{Su2020}. 

On the other hand, it has been proved in \cite{Schmidt2008} that the algebraic degree of any $n$-variable ($n$ even and $n > 6$) bent-negabent function in the Maiorana-McFarland class is at most $\frac{n}{2}-1$. 
Hence, our newly constructed bent-negabent functions with the maximum algebraic degree cannot be in the Maiorana-McFarland class. 
%The authors are not clear whether there exist bent-negabent functions outside the completed Maiorana-McFarland class from our construction. 
Are they in or outside the completed Maiorana-McFarland class? 
%The classification of our newly constructed bent-negabent functions is an interesting problem. 
It is an interesting problem deserving further research. 

\section*{Appendix} 

\textbf{Proof of Lemma \ref{lemma:4k+2_WHT&NHT}}: 
	%We first prove that $h$ is bent. 
	%	It is easy to verify that the inverse of $(y_0 + y_m, y_1, \cdots, y_m)$ is itself. 
	%	By (\ref{align:dual_MM}) and (\ref{align:Dual_def}), we know the Walsh-Hadamard transform of $h_0$ at $(\bm{\mathrm{U}}, \bm{\mathrm{V}}) \in \mathbb{F}_2^{4k+2}$ is given by 
	%	\begin{align}
		%	\W_{h_0}(\bm{\mathrm{U}}, \bm{\mathrm{V}}) =& 2^{2k+1} (-1)^{\bm{\mathrm{V}} \cdot (u_0 + u_m, u_1, \cdots, u_m) + (u_0+ u_m, u_1, \cdots, u_{k-1}) \cdot \bm{\mathrm{u}}''} \notag \\ 
		%	=& 2^{2k+1} (-1)^{\bm{\mathrm{u}}'\cdot \bm{\mathrm{u}}'' + \bm{\mathrm{U}}\cdot \bm{\mathrm{V}} + u_m(v_0 + u_k)}. 
		%	\label{align:WHT_h0}
		%	\end{align} 
	By (\ref{align:PWHT}), the fragmentary Walsh-Hadamard transform of $h_0$ over $S_3$ at $(\bm{\mathrm{U}}, \bm{\mathrm{V}}) \in \mathbb{F}_2^{4k+2}$ is given by 
	\begin{align*}
		\W_{h_0, S_3}(\bm{\mathrm{U}}, \bm{\mathrm{V}}) =& \sum_{\gmbu \in \Gamma} \sum_{(\bm{\mathrm{X}}, \bm{\mathrm{Y}}) \in L_{\gmbu, E_{\gmbu}}} (-1)^{\bm{\mathrm{X}} \cdot \bm{\mathrm{Y}}+ x_0y_m + \bm{\mathrm{y}}'\cdot \bm{\mathrm{y}}'' + \bm{\mathrm{U}}\cdot \bm{\mathrm{X}}+ \bm{\mathrm{V}} \cdot \bm{\mathrm{Y}}} \\ 
		%=& \W_{h_0}(u, v) - 2\sum_{\mbox{\tiny $\begin{array} {c} \bm{\gamma}_1 \in \Gamma_1 \\ \bm{\gamma}_2 \in \Gamma_2 \end{array}$}} \sum_{x' \in \mathbb{F}_2^k} \sum_{y' \in \mathbb{F}_2^k}  \\ 
		%& \hspace{0.5cm} \sum_{x_{m-1} \in \mathbb{F}_2} \sum_{y_{m-1} \in E_{\gamma_2}} (-1)^{(x', x'+ \gamma_1, x_{m-1})\cdot (y', y'+ \gamma_2, y_{m-1}) + x_0y_{m-1} + y'\cdot (y' + \gamma_2) + u\cdot(x', x'+ \gamma_1, x_{m-1}) + v\cdot(y', y'+ \gamma_2, y_{m-1})} \\ 
		=& \sum_{\gmbu \in \Gamma}(-1)^{\bm{\mathrm{u}}'' \cdot \bm{\gamma}_1 + \bm{\mathrm{v}}''\cdot \bm{\gamma}_2 + \bm{\gamma}_1 \cdot \bm{\gamma}_2} \sum_{\bm{\mathrm{x}}' \in \mathbb{F}_2^k} (-1)^{(\bm{\mathrm{u}}'+ \bm{\mathrm{u}}''+ \bm{\gamma}_2)\cdot \bm{\mathrm{x}}'} \sum_{\bm{\mathrm{y}}' \in \mathbb{F}_2^k} (-1)^{(\bm{\mathrm{v}}'+ \bm{\mathrm{v}}'' + \bm{1}_k + \bm{\gamma}_1 + \bm{\gamma}_2) \cdot \bm{\mathrm{y}}'} \\ 
		& \hspace{1.0cm} \sum_{y_m \in E_{\bm{\gamma}}} (-1)^{(x_0 + v_m) \cdot y_m} \sum_{x_m \in \mathbb{F}_2}(-1)^{(u_m + y_m) \cdot x_m}.  
		%=& \W_{h_0}(u, v) - 2\sum_{\gamma \in \Gamma}(-1)^{v''\cdot \gamma} \sum_{x' \in \mathbb{F}_2^k} (-1)^{(u'+ u''+ \gamma)\cdot x'} \sum_{y' \in \mathbb{F}_2^k} (-1)^{(v'+ v'' + \gamma + 1_k) \cdot y'} \\ 
		%& \hspace{7.0cm} \left[1 + (-1)^{u_{m-1}} + (-1)^{x_0 + v_{m-1}} - (-1)^{u_{m-1} + v_{m-1}+ x_0} \right]. 
		%	=& \begin{cases} 
			%	\W_{h_0}(u, v) - 2^{k+1}\sum_{\gamma \in \Gamma}(-1)^{v''\cdot \gamma} \sum_{x' \in \mathbb{F}_2^k} (-1)^{(u'+ u''+ \gamma)\cdot x'} \left[1 + (-1)^{u_{m-1}} + (-1)^{x_0 + v_{m-1}} - (-1)^{u_{m-1} + v_{m-1}+ x_0} \right], \\ 
			%	\hspace{11.0cm} \text{if}\ v' + v'' + \gamma = 1_k\ \text{for a}\ \gamma\in \Gamma, \\  
			%	\W_{h_0}(u, v),\ \hspace{9.2cm}\text{otherwise}
			%	\end{cases}
	\end{align*} 
	From Lemma \ref{lemma:ex_sum} we know
	\begin{align}
		\sum_{y_m \in E_{\bm{\gamma}}} (-1)^{(x_0 + v_m) \cdot y_m} \sum_{x_m \in \mathbb{F}_2}(-1)^{(u_m + y_m) \cdot x_m} = \begin{cases} 
			2(-1)^{(x_0 + v_m) \cdot u_m},\ u_m \in E_{\bm{\gamma}}, \\ 
			0,\ \hspace{2.3cm} u_m \notin E_{\bm{\gamma}}. 
		\end{cases} 
		\label{align:inter_Res2} 
	\end{align} 
	Let $\Theta(\bm{\mathrm{U}}, \bm{\mathrm{V}})$ be a subset of $\Gamma$ defined by $\Theta(\bm{\mathrm{U}}, \bm{\mathrm{V}}) = \{\bm{\gamma} \in \Gamma : u_m \in E_{\bm{\gamma}} \}$. 
	Then, we have 
	\begin{align*}
		\W_{h_0, S_3}(\bm{\mathrm{U}}, \bm{\mathrm{V}}) 
		=& 2 \sum_{\gmbu \in \Theta(\bm{\mathrm{U}}, \bm{\mathrm{V}})}(-1)^{\bm{\mathrm{u}}'' \cdot \bm{\gamma}_1 + \bm{\mathrm{v}}''\cdot \bm{\gamma}_2 + \bm{\gamma}_1 \cdot \bm{\gamma}_2 + u_m\cdot v_m} \sum_{\bm{\mathrm{x}}' \in \mathbb{F}_2^k} (-1)^{(\bm{\mathrm{u}}'+ \bm{\mathrm{u}}'' + \bm{\mathrm{e}}_{k}^{u_m} + \bm{\gamma}_2)\cdot \bm{\mathrm{x}}'} \\ 
		& \hspace{1.0cm} \sum_{\bm{\mathrm{y}}' \in \mathbb{F}_2^k} (-1)^{(\bm{\mathrm{v}}'+ \bm{\mathrm{v}}'' + \bm{1}_k + \bm{\gamma}_1 + \bm{\gamma}_2) \cdot \bm{\mathrm{y}}'}. 
		%&\hspace{2.0cm} \sum_{\bm{\mathrm{x}}' \in \mathbb{F}_2^k} (-1)^{(\bm{\mathrm{u}}'+ \bm{\mathrm{u}}''+ (u_m, \bm{0}_{k-1}) + \bm{\gamma}_2)\cdot \bm{\mathrm{x}}'} \sum_{\bm{\mathrm{y}}' \in \mathbb{F}_2^k} (-1)^{(\bm{\mathrm{v}}'+ \bm{\mathrm{v}}'' + \bm{1}_k + \bm{\gamma}_1 + \bm{\gamma}_2) \cdot \bm{\mathrm{y}}'} \\ 
		%	=& \begin{cases}
			%		\W_{h_0}(\bm{\mathrm{U}}, \bm{\mathrm{V}}) - 2^{2k+2} (-1)^{\bm{\mathrm{u}}'' \cdot \bm{\gamma}_1 + \bm{\mathrm{v}}''\cdot \bm{\gamma}_2 + \bm{\gamma}_1 \cdot \bm{\gamma}_2 + u_m\cdot v_m},\ \bm{\mathrm{u}}'+ \bm{\mathrm{u}}'' + (u_m, \bm{0}_{k-1}) + \bm{\gamma}_2 = \bm{0}_k,\\ 
			%		\hspace{3.5cm} \bm{\mathrm{v}}' + \bm{\mathrm{v}}'' + \bm{1}_k + \bm{\gamma}_1 + \bm{\gamma}_2 = \bm{0}_k, \text{and}\ u_m \in E_{\bm{\gamma}}\ \text{for}\ (\bm{\gamma}_1, \bm{\gamma}_2) \in \Gamma, \\ 
			%		\W_{h_0}(\bm{\mathrm{U}}, \bm{\mathrm{V}}),\ \hspace{1.5cm} \text{otherwise}, 
			%	\end{cases} 
	\end{align*}
	%where the second identity holds by Lemma \ref{lemma:ex_sum}.  
	
	(1) If there does not exist a $\gmbu$ in $\Theta(\bm{\mathrm{U}}, \bm{\mathrm{V}})$ such that $\bm{\gamma}_2 = \bm{\mathrm{u}}'+ \bm{\mathrm{u}}'' + \bm{\mathrm{e}}_{k}^{u_m}$ and $\bm{\gamma}_1 + \bm{\gamma}_2 = \bm{\mathrm{v}}' + \bm{\mathrm{v}}'' + \bm{1}_k$, 
	then we have $\W_{h_0, S_3}(\bm{\mathrm{U}}, \bm{\mathrm{V}}) = 0$. 
	
	(2) If there exists a $\gmbu$ in $\Theta(\bm{\mathrm{U}}, \bm{\mathrm{V}})$ such that $\bm{\gamma}_2 = \bm{\mathrm{u}}'+ \bm{\mathrm{u}}'' + \bm{\mathrm{e}}_{k}^{u_m}$ and $\bm{\gamma}_1 + \bm{\gamma}_2 = \bm{\mathrm{v}}' + \bm{\mathrm{v}}'' + \bm{1}_k$, 
	it holds that  
	\begin{align*} 
		\bm{\mathrm{u}}''\cdot \bm{\gamma}_1 + \bm{\mathrm{v}}'' \cdot \bm{\gamma}_2 + \bm{\gamma}_1 \cdot \bm{\gamma}_2 + u_m \cdot v_m =& \bm{\mathrm{u}}'' \cdot \bm{\gamma}_1 + (\bm{\mathrm{v}}' + \bm{\gamma}_2 + \bm{1}_k) \cdot \bm{\gamma}_2 + u_m \cdot v_m \\ 
		=& \bm{\mathrm{u}}''\cdot \bm{\gamma}_1 + \bm{\mathrm{v}}' \cdot \bm{\gamma}_2 + u_m \cdot v_m \\ 
		=& \bm{\mathrm{u}}'' \cdot (\bm{\mathrm{v}}' + \bm{\mathrm{v}}'' + \bm{1}_k) + (\bm{\mathrm{u}}''+ \bm{\mathrm{v}}')\cdot \bm{\gamma}_2 + u_m \cdot v_m \\ 
		=& \bm{\mathrm{u}}'\cdot \bm{\mathrm{u}}'' + \bm{\mathrm{u}}'\cdot \bm{\mathrm{v}}' + \bm{\mathrm{u}}''\cdot \bm{\mathrm{v}}'' + u_m\cdot v_m + u_m \cdot (u_k + v_0) \\ 
		=& \bm{\mathrm{u}}'\cdot \bm{\mathrm{u}}'' + \bm{\mathrm{U}} \cdot \bm{\mathrm{V}} + u_m \cdot (u_k + v_0). 
	\end{align*} 
	Then we have $\W_{h_0, S_3}(\bm{\mathrm{U}}, \bm{\mathrm{V}}) = 2^{2k+1} (-1)^{\bm{\mathrm{u}}'\cdot \bm{\mathrm{u}}'' + \bm{\mathrm{U}} \cdot \bm{\mathrm{V}} + u_m \cdot (u_k + v_0)} = \W_{h_0} (\bm{\mathrm{U}}, \bm{\mathrm{V}})$, by (\ref{align:WHT_h0}). 
	
	Then (\ref{align:WHT_4k+2}) follows from the two cases discussed above. 
%	the Walsh-Hadamard transform of $h$ at $(\bm{\mathrm{U}}, \bm{\mathrm{V}}) \in \mathbb{F}_2^{4k+2}$ is given by 
%	\begin{align}
%		\W_{h}(\bm{\mathrm{U}}, \bm{\mathrm{V}}) =& \begin{cases}
%			-\W_{h_0}(\bm{\mathrm{U}}, \bm{\mathrm{V}}),\ \bm{\mathrm{u}}'+ \bm{\mathrm{u}}'' + (u_m, \bm{0}_{k-1}) + \bm{\gamma}_2 = \bm{0}_k, \bm{\mathrm{v}}' + \bm{\mathrm{v}}'' + \bm{1}_k + \bm{\gamma}_1 + \bm{\gamma}_2 = \bm{0}_k, \\ 
%			\hspace{2.4cm} \text{and}\ u_m \in E_{\bm{\gamma}}\ \text{for}\ (\bm{\gamma}_1, \bm{\gamma}_2) \in \Gamma, \\ 
%			\W_{h_0}(\bm{\mathrm{U}}, \bm{\mathrm{V}}),\ \hspace{0.4cm} \text{otherwise}, 
%		\end{cases}
%		\label{align:WHT_4k+2} 
%	\end{align}
%	Hence, we have $|\W_h(\bm{\mathrm{U}}, \bm{\mathrm{V}})| = |\W_{h_0}(\bm{\mathrm{U}}, \bm{\mathrm{V}})| = 2^{2k+1}$ for all $(\bm{\mathrm{U}}, \bm{\mathrm{V}}) \in \mathbb{F}_2^{4k+2}$. 
%	That is to say, $h$ is a bent function. 
	
	%Next, we prove that $h$ is negabent. 
	By (\ref{align:PNHT}), the fragmentary nega-Hadamard transform of $h_0$ over $S_3$ at $(\bm{\mathrm{U}}, \bm{\mathrm{V}}) \in \mathbb{F}_2^{4k+2}$ is given by 
	\begin{align*} 
		\N_{h_0, S_3} (\bm{\mathrm{U}}, \bm{\mathrm{V}}) =& \sum_{\gmbu \in \Gamma} \sum_{(\bm{\mathrm{X}}, \bm{\mathrm{Y}}) \in L_{\gmbu, E_{\gmbu}}} (-1)^{\bm{\mathrm{X}} \cdot \bm{\mathrm{Y}} + x_0y_m + \bm{\mathrm{y}}'\cdot \bm{\mathrm{y}}'' + \bm{\mathrm{U}}\cdot \bm{\mathrm{X}} + \bm{\mathrm{V}} \cdot \bm{\mathrm{Y}}} \imath^{\wt(\bm{\mathrm{X}}, \bm{\mathrm{Y}})} \\  
		=& \sum_{\gmbu \in \Gamma} \sum_{\bm{\mathrm{x}}' \in \mathbb{F}_2^k} (-1)^{\ubu \cdot (\bm{\mathrm{x}}', \bm{\mathrm{x}}'+ \bm{\gamma}_1)} \imath^{\wt(\bm{\mathrm{x}}', \bm{\mathrm{x}}'+ \bm{\gamma}_1)} \\ 
		& \hspace{1.0cm} \sum_{\bm{\mathrm{y}}' \in \mathbb{F}_2^k} (-1)^{(\bm{\mathrm{x}}', \bm{\mathrm{x}}'+ \bm{\gamma}_1)\cdot (\bm{\mathrm{y}}', \bm{\mathrm{y}}'+ \bm{\gamma}_2)+ \bm{\mathrm{y}}'\cdot (\bm{\mathrm{y}}' + \bm{\gamma}_2) + \vbu \cdot (\bm{\mathrm{y}}', \bm{\mathrm{y}}' + \bm{\gamma}_2)} \imath^{\wt(\bm{\mathrm{y}}', \bm{\mathrm{y}}' + \bm{\gamma}_2)} \\ 
		& \hspace{1.0cm} \sum_{y_m \in E_{\gamma}}(-1)^{(x_0 + v_m)\cdot y_m} \imath^{\wt(y_m)} \sum_{x_m \in \mathbb{F}_2}(-1)^{(u_m+ y_m) \cdot x_m} \imath^{\wt(x_m)} \\ 
		=& \sum_{\gmbu \in \Gamma}(-1)^{\bm{\mathrm{u}}'' \cdot \bm{\gamma}_1 + \bm{\mathrm{v}}''\cdot \bm{\gamma}_2 + \bm{\gamma}_1 \cdot \bm{\gamma}_2} \imath^{\wt(\bm{\gamma}_1) + \wt(\bm{\gamma}_2)} \sum_{\bm{\mathrm{x}}' \in \mathbb{F}_2^k} (-1)^{(\bm{\mathrm{u}}'+ \bm{\mathrm{u}}''+ \bm{1}_k + \bm{\gamma}_1 + \bm{\gamma}_2)\cdot \bm{\mathrm{x}}'} \\ 
		& \hspace{1.0cm} \sum_{\bm{\mathrm{y}}' \in \mathbb{F}_2^k} (-1)^{(\bm{\mathrm{v}}'+ \bm{\mathrm{v}}'' + \bm{\gamma}_1) \cdot \bm{\mathrm{y}}'} \sum_{y_m \in E_{\bm{\gamma}}} [1 + \imath (-1)^{u_m + y_m} ](-1)^{(x_0 + v_m)\cdot y_m} \imath^{\wt(y_m)} \\ 
		=& \sum_{\gmbu \in \Gamma} (-1)^{\bm{\mathrm{u}}'' \cdot \bm{\gamma}_1 + \bm{\mathrm{v}}''\cdot \bm{\gamma}_2 + \bm{\gamma}_1 \cdot \bm{\gamma}_2} \imath^{\wt(\bm{\gamma}_1) + \wt(\bm{\gamma}_2)} \sum_{\bm{\mathrm{y}}' \in \mathbb{F}_2^k} (-1)^{(\bm{\mathrm{v}}'+ \bm{\mathrm{v}}'' + \bm{\gamma}_1) \cdot \bm{\mathrm{y}}'} \\ 
		& \hspace{1.0cm} \sum_{y_m \in E_{\bm{\gamma}}} [1+ \imath (-1)^{u_m + y_m} ](-1)^{v_m \cdot y_m} \imath^{\wt(y_m)} \sum_{\bm{\mathrm{x}}' \in \mathbb{F}_2^k} (-1)^{(\bm{\mathrm{u}}'+ \bm{\mathrm{u}}''+ \bm{1}_k + \bm{\gamma}_1 + \bm{\gamma}_2 + \bm{\mathrm{e}}_{k}^{y_m})\cdot \bm{\mathrm{x}}'}.  
		%	=& \begin{cases}
			%	\N_{h_0}(u,v) - 2^{2k+1} (-1)^{u''\cdot \gamma_1 + v''\cdot \gamma_2 + \gamma_1 \cdot \gamma_2 + v_{m-1}\cdot \varepsilon} \imath^{\wt(\gamma_1) + \wt(\gamma_2) + \wt(\varepsilon)} [1+ \imath (-1)^{u_{m-1} + \varepsilon} ], \\ 
			%	\hspace{1.5cm}\ v' + v'' = \gamma_1, \varepsilon \in E_{\gamma_2}, u'+ u''+ 1_k + \gamma_1 + \gamma_2 + (\varepsilon, 0_{k-1}) = 0_k\ \text{for}\ \gamma_1 \in \Gamma_1, \gamma_2 \in \Gamma_2, \\ 
			%	\N_{h_0}(u,v),\ \text{otherwise}
			%	\end{cases} \\ 
		%=& \N_{h_0}(u,v) - 2 \sum_{\mbox{\tiny $\begin{array} {c} \bm{\gamma}_1 \in \Gamma_1 \\ \bm{\gamma}_2 \in \Gamma_2 \end{array}$}} T(u, v, \gamma_1, \gamma_2)
	\end{align*}
	
	For simplicity, we shall denote $(-1)^{\bm{\mathrm{u}}'' \cdot \bm{\gamma}_1 + \bm{\mathrm{v}}''\cdot \bm{\gamma}_2 + \bm{\gamma}_1 \cdot \bm{\gamma}_2} \imath^{\wt(\bm{\gamma}_1) + \wt(\bm{\gamma}_2)} \cdots$ by $T(\bm{\mathrm{U}}, \bm{\mathrm{V}}, \gmbu)$. 
	That is to say, 
	\[
	\N_{h_0, S_3}(\bm{\mathrm{U}}, \bm{\mathrm{V}}) = \sum_{\gmbu \in \Gamma} T(\bm{\mathrm{U}}, \bm{\mathrm{V}}, \gmbu). 
	\]
	
	For $(\bm{\mathrm{U}}, \bm{\mathrm{V}}) \in \mathbb{F}_2^{4k+2}$ satisfying $\bm{\mathrm{v}}' + \bm{\mathrm{v}}'' = \bm{\gamma}_1$, 
	we know that there exist at most two different vectors $\gmbu = (\gmbu_1, \gmbu_2)$ and $\hat{\gmbu} = (\gmbu_1, \hat{\bm{\gamma}}_2)$ such that $\bm{\mathrm{u}}'+ \bm{\mathrm{u}}''+ \bm{1}_k + \bm{\gamma}_1 + \bm{\gamma}_2 + \bm{\mathrm{e}}_{k}^{\varepsilon} = \bm{0}_k$ and $\bm{\mathrm{u}}'+ \bm{\mathrm{u}}''+ \bm{1}_k + \bm{\gamma}_1 + \hat{\bm{\gamma}}_2 + \bm{\mathrm{e}}_{k}^{\hat{\varepsilon}} = \bm{0}_k$, where $\varepsilon \in E_{\gmbu}$ and $\hat{\varepsilon} \in E_{\hat{\gmbu}}$. 
	Moreover, if both $\bm{\gamma}, \hat{\bm{\gamma}}$ exist, then $\varepsilon + \hat{\varepsilon} = 1$. 
	We consider the following three cases. 
	
	(1) If there does not exist a $\bm{\gamma} \in \Gamma$ such that $\bm{\mathrm{v}}' + \bm{\mathrm{v}}'' = \bm{\gamma}_1, \varepsilon \in E_{\bm{\gamma}}$, and $\bm{\mathrm{u}}'+ \bm{\mathrm{u}}''+ \bm{1}_k + \bm{\gamma}_1 + \bm{\gamma}_2 + \bm{\mathrm{e}}_{k}^{\varepsilon} = \bm{0}_k$, then $T(\bm{\mathrm{U}}, \bm{\mathrm{V}}, \gmbu) = 0$ for all $\gmbu \in \Gamma$, by Lemma \ref{lemma:ex_sum}. 
	Then we have $\N_{h_0, S_3}(\bm{\mathrm{U}}, \bm{\mathrm{V}}) = 0$. 
	
	(2) If there exists only one $\bm{\gamma} \in \Gamma$ such that $\bm{\mathrm{v}}' + \bm{\mathrm{v}}'' = \bm{\gamma}_1$ and $\bm{\mathrm{u}}'+ \bm{\mathrm{u}}''+ \bm{1}_k + \bm{\gamma}_1 + \bm{\gamma}_2 + \bm{\mathrm{e}}_{k}^{\varepsilon} = \bm{0}_k$, where $\varepsilon \in E_{\bm{\gamma}}$, 
	then we have 
	\[
	T(\bm{\mathrm{U}}, \bm{\mathrm{V}}, \bm{\gamma}) = 2^{2k}(-1)^{\bm{\mathrm{u}}''\cdot \bm{\gamma}_1 + \bm{\mathrm{v}}''\cdot \bm{\gamma}_2 + \bm{\gamma}_1 \cdot \bm{\gamma}_2 + v_m \cdot \varepsilon} \imath^{\wt(\bm{\gamma}_1) + \wt(\bm{\gamma}_2) + \wt(\varepsilon)} [1+ \imath (-1)^{u_m + \varepsilon} ]. 
	\]
	By (\ref{align:inter_Res1}), it holds that 
	\begin{align*}
		(-1)^{(\bm{\mathrm{u}}'+ \bm{\mathrm{v}}')\cdot (\bm{\mathrm{u}}''+ \bm{\mathrm{v}}'')} \imath^{k - \wt(\bm{\mathrm{u}})} 
		=& (-1)^{\bm{\mathrm{u}}''\cdot \bm{\gamma}_1 + \bm{\mathrm{v}}'' \cdot (\bm{\gamma}_2 + \bm{\mathrm{e}}_{k}^{\varepsilon}) + \bm{\gamma}_1 \cdot (\bm{\gamma}_2 + \bm{\mathrm{e}}_{k}^{\varepsilon})} \imath^{\wt(\bm{\gamma}_1) + \wt(\bm{\gamma}_2 + \bm{\mathrm{e}}_{k}^{\varepsilon})} \\ 
		=& (-1)^{\bm{\mathrm{u}}''\cdot \bm{\gamma}_1 + \bm{\mathrm{v}}'' \cdot \bm{\gamma}_2 + \bm{\gamma}_1 \cdot \bm{\gamma}_2 + (\gamma_{1, 0} + v_k)\cdot \varepsilon} \imath^{\wt(\bm{\gamma}_1) + \wt(\bm{\gamma}_2) + \wt(\varepsilon) - 2 \wt(\gamma_{2, 0} \cdot \varepsilon)} \\ 
		=& (-1)^{\bm{\mathrm{u}}''\cdot \bm{\gamma}_1 + \bm{\mathrm{v}}'' \cdot \bm{\gamma}_2 + \bm{\gamma}_1 \cdot \bm{\gamma}_2 + (\gamma_{1, 0} + \gamma_{2, 0} + v_k) \cdot \varepsilon} \imath^{\wt(\bm{\gamma}_1) + \wt(\bm{\gamma}_2) + \wt(\varepsilon)} \\ 
		=& (-1)^{\bm{\mathrm{u}}''\cdot \bm{\gamma}_1 + \bm{\mathrm{v}}'' \cdot \bm{\gamma}_2 + \bm{\gamma}_1 \cdot \bm{\gamma}_2 + (u_0 + u_k + v_k) \cdot \varepsilon} \imath^{\wt(\bm{\gamma}_1) + \wt(\bm{\gamma}_2) + \wt(\varepsilon)}, 
	\end{align*} 
	where $\bm{\gamma_1} = (\gamma_{1, 0}, \gamma_{1, 1}, \cdots, \gamma_{1, k-1}), \bm{\gamma_2} = (\gamma_{2, 0}, \gamma_{2, 1}, \cdots, \gamma_{2, k-1}) \in \mathbb{F}_2^k$. 
	Let $g_0 \in \mathcal{B}_{4k}$ be defined in (\ref{align:g_0}). 
	Together with (\ref{align:Nega_{g_0}}) we have 
	\begin{align}
		\N_{h_0, S_3} (\Ubu, \Vbu) =& T(\bm{\mathrm{U}}, \bm{\mathrm{V}}, \gmbu) = (-1)^{(u_0 + u_k + v_k+ v_m) \cdot \varepsilon} [1+ \imath (-1)^{u_m + \varepsilon} ] \N_{g_0}(\bm{\mathrm{u}}, \bm{\mathrm{v}}) \notag \\ 
		=& \frac{1}{2}[(1 + (-1)^\varepsilon)(1+\imath(-1)^{u_m}) + (1-(-1)^\varepsilon) (1-\imath(-1)^{u_m}) (-1)^{u_0+u_k+v_k+v_m}]\N_{g_0}(\ubu, \vbu) \notag \\ 
		=& \frac{1}{2}(1+\imath (-1)^{u_0+u_k+v_k+v_m+u_m+\varepsilon}) \N_{h_0} (\Ubu, \Vbu).
		\label{align:T(UV)}
	\end{align}
%	Hence, the nega-Hadamard transform of $h$ is given by 
%	\begin{align*} 
%		\N_h(\bm{\mathrm{U}}, \bm{\mathrm{V}}) 
%		=& \N_{h_0}(\bm{\mathrm{U}}, \bm{\mathrm{V}}) - 2 \N_{g_0}(\bm{\mathrm{u}}, \bm{\mathrm{v}}) (-1)^{(u_0 + u_k + v_k+ v_m) \cdot \varepsilon} [1+ \imath (-1)^{u_m + \varepsilon} ].
%	\end{align*} 
	
	(3) If there exist two vectors $\bm{\gamma} = (\bm{\gamma}_1, \bm{\gamma}_2), \hat{\bm{\gamma}} = (\bm{\gamma}_1, \hat{\bm{\gamma}}_2)$ in $\Gamma$ such that $\bm{\mathrm{v}}' + \bm{\mathrm{v}}'' = \bm{\gamma}_1$, $\bm{\mathrm{u}}'+ \bm{\mathrm{u}}''+ \bm{1}_k + \bm{\gamma}_1 + \bm{\gamma}_2 + \bm{\mathrm{e}}_{k}^{\varepsilon} = \bm{0}_k$ and $\bm{\mathrm{u}}'+ \bm{\mathrm{u}}''+ \bm{1}_k + \bm{\gamma}_1 + \hat{\bm{\gamma}}_2 + \bm{\mathrm{e}}_{k}^{\hat{\varepsilon}} = \bm{0}_k$, where $\varepsilon \in E_{\bm{\gamma}}$ and $\hat{\varepsilon} \in E_{\hat{\bm{\gamma}}}$, then the fragmentary nega-Hadamard transform of $h_0$ over $S_3$ at $(\Ubu, \Vbu) \in \F_2^{4k+2}$ is 
	\begin{align*}
		\N_{h_0, S_3}(\bm{\mathrm{U}}, \bm{\mathrm{V}}) = T(\bm{\mathrm{U}}, \bm{\mathrm{V}}, \gmbu) + T(\bm{\mathrm{U}}, \bm{\mathrm{V}}, \hat{\gmbu}). 
	\end{align*} 
	Together with (\ref{align:T(UV)}) and $\varepsilon + \hat{\varepsilon} = 1$, we have 
	\begin{align*} 
		\N_{h_0, S_3}(\bm{\mathrm{U}}, \bm{\mathrm{V}}) = \N_{h_0}(\Ubu, \Vbu).  
	\end{align*} 
	
	Hence, (\ref{align:NHT_4k+2}) follows immediately from the three cases discussed above. 
{\hfill $\square$\par} 

To prove Lemma \ref{lemma:8k+2_WHT&NHT}, we need the following lemma. 

\begin{lemma} \rm 
	\label{lemma:Gamma_4}
	Let the notations $\Gamma$ and $E_{\gmbu}$ be the same as those in Subsection \ref{subsection:8k+2}. 
	For $(\ubu, \vbu) \in \F_2^{8k}$, there is at most one $\gmbu \in \Gamma$ satisfying $\ubu + \gmbu + \bm{\mathrm{e}}_{4k}^{\varepsilon} \in B_2^{2k}$ and $\vbu + \gmbu + (\gmbu_2, \gmbu_1) \in B_2^{2k}$, where $\varepsilon \in E_{\bm{\gamma}}$. 
\end{lemma}

\begin{proof}
	Suppose that there are two elements $\gmbu, \thbu \in \Gamma$ satisfying  
	$\begin{cases}
		\ubu + \gmbu + \bm{\mathrm{e}}_{4k}^{\varepsilon} \in B_2^{2k}, \\ 
		\ubu + \thbu + \bm{\mathrm{e}}_{4k}^{\kappa} \in B_2^{2k}, 
	\end{cases}$ 
	and $\begin{cases}
		\vbu + \gmbu + (\gmbu_2, \gmbu_1) \in B_2^{2k}, \\ 
		\vbu + \thbu + (\thbu_2, \thbu_1) \in B_2^{2k}, 
	\end{cases}$ 
	where $\varepsilon \in E_{\bm{\gamma}}, \kappa \in E_{\thbu}$. 
	Let $\xibu$ be an arbitrary element in $B_2^{2k}$. 
	Then there are $\ldbu_1, \ldbu_2, \ldbu_3, \ldbu_4$ in $A_2^{2k}$ such that $\begin{cases}
		\ubu + \gmbu + \bm{\mathrm{e}}_{4k}^{\varepsilon} = \xibu + \ldbu_1, \\ 
		\ubu + \thbu + \bm{\mathrm{e}}_{4k}^{\kappa} = \xibu + \ldbu_2, 
	\end{cases}$ 
	and $\begin{cases}
		\vbu + \gmbu + (\gmbu_2, \gmbu_1) = \xibu + \ldbu_3, \\ 
		\vbu + \thbu + (\thbu_2, \thbu_1) = \xibu + \ldbu_4. 
	\end{cases}$ 
	We consider the following two cases. 
	
	(1) Supposing $\varepsilon = \kappa$, then we have $\gmbu + A_2^{2k} = \ubu + \bm{\mathrm{e}}_{4k}^{\varepsilon} + \xibu + A_2^{2k} = \thbu + A_2^{2k}$, which indicates $\gmbu = \thbu$ by the definition of $\Gamma$. 
	
	(2) Supposing $\varepsilon \ne \kappa$, i.e., $\varepsilon + \kappa = 1$, then we have $\begin{cases}
		\gmbu + \thbu + \bm{\mathrm{e}}_{4k}^1 \in A_2^{2k}, \\ 
		\gmbu + \thbu + (\gmbu_2, \gmbu_1)  + (\thbu_2, \thbu_1) \in A_2^{2k}, 
	\end{cases}
	\Rightarrow (\gmbu_2, \gmbu_1)  + (\thbu_2, \thbu_1) + \bm{\mathrm{e}}_{4k}^1 \in A_2^{2k} \Rightarrow \gmbu_1 + \thbu_1 \in A_2^k$. 
	This causes a contradiction with $\gmbu + \thbu + \bm{\mathrm{e}}_{4k}^1 \in A_2^{2k}$. 
	Hence, this cases will not happen. 
	
	Then the expected result follows from the cases discussed above. 
\end{proof}

\textbf{Proof of Lemma \ref{lemma:8k+2_WHT&NHT}}:  
	%We first prove that $h$ is a bent function. 
	By (\ref{align:PWHT}), the fragmentary Walsh-Hadamard transform of $h_0$ over $S_4$ at $(\bm{\mathrm{U}}, \bm{\mathrm{V}}) \in \mathbb{F}_2^{8k+2}$ is given by 
	\begin{align*}
		\W_{h_0, S_4}(\bm{\mathrm{U}}, \bm{\mathrm{V}}) 
		=& \sum_{\gmbu \in \Gamma} \sum_{(\bm{\mathrm{X}}, \bm{\mathrm{Y}}) \in C_{\gmbu, A_2^{2k}, E_{\gmbu}}} (-1)^{\bm{\mathrm{X}}\cdot \bm{\mathrm{Y}}+ x_0y_m + \bm{\mathrm{y}}'\cdot \bm{\mathrm{y}}'' + \bm{\mathrm{U}} \cdot \bm{\mathrm{X}} + \bm{\mathrm{V}} \cdot \bm{\mathrm{Y}}} \\ 
		=& \sum_{\gmbu \in \Gamma} \sum_{\xbu \in A_2^{2k} } \sum_{\ztbu \in A_2^{2k} } (-1)^{\xbu \cdot (\gmbu + \ztbu) + (\gmbu_1 + \ztbu_1)\cdot (\gmbu_2 + \ztbu_2) + \ubu \cdot \xbu + \vbu \cdot (\gmbu + \ztbu) } \\ 
		& \hspace{1.0cm} \sum_{x_m \in \mathbb{F}_2} \sum_{y_m \in E_{\bm{\gamma}}} (-1)^{(x_0 + x_m) \cdot y_m+ u_m\cdot x_m + v_m \cdot y_m} \\ 
		=& \sum_{\gmbu \in \Gamma} (-1)^{\vbu \cdot \gmbu + \bm{\gamma}_1 \cdot \bm{\gamma}_2} \sum_{\ztbu \in A_2^{2k}}(-1)^{(\vbu + (\gmbu_2, \gmbu_1)) \cdot \ztbu}  \sum_{\xbu \in A_2^{2k}}(-1)^{(\ubu + \gmbu ) \cdot \xbu} \\ 
		& \hspace{1.0cm} \sum_{y_m \in E_{\bm{\gamma}}} (-1)^{(x_0 + v_m) \cdot y_m} \sum_{x_m \in \mathbb{F}_2}(-1)^{(u_m + y_m) \cdot x_m},  
	\end{align*} 
	where $\ztbu \in \F_2^{2k}$ for $i=1, 2$ and $\ztbu = (\ztbu_1, \ztbu_2)$, and the second identity holds since $\ybu \in C_{\gmbu} (A_2^{2k})$ if and only if $\ybu= \gmbu + \bm{\zeta}$ for $\bm{\zeta} \in A_2^{2k}$, 
	and the last identity holds by the fact that $\ztbu \cdot \bm{\mathrm{x}} = \bm{\zeta}_1 \cdot \bm{\zeta}_2 = 0$ for $\bm{\mathrm{x}}, \bm{\zeta} \in A_2^{2k}$. 
	
	Let $\Phi(\bm{\mathrm{U}}, \bm{\mathrm{V}})$ be a subset of $\Gamma$ defined by $\Phi(\bm{\mathrm{U}}, \bm{\mathrm{V}}) = \{\bm{\gamma} \in \Gamma : u_m \in E_{\bm{\gamma}} \}$. 
	By (\ref{align:inter_Res2}), we have 
	\begin{align*}
		\W_{h_0, S_4}(\bm{\mathrm{U}}, \bm{\mathrm{V}}) =& 2\sum_{\bm{\gamma} \in \Phi(\bm{\mathrm{U}}, \bm{\mathrm{V}})} (-1)^{\vbu \cdot \gmbu + \bm{\gamma}_1 \cdot \bm{\gamma}_2 + u_m \cdot v_m} \sum_{\ztbu \in A_2^{2k}}(-1)^{(\vbu + (\gmbu_2, \gmbu_1)) \cdot \ztbu} \sum_{\xbu \in A_2^{2k}}(-1)^{(\ubu + \gmbu + \bm{\mathrm{e}}_{4k}^{u_m}) \cdot \xbu} \\ 
		=& 2^{4k+1} \sum_{\gmbu \in \Gamma_3(\bm{\mathrm{U}}, \bm{\mathrm{V}})} (-1)^{\vbu \cdot \bm{\gamma} + \bm{\gamma}_1 \cdot \bm{\gamma}_2 + u_m \cdot v_m}, 
	\end{align*}
	where $\Gamma_3(\bm{\mathrm{U}}, \bm{\mathrm{V}})$ is a subset of $\Phi(\bm{\mathrm{U}}, \bm{\mathrm{V}})$ defined by 
	\begin{align*}
	\Gamma_3(\bm{\mathrm{U}}, \bm{\mathrm{V}}) = \{\gmbu \in \Phi(\bm{\mathrm{U}}, \bm{\mathrm{V}}) : \ubu + \bm{\mathrm{e}}_{4k}^{u_m} \in C_{\gmbu} (A_2^{2k}), \bm{\mathrm{v}} \in C_{(\gmbu_2, \gmbu_1)} (A_2^{2k}) \}, 
	\end{align*}
	and the second identity holds by (\ref{align:ex_sum_A^k}). 
	
	For $\gmbu \in \Gamma_3(\bm{\mathrm{U}}, \bm{\mathrm{V}})$, by (\ref{align:inter_Res3}), we know 
	\begin{align*}
		\vbu \cdot \gmbu + \bm{\gamma}_1 \cdot \bm{\gamma}_2 + u_m \cdot v_m
		=& (\bm{\mathrm{u}}'+ \bm{\mathrm{e}}_{2k}^{u_m}) \cdot \bm{\mathrm{u}}'' + (\ubu + \bm{\mathrm{e}}_{4k}^{u_m}) \cdot \bm{\mathrm{v}} + u_m \cdot v_m \\ 
		=& \bm{\mathrm{u}}'\cdot \bm{\mathrm{u}}'' + \bm{\mathrm{U}} \cdot \bm{\mathrm{V}} + u_m \cdot (u_{2k} + v_0). 
	\end{align*} 
%	Then, we have 
%	\[
%	\bm{\mathrm{v}} \cdot \bm{\gamma} + \bm{\gamma}_1 \cdot \bm{\gamma}_2 + u_m \cdot v_m = \bm{\mathrm{u}}'\cdot \bm{\mathrm{u}}'' + \bm{\mathrm{U}} \cdot \bm{\mathrm{V}} + u_m \cdot (u_{2k} + v_0). 
%	\]
	Together with (\ref{align:WHT_h0}), it holds that 
	\[
	\W_{h_0, S_4}(\bm{\mathrm{U}}, \bm{\mathrm{V}}) = |\Gamma_3(\bm{\mathrm{U}}, \bm{\mathrm{V}})| \W_{h_0} (\bm{\mathrm{U}}, \bm{\mathrm{V}}). 
	\]
	Similarly to the proof of $|\Gamma_1(\bm{\mathrm{u}}, \bm{\mathrm{v}})| \le 1$ in Lemma \ref{lemma:8k_WHT&NHT}, we can easily prove $|\Gamma_3(\bm{\mathrm{U}}, \bm{\mathrm{V}})| \le 1$. 
	%Then, the Walsh-Hadamard transform of $h$ at $(\bm{\mathrm{U}}, \bm{\mathrm{V}}) \in \mathbb{F}_2^{4k+2}$ is given by 
%	\begin{align} 
%		\W_h(\bm{\mathrm{U}}, \bm{\mathrm{V}}) = \begin{cases} 
%			- \W_{h_0}(\bm{\mathrm{U}}, \bm{\mathrm{V}}),\ u_m \in E_{\bm{\gamma}}, \bm{\mathrm{u}}' + (u_m, \bm{0}_{2k-1}) \in A^k, \bm{\mathrm{u}}''\in C_{\bm{\gamma}_2} (A_2^k), \bm{\mathrm{v}}' \in C_{\bm{\gamma}_2} (A_2^k), \\ 
%			\hspace{2.4cm} \bm{\mathrm{v}}'' \in C_{\bm{\gamma}_1} (A_2^k)\ \text{for}\ (\bm{\gamma}_1, \bm{\gamma}_2) \in \Gamma, \\ 
%			\W_{h_0}(\bm{\mathrm{U}}, \bm{\mathrm{V}}),\ \hspace{0.4cm} \text{otherwise}. 
%		\end{cases}
%		\label{align:WHT_8k+2}
%	\end{align}
%	Hence, we have $|\W_h(\bm{\mathrm{U}}, \bm{\mathrm{V}})| = |\W_{h_0}(\bm{\mathrm{U}}, \bm{\mathrm{V}})| = 2^{4k+1}$ for all $(\bm{\mathrm{U}}, \bm{\mathrm{V}}) \in \mathbb{F}_2^{8k+2}$. 
%	That is to say, $h$ is a bent function.  
	Then (\ref{align:WHT_8k+2}) follows immediately. 
	
	%Next, we prove that $h$ is negabent. 
	By (\ref{align:PNHT}), the fragmentary nega-Hadamard transform of $h_0$ over $S_4$ at $(\bm{\mathrm{U}}, \bm{\mathrm{V}}) \in \mathbb{F}_2^{8k+2}$ is given by 
	\begin{align*} 
		\N_{h_0, S_4}(\bm{\mathrm{U}}, \bm{\mathrm{V}}) 
		=& \sum_{\gmbu \in \Gamma} \sum_{(\bm{\mathrm{X}}, \bm{\mathrm{Y}}) \in C_{\gmbu, A_2^{2k}, E_{\gmbu}}} (-1)^{\bm{\mathrm{X}}\cdot \bm{\mathrm{Y}}+ x_0y_m + \bm{\mathrm{y}}'\cdot \bm{\mathrm{y}}'' + \bm{\mathrm{U}}\cdot \bm{\mathrm{X}}+ \bm{\mathrm{V}} \cdot \bm{\mathrm{Y}}} \imath^{\wt(\bm{\mathrm{X}}, \bm{\mathrm{Y}})} \\  
		=& \sum_{\gmbu \in \Gamma} \sum_{\ztbu \in A_2^{2k} } \sum_{\xbu \in A_2^{2k}} (-1)^{\xbu \cdot (\gmbu + \ztbu) + (\gmbu_1 + \ztbu_1) \cdot (\gmbu_2 + \ztbu_2) + \ubu \cdot \xbu + \vbu \cdot (\gmbu + \ztbu)} \imath^{\wt(\xbu) + \wt(\gmbu + \ztbu)} \\ 
		& \hspace{1.0cm} \sum_{y_m \in E_{\bm{\gamma}}}(-1)^{(x_0 + v_m)\cdot y_m} \imath^{\wt(y_m)} \sum_{x_m \in \mathbb{F}_2}(-1)^{(u_m+ y_m) \cdot x_m} \imath^{\wt(x_m)} \\
		=& \sum_{\gmbu \in \Gamma} (-1)^{\vbu \cdot \gmbu + \bm{\gamma}_1\cdot \bm{\gamma}_2} \imath^{\wt(\bm{\gamma})} \sum_{\ztbu \in A_2^{2k}}(-1)^{(\vbu + \gmbu + (\gmbu_2, \gmbu_1)) \cdot \ztbu} \imath^{\wt(\ztbu)}  \sum_{\xbu \in A_2^{2k}}(-1)^{(\ubu + \gmbu) \cdot \xbu} \imath^{\wt(\xbu)} \\ 
		& \hspace{1.0cm} \sum_{y_m \in E_{\bm{\gamma}}} [1 + \imath(-1)^{u_m + y_m}] (-1)^{(x_0 + v_m)\cdot y_m} \imath^{\wt(y_m)} \\ 
		=& \sum_{\gmbu \in \Gamma} (-1)^{\vbu \cdot \gmbu + \bm{\gamma}_1\cdot \bm{\gamma}_2} \imath^{\wt(\bm{\gamma})} \sum_{\ztbu \in A_2^{2k}}(-1)^{(\vbu + \gmbu + (\gmbu_2, \gmbu_1)) \cdot \ztbu} \imath^{\wt(\ztbu)} \\ 
		& \hspace{1.0cm} \sum_{y_m \in E_{\bm{\gamma}}} [1 + \imath(-1)^{u_m + y_m}] (-1)^{v_m\cdot y_m} \imath^{\wt(y_m)} \sum_{\xbu \in A_2^{2k}}(-1)^{(\ubu + \gmbu + \bm{\mathrm{e}}_{4k}^{y_m}) \cdot \xbu} \imath^{\wt(\xbu)}, 
	\end{align*} 
	where $\ztbu \in \F_2^{2k}$ for $i=1, 2$ and $\ztbu = (\ztbu_1, \ztbu_2)$, and the second identity holds since $\ybu \in C_{\gmbu} (A_2^{2k})$ if and only if $\ybu= \gmbu + \bm{\zeta}$ for $\bm{\zeta} \in A_2^{2k}$, 
	and the third identity holds by the fact that $\ztbu \cdot \bm{\mathrm{x}} = \bm{\zeta}_1 \cdot \bm{\zeta}_2 = 0$ for $\bm{\mathrm{x}}, \bm{\zeta} \in A_2^{2k}$. 
	
%	For a fixed $\bm{\mathrm{u}} \in \mathbb{F}_2^{4k}$, we know that there exists at most one $\varepsilon \in \mathbb{F}_2$ such that $\bm{\mathrm{u}}' + \bm{\mathrm{e}}_{2k}^{\varepsilon} \in B_2^k$. 
%	The following cases are considered. 
%	
%	(1) Suppose $\bm{\mathrm{u}} + \gmbu + \bm{\mathrm{e}}_{2k}^{\varepsilon} \notin B_2^{2k}$ for any $\gmbu \in \Gamma$ and $\varepsilon \in E_{\gmbu}$. 
%	By (\ref{align:nega_ex_sum_A^k}), it always holds that  
%	\[
%	\sum_{y_m \in E_{\bm{\gamma}}} [1 + \imath(-1)^{u_m + y_m}] (-1)^{v_m\cdot y_m} \imath^{\wt(y_m)} \sum_{\xbu \in A_2^{2k}}(-1)^{(\ubu + \gmbu + \bm{\mathrm{e}}_{4k}^{y_m}) \cdot \xbu} \imath^{\wt(\xbu)} = 0. 
%	\]
%	Hence, we have $\N_h(\bm{\mathrm{U}}, \bm{\mathrm{V}}) = \N_{h_0}(\bm{\mathrm{U}}, \bm{\mathrm{V}})$. 
%	
%	(2) Suppose there is only one $\gmbu \in \Gamma$ such that $\bm{\mathrm{u}} + \gmbu + \bm{\mathrm{e}}_{2k}^{\varepsilon} \in B_2^{2k}$, where $\varepsilon \in E_{\gmbu}$. 
	Let $\Gamma_4(\bm{\mathrm{U}}, \bm{\mathrm{V}})$ be a subset of $\Gamma$ defined by 
	\[
	\Gamma_4(\bm{\mathrm{U}}, \bm{\mathrm{V}}) = \{\gmbu \in \Gamma :  \varepsilon \in E_{\bm{\gamma}}, \ubu + \gmbu + \bm{\mathrm{e}}_{4k}^{\varepsilon} \in B_2^{2k}, \vbu + \gmbu + (\gmbu_2, \gmbu_1) \in B_2^{2k} \}. 
	\] 
	Then, by (\ref{align:nega_ex_sum_A^k}) we have 
	\[
	\N_{h_0, S_4}(\bm{\mathrm{U}}, \bm{\mathrm{V}}) = 2^{4k} \sum_{\gmbu \in \Gamma_4(\bm{\mathrm{U}}, \bm{\mathrm{V}})} (-1)^{\bm{\mathrm{v}} \cdot \bm{\gamma} + \bm{\gamma}_1\cdot \bm{\gamma}_2 + v_m \cdot \varepsilon} \imath^{wt(\bm{\gamma}) + \wt(\varepsilon)}[1 + \imath(-1)^{u_m + \varepsilon}]. 
	\] 
	From Lemma \ref{lemma:Gamma_4}, we know $|\Gamma_4(\Ubu, \Vbu)| \le 1$ for all $(\Ubu, \Vbu) \in \F_2^{8k+2}$. 
	
	(1) If $|\Gamma_4(\Ubu, \Vbu)| = 0$, then $\N_{h_0, S_4}(\Ubu, \Vbu) = 0$. 
	
	(2) If $|\Gamma_4(\Ubu, \Vbu)| = 1$, 
	for $\gmbu \in \Gamma_4(\bm{\mathrm{U}}, \bm{\mathrm{V}})$, from (\ref{align:inter_Res4}) we have 
	\begin{align*} 
		(-1)^{\bm{\mathrm{v}} \cdot \bm{\gamma} + \bm{\gamma}_1 \cdot \bm{\gamma}_2} \imath^{\wt(\gmbu)}  
		=& (-1)^{(\bm{\mathrm{u}}'+ \bm{\mathrm{v}}' + \bm{\mathrm{e}}_{2k}^{\varepsilon}) \cdot (\bm{\mathrm{u}}''+ \bm{\mathrm{v}}'')} \imath^{2k -\wt(\bm{\mathrm{u}} + \bm{\mathrm{e}}_{4k}^{\varepsilon})} \\ 
		=& (-1)^{(\bm{\mathrm{u}}' + \bm{\mathrm{v}}') \cdot (\bm{\mathrm{u}}'' + \bm{\mathrm{v}}'') + (u_{2k} + v_{2k})\cdot \varepsilon} \imath^{2k - \wt(\bm{\mathrm{u}}) - \wt(\varepsilon) + 2\wt(u_0 * \varepsilon)} \\ 
		=& (-1)^{(\bm{\mathrm{u}}' + \bm{\mathrm{v}}') \cdot (\bm{\mathrm{u}}'' + \bm{\mathrm{v}}'') + (u_0 + u_{2k} + v_{2k})\cdot \varepsilon} \imath^{2k - \wt(\bm{\mathrm{u}}) - \wt(\varepsilon)}. 
	\end{align*} 
	Then the fragmentary nega-Hadamard transform of $h_0$ over $S_4$ at $(\Ubu, \Vbu) \in \F_2^{8k+2}$ is given by  
	\begin{align*} 
		\N_{h_0, S_4}(\bm{\mathrm{U}}, \bm{\mathrm{V}}) =& 2^{4k} (1 + \imath(-1)^{u_m + \varepsilon}) (-1)^{(\bm{\mathrm{u}}' + \bm{\mathrm{v}}') \cdot (\bm{\mathrm{u}}'' + \bm{\mathrm{v}}'') + (u_0 + u_{2k} + v_{2k} + v_m)\cdot \varepsilon } \imath^{2k - \wt(\bm{\mathrm{u}})} \\ 
		=& (-1)^{(u_0 + u_{2k} + v_{2k} + v_m)\cdot \varepsilon } (1 + \imath(-1)^{u_m + \varepsilon}) \N_{g_0}(\bm{\mathrm{u}}, \bm{\mathrm{v}}) \\ 
		=& \frac{1}{2}(1+\imath (-1)^{u_0+u_{2k}+v_{2k}+v_m+u_m+\varepsilon}) \N_{h_0} (\Ubu, \Vbu), 
	\end{align*} 
	where $g_0 \in B_{8k}$ is the function defined in (\ref{align:g_0}), and the second identity holds by (\ref{align:Nega_{g_0}}). 
	
	Therefore, (\ref{align:NHT_8k+2}) follows immediately from the two cases discussed above. 
{\hfill $\square$\par}

\end{document}